\documentstyle[oldgerm,amsfonts,amssymb,eqsecnum,pre,aps,preprint]{revtex}

\newcommand{\beq}{\begin{eqnarray}}
\newcommand{\enq}{\end{eqnarray}}
\newcommand{\esp}{\enspace}
\newcommand{\te}{\tilde \epsilon}
\newcommand{\ri}{{\mbox{\tiny I}}}
\newcommand{\rii}{{\mbox{\tiny II}}}
\newcommand{\riii}{{\mbox{\tiny III}}}
\newcommand{\riv}{{\mbox{\tiny IV}}}
\newcommand{\rv}{{\mbox{\tiny V}}}

\begin{document}

\draft

\title{Structure factor of thin films near continuous phase transitions}
\author{R. Klimpel and S. Dietrich}
\address{Fachbereich Physik, Bergische Universit\"at Wuppertal,\\
D-42097 Wuppertal, Federal Republic of Germany}
\maketitle
\begin{abstract}
The two-point correlation function in thin films is studied near the
critical point of the corresponding bulk system. Based on
fieldtheoretic renormalization group theory the dependences of this
correlation function on the lateral momentum, the two distances normal
to the free surfaces, temperature, and film thickness are
determined. The corresponding scattering cross section of X-rays and
neutrons under grazing incidence is calculated. It reveals the various
singularities of the two-point correlation function. 
\end{abstract}
\bigskip


\narrowtext

\section{Introduction}
\label{secI}
Structural properties of condensed matter depend sensitively on the
space dimension $d$. Thin films offer the opportunity to reveal this
dependence. By varying the film thickness $L$ one can interpolate
smoothly between $d=2$ and $d=3$. For crystalline materials this
variation can be accomplished with atomic resolution by using
molecular beam epitaxy \cite{Herman:96}. As an alternative, which is
applicable also for fluids, thin films can be built up via wetting
phenomena where the film thickness is controlled by temperature or
chemical potentials \cite{Dietrich:88}. 
Once such films are prepared the dependence of their structural
properties on the space dimension can be studied particularly clearly
close to phase transitions. For first-order phase transitions the main
influence of a variation of the film thickness is to shift the phase
boundaries in the phase diagram (see, e.g., capillary condensation
\cite{Evans:90} or the shift of the melting curve \cite{Duffy:95}) 
without changing 
much the {\em local} properties of condensed matter. In rare cases,
however, even the {\em character} of the phase transition can change
as function of $L$; see, e.g., the possibility of continuous melting
in $d=2$ \cite{Marcus:96} as opposed to $d=3$ or the
crossover from a first-order phase transition in $d=3$ to a
second-order phase transition in $d=2$ at a certain thickness of a slab 
of the 3-states Potts model \cite{An:88}. 

In the case of {\em first-order} phase transitions the robustness of
the local structural properties with respect to changes of the film
thickness is due to the smallness of the correlation lengths which
characterize these systems and $-$ putting aside possible wetting
phenomena $-$ thus severely limit the propagation of the structural
changes, which necessarily occur near the confining surfaces of the
film, into the interior of the films. In contrast, {\em second-order}
phase transitions are characterized by diverging correlation lengths
which affect not only the location of phase boundaries but in addition
lead to pronounced changes in the local properties even deep in the
interior of the films if the critical point is approached. These
effects are thus not only particularly suitable to shed light on the
aforementioned dependence of the structural properties on space
dimension but they offer an additional advantage: the divergence of
the correlation length as function of temperature upon approaching the
critical point leads to {\em universal} behavior which makes a {\em
quantitative} comparison between theoretical predictions and
experiments much easier as compared with systems exhibiting
first-order phase transitions which are characterized be several
competing length scales of comparable, atomic size which are
difficult to determine accurately and to vary systematically and
independently. 

A sizable body of theoretical research has emerged describing
continuous phase transitions in thin films (see, e.g.,
Refs.
\cite{Fisher:70,Kaganov:72:a,Nauenberg:75,Suzuki:77,Berker:79,Dunfield:80,Brezin:82,Nightingale:82,Barber:83,Brezin:85,Rudnick:85}).
Initiated by the theory of finite-size scaling (see, e.g.,
Refs.
\cite{Binder:72,Fisher:72,Barber:73,Ritchie:73,Capehart:76,Callaway:81}),
inter alia the shift $T_c(L)$ of the critical temperature with respect
to its bulk value $T_c \equiv T_c(L = \infty)$
\cite{Allan:70,Bray:78,Nakanishi:83}, the magnetization
\cite{Kaganov:72:b,Rudnick:82:a} as well as the free energy, the
Casimir force, and the specific heat
\cite{Krech:91,Krech:92:a,Eisenriegler:93,Krech:94,Sutter:94,Chen:95,Krech:96}
have been analyzed. Here we emphasize that in order to observe
universal film behavior the thicknesses $L$ of the films have still to
be large on an atomic scale. This is assumed to be the case throughout
our analysis. The analytic description of the dimensional crossover
between $d=3$ critical behavior near $T_c$ and the $d=2$ critical
behavior near $T_c(L)$ poses still a challenge
\cite{Schmeltzer:85,OConnor:91:a} which has not yet been overcome with
satisfactory quantitative accuracy. Numerous experiments (see, e.g.,
Refs.
\cite{Gasparini:92,Ledermann:93,Nissen:93,Mehta:97,Andrieu:98,Henkel:98})
and simulations (see, e.g. Refs. \cite{Binder:88,Binder:95,Binder:97})
have been carried out to test these theoretical predictions. They lend
support to the finite size scaling theory but still pose a puzzle as
far as detailed quantitative agreement is concerned.

The vast majority of these studies is devoted to integral or excess
quantities without spatial resolution. However, the studies of {\em
local} critical properties, such as of one- and two-point correlation
functions, near a {\em single} surface have revealed a wealth of
universal phenomena featuring numerous surface critical exponents and
interesting crossover phenomena $-$ on the scale of the bulk correlation
length $\xi$ $-$ between surface and bulk critical behavior
\cite{Binder:83,Diehl:86}; the integral and excess quantities offer
either no or only very limited access to these local properties. 

The successful development of surface specific X-ray and neutron
scattering techniques based on exploiting total external reflection at
grazing incidence has proven to be very fruitful, inter alia, for
facilitating the quantitative comparison between experiments and
theoretical predictions of the local critical behavior near interfaces
\cite{Dietrich:95,Dosch:92}. These scattering techniques allow one to
determine order parameter profiles normal to the surface and the depth
resolved lateral two-point correlation function. In the present
context such experiments have been carried out successfully for the binary
alloy $Fe_3 Al$ \cite{Mailaender:90,Dosch:92:a,Dosch:92:b} and, by
using truncation rod scattering, for $Fe Co$ \cite{Krimmel:97:b} which
exhibit continuous order-disorder transitions in the bulk. In the case
of $Fe_3 Al$ the cusplike surface singularities of the momentum and
temperature dependence of the two-point correlation function turned
out to be in excellent agreement with the theoretical predictions
\cite{Dietrich:83:b,Dietrich:84}. The fact, that due to the occurrence
of surface segregation suitable choices for the crystallographic
orientation of the surface allows one to switch between the different
surface universality classes corresponding to free boundary conditions
and boundary conditions with surface fields, respectively, of the same
bulk sample \cite{Drewitz:97,Leidl:98}, offers wide ranges of
interesting comparative studies.

In view of these developments and in view of the increasing
availability of powerful synchrotron and neutron sources it appears
promising to extend these studies of local critical properties to thin
films. There are several predictions concerning the behavior of
one-point correlation functions in thin films such as order-parameter
profiles \cite{Kaganov:72:b,Rudnick:82:a,Krech:96} and energy-density
profiles \cite{Eisenriegler:96}. However, on the level of the two-point
correlation function so far only very little is known. This function
depends on the lateral distance ${\bf x}_\| = {\bf x}^{(2)}_\| - {\bf
x}^{(1)}_\|$ between the two points ${\bf x}_1 = ({\bf x}^{(1)}_\|,
z_1)$ and ${\bf x}_2 = ({\bf x}^{(2)}_\|, z_2)$ (or equivalently the
lateral momentum ${\bf p}$ corresponding to the $d-1$ translationally
invariant 
directions), the coordinates $z_1$ and $z_2$ perpendicular to the
parallel surfaces of the film, the film thickness $L$, and temperature
$t = (T-T_c)/T_c$ (or equivalently the bulk correlation length $\xi =
\xi_0 t^{-\nu}$). Since a full sweep of this large parameter space is
practically not possible for computer simulations, we have applied
fieldtheoretic techniques which provide analytic access to the full
parameter space. This approach encompasses nonperturbative features
such as scaling properties and short-distance expansions as well as an
explicit and systematic perturbative result to first order
in $\epsilon = 4 - d$. The latter serves as to corroborate the
nonperturbative results and to provide numerical results which are not
accessible by general arguments. These explicit calculations are
carried out for the fixed point of the so-called ordinary transition
for both confining surfaces in the classification scheme of surface
critical phenomena \cite{Diehl:86} corresponding to free boundary
conditions on both sides. This is applicable to thin antiferromagnetic
films near their N\'eel temperature, to ferromagnetic films near their
Curie temperature in the absence of external bulk and surface fields,
and to thin films of binary alloys near their continuous
order-disorder transitions. Among the numerous order-disorder phase
transitions in binary alloys only a few are of second order including
$Fe_3 Al$ \cite{Guttmann:56,Guttmann:69:a,Guttmann:69:b}, $FeCo$
\cite{Oyedele:77}, $CuZn$ \cite{AlsNielsen:67}, and $FeAl$
\cite{Guttmann:69:a}. 
Both the $B2-DO_3$ transition in $Fe_3Al$ and the $A2-B2$ transitions
in $FeCo$, $CuZn$, and $FeAl$ belong to the Ising universality class
\cite{Diehl:97}. For the $A2-B2$ transitions it is predicted
theoretically that the $(110)$ surface belongs to the surface universality
class of the ordinary transition whereas the $(100)$ surface exhibits
the so-called normal transition associated with the presence of an
effective surface field \cite{Drewitz:97,Leidl:98}. Indeed truncation
rod scattering at the $FeCo$ $(100)$ surface has provided clear
evidence for the presence of an effective surface field
\cite{Krimmel:97:b} above $T_c$ although the expected
associated crossover from ordinary to normal critical behavior
\cite{Ritschel:96} could not yet been resolved experimentally in an
unequivocal way. The results of the diffuse scattering of X-rays under
grazing incidence from the $(1 \bar 1 0)$ surface (equivalent to the
$(110)$ surface) of $Fe_3Al$
\cite{Mailaender:90} are in excellent agreement with the theoretical
predictions \cite{Dietrich:83:b,Dietrich:84} for the ordinary
transition. But even for $Fe_3Al$ $(1 \bar 1 0)$ a residual order
parameter above $T_c$ has been reported
\cite{Mailaender:90,Dosch:92:a}. Thus it 
still remains to be seen theoretically whether for the $B2-DO_3$
transition in $Fe_3Al$, in contrast to the $A2-B2$ transition in
$FeCo$, the $(1 \bar 1 0)$ surface can support a weak effective
surface field. In view of this state of affairs our present result are
expected to be closely applicable to thin films of $Fe_3Al$, $FeCo$,
$CuZn$, and $FeAl$ bounded by $(110)$ surfaces on both sides. Among
them $Fe_3Al$ and $FeCo$ appear to be the most promising candidates
because the others exhibit strong surface segregation. For an
assessment of the possibilities to probe critical magnetic surface
transitions by grazing incidence of neutrons see
Ref. \cite{Dosch:93}. 

In view of
the aforementioned difficulties concerning the analytic description of
the dimensional crossover we confine our analysis to the temperature
range $T \ge T_c$. We note that elements of the perturbation theory
for thin critical films can be found in
Ref. \cite{Nemirovsky:86:a}. We had, however, to carry out our own
approach because the representation given in
Ref. \cite{Nemirovsky:86:a} is not suited for making predictions for
the scattering experiments and because Ref. \cite{Nemirovsky:86:a}
contains errors. Finally we note that experience tells that
calculations carried out for the spherical model, as have been done
for the present system \cite{Allen:94}, lack the quantitative
reliability needed for comparison with experiments and simulations.

In order to encourage future scattering experiments for critical thin
films and to facilitate an explicit quantitative comparison of such
data with the present theoretical predictions, we have calculated the
singular contributions to the scattering cross section for X-ray and
neutron scattering under the condition of grazing incidence based on
our results for the critical two-point correlation function in thin
films. This allows us to describe the conditions under which the
various singularities of the two-point correlation function become
visible in scattering data.

This introduction is followed be three sections, the Summary, and four
Appendices. In Sec. \ref{model} we introduce the fieldtheoretical model.
The two-point correlation function is discussed in
Sec. \ref{correlation} and in Sec. \ref{crosssection} we investigate
the scattering cross section. Relations between bulk and film
amplitudes are derived in Appendix \ref{amplitudes}, explicit one-loop
results are presented in Appendix \ref{one_loop_results}, and
Appendices \ref{details_crosssection} and \ref{ssss} contain details
required for the calculation of the scattering cross section.   


\section{Field-theoretical model}
\label{model}
The leading critical behavior in a film follows from the statistical
weight $\exp (-{\cal H} \{ \Phi \})$ for the configuration $\Phi ({\bf
x}) = (\phi_a ({\bf x}), a=1, \dots , n)$ of a $n$-component
field, which is proportional to the order parameter, where
\cite{Diehl:86,Krech:92:a,Krech:94} 
\beq
\label{free_energy}
{\cal H} \{ {\Phi} ({\bf x}) \} & = & \int d^{d-1}x_\Vert \int_0^L dz \left( 
 \frac{1}{2} (\nabla \Phi)^2 + \frac{\tau}{2} \Phi^2 
+ \frac{g}{4!} (\Phi^2)^2 - {\bf h} \cdot \Phi \right) \\
&+& \int d^{d-1}x_\Vert \left( \frac{c}{2} \Phi^2 (z=0) - {\bf h}_1 \cdot
\Phi(z=0) 
+\frac{c}{2} \Phi^2 (z=L) - {\bf h}_1 \cdot
\Phi(z=L) \right) \nonumber
\enq
with space dimension $d$ and position vector ${\bf x}=({\bf
x_\Vert},z)$ of $d-1$ parallel and one perpendicular components. The
$z$ integration extends over the interval $[0,L]$, where $z=0$ and
$z=L$ give the positions of the film surfaces. $\tau$ is the
temperature parameter such that in the bulk $\tau = 0$ marks the
transition temperature within mean-field theory. The coupling constant
$g>0$ ensures the stability of the statistical weight below the
transition temperature, i.e., for $\tau < 0$. $c$ denotes the surface 
enhancement, ${\bf h}$ and ${\bf h}_1$ are bulk and surface fields,
respectively. We focus on the ordinary transition at zero fields, 
i.e., we adopt the fixed point value $c=\infty$ for the surface
enhancement and set ${\bf h}={\bf h}_1=0$. After carrying out a Fourier
transformation with respect to the $d-1$ directions exhibiting
translational invariance parallel to the surfaces the mean-field
propagator for the disordered phase  $(\tau > 0)$ in
$p$-$z$-representation is given by \cite{Diehl:86,Krech:95}  
\beq
\label{propagator_dirichlet}
G_D(p,z_1,z_2,L,\tau) & = & \int d^{d-1}x_\| \esp e^{i {\bf p} \cdot
{\bf x}_\|} \esp \langle \Phi ({\bf x}_\|,z_1) \Phi (0,z_2 ) \rangle \\
& = & \frac{1}{2 b} \Big( e^{-b|z_1-z_2|} -
e^{-b(z_1+z_2)} \nonumber \\
&& \quad + \left. \frac{e^{-b(z_1-z_2)} +
e^{-b(z_2-z_1)} - e^{-b(z_1+z_2)} - e^{b(z_1+z_2)}}{e^{2bL}-1} \right)
, \esp b=\sqrt{p^2+\tau}.
\nonumber 
\enq
The first exponential function corresponds to the bulk part followed
by the contribution from the surface at $z=0$. Both exponentials
together give the propagator for the ordinary transition of the
semi-infinite system $(L=\infty)$. The remaining ratio carries the $L$
dependence. The propagator satisfies the Dirichlet boundary conditions
$G_D(z=0)=0=G_D(z=L)$. Equation (\ref{propagator_dirichlet})
represents the mean-field approximation for the two-point correlation
function in the film corresponding to the critical behavior in $d=4$. 
The non-Gaussian fluctuations in $d=3$ are taken into account
approximately by the one-loop correction which amounts to the first
term in a systematic expansion in terms of $\epsilon = 4 - d$:
\beq
\label{one_loop}
&&G_{bare}(p,z_1,z_2,L,\tau,g) = G_D(p,z_1,z_2,L,\tau) \\
& - & \frac{g}{2} \frac{n+2}{3}
\int \frac{d^{d-1}q}{(2\pi)^{d-1}} \int\limits_0^L dz \esp
G_D(p,z_1,z,L,\tau) \esp G_D(q,z,z,L,\tau) \esp G_D(p,z,z_2,L,\tau) +
{\cal O}(g^2). \nonumber 
\enq
As regularization scheme we use dimensional regularization by analytic
continuation in the space dimension $d=4-\epsilon $. As long as $z_1$
and $z_2$ are both off the surfaces only bulk singularities occur. We
absorb the corresponding poles in $\epsilon$ by minimal subtraction
through the standard $Z$ factors:
\beq
\label{renormalized_parameters}
\phi = Z^{1/2}_{\phi} \phi^R , \quad g = \mu^{\epsilon} 2^d \pi^{d/2}
Z_u u , \quad \tau = \mu^2 Z_t t , 
\enq
where $\mu$ is the momentum scale and the bulk Z-factors are
\cite{Amit:78} 
\beq
\label{bulk_z_factors}
Z_{\phi} = 1 + {\cal O} (u^2), \quad Z_u = 1 + \frac{n+8}{3}
\frac{u}{\epsilon} + {\cal O} (u^2), \quad Z_t = 1 + \frac{n+2}{3}
\frac{u}{\epsilon} + {\cal O} (u^2). 
\enq
The renormalized correlation function reads (see
Eq. (\ref{renormalized_parameters})) 
\beq
\label{renormalization}
G(p,z_1,z_2,L,t,u;\mu) & = & Z_{\phi}^{-1}
G_{bare}(p,z_1,z_2,L,\tau,g)
\enq
which is valid in all orders of perturbation theory. The solution of
the corresponding renormalization group equation leads to the
following scaling property:
\beq
\label{scaling_form}
G(p,z_1,z_2,L,t;\mu) & = & {\cal G}_\ri p^{-1+\eta} g_\ri
(p \xi, z_1/\xi, z_2/\xi, L/\xi) .
\enq
This holds at the fixed point $u^\ast = \frac{3}{n+8}\epsilon +
{\cal O} (\epsilon^2)$ and involves the bulk correlation length
$\xi=\xi^+_0 t^{-\nu}$, the exponents $\eta = {\cal O} (\epsilon^2)$,
and $\nu = \frac{1}{2} + \frac{1}{4} \frac{n+2}{n+8} \epsilon + 
{\cal O} (\epsilon^2)$. With suitable
normalization (see, c.f., Eq. (\ref{norm_ri})) the scaling
function $g_\ri$ is universal. The amplitude ${\cal G}_\ri$, which is
fixed by this normalization, and the amplitude $\xi^+_0$ carry the 
nonuniversal scaling factors. We fix $\xi^+_0$ by defining $\xi$ as
the so-called true correlation length \cite{Tarko:75} so that $\xi^+_0
= \mu^{-1} (1+ \frac{1}{4} \frac{n+2}{n+8} (1-C_E) \epsilon +
{\cal O} (\epsilon^2))$. This expression for $\xi^+_0$ allows one to
express the momentum scale $\mu$ introduced in
Eq. (\ref{renormalized_parameters}) in terms of the experimentally
accessible, nonuniversal amplitude $\xi^+_0$:
\beq
\label{mu_fixing}
\mu & = & (\xi^+_0)^{-1} \Big( 1 + \frac{1}{4} \frac{n+2}{n+8} (1-C_E)
\epsilon + {\cal O} (\epsilon^2) \Big) .
\enq
If subsequent formulae contain the momentum scale $\mu$ explicitly it
is to be replaced by Eq. (\ref{mu_fixing});
moreover we omit $\mu$ from the explicit list of variables of $G$.  

Depending on the problem under consideration it is often
advantageous to use different but equivalent representations of the
correlation function such as
\beq
\label{alternatives_1}
G(p,z_1,z_2,L,t) & = & {\cal G}_\rii z_1^{1-\eta} g_\rii (p z_2,
z_1/\xi, z_2/\xi, z_2/L) ,
\enq
\beq
\label{alternatives_2}
G(p,z_1,z_2,L,t) & = & {\cal G}_\riii L^{1-\eta} g_\riii (p z_1, z_1/L,
z_2/L, L/\xi) ,
\enq
\beq
\label{alternatives_3}
G(p,z_1,z_2,L,t) & = & {\cal G}_\riv \xi^{1-\eta} g_\riv (p L,
p z_1, p z_2, \xi/L) ,
\enq
and
\beq
\label{alternatives_4}
G(p,z_1,z_2,L,t) & = & {\cal G}_\rv p^{-1+\eta} g_\rv (p \xi,
p (z_1-z_2), p (z_1+z_2), L/\xi) . 
\enq
The nonuniversal amplitudes ${\cal G}_x$ and the universal scaling
functions $g_x$, $x={\mbox{\tt I,II,III,IV,V}}$, are fixed by the
following normalizations: 
\beq
\label{norm_ri}
\lim_{\alpha \to \infty} \lim_{\beta \to \infty} \lim_{\delta \to \infty}
g_\ri (\alpha,\beta,\gamma=\beta,\delta) =1,
\enq
\beq
\label{norm_rii}
\lim_{\alpha \to 0} \lim_{\beta \to 0} \lim_{\delta \to 0} 
g_\rii(\alpha,\beta,\gamma = \beta,\delta) =: g_\rii(0,0,0,0)=1,
\enq
\beq
\label{norm_riii}
\lim_{\alpha \to 0} \lim_{\delta \to 0}  
g_\riii (\alpha,\beta=1/2,\gamma=\beta=1/2,\delta) =1,
\enq
\beq
\label{norm_riv}
\lim_{\delta \to 0} \lim_{\alpha \to 0}  
g_\riv(\alpha,\beta=\alpha/2,\gamma=\beta=\alpha/2,\delta) =1, 
\enq
and 
\beq
\label{norm_rv}
\lim_{\beta \to 0} \lim_{\alpha \to \infty} 
\lim_{\gamma \to \infty} \lim_{\delta \to \infty}
g_\rv(\alpha,\beta,\gamma,\delta) =: g_\rv(\infty,0,\infty,\infty)=1.
\enq
The universal scaling functions $g_x$ can be expressed in terms of 
each other because in Eqs. (\ref{scaling_form}) and
(\ref{alternatives_1}) - (\ref{alternatives_4})
the left hand side 
is the same quantity and the sets of scaling variables are complete,
i.e., from each set one can form any of the others by a suitable
combination of variables. 

Since the nonuniversal amplitudes ${\cal G}_x$ correspond to the same
correlation function $G(p,z_1,z_2,L,t)$ and because the scaling
functions fixed by the normalizations in Eqs. (\ref{norm_ri}) -
(\ref{norm_rv}) are universal, their ratios ${\cal G}_x/{\cal G}_{x'}$
are universal numbers. Thus the knowledge of one of them and of the
corresponding universal scaling functions determines all the others.

Moreover, as discussed in Appendix \ref{amplitudes}, all nonuniversal
amplitudes ${\cal G}_x$ are determined by any pair of nonuniversal
scale factors which characterize the critical {\em bulk} properties. A
transparent and experimentally directly accessible choice for the
latter is the nonuniversal amplitude $B$ of the leading temperature
singularity of the field $\langle \phi ({\bf x}) \rangle$ in the bulk
below $T_c$, 
\beq
\label{bulk_op}
\langle \phi ({\bf x}) \rangle & = & B (-t)^\beta ,
\enq
and the amplitude $\xi_0^+$ of the true correlation length above
$T_c$. In terms of these quantities one has 
\beq
\label{G_e_B_xi0}
{\cal G}_\rv & = & B^2 (\xi_0^+)^{d-2+\eta} {\cal U}
\enq
where ${\cal U}$ is a universal number, whose value ${\cal U} \simeq
1.58$ is derived in Appendix \ref{amplitudes} based on
Eq. (\ref{norm_rv}). In the following most of our analysis focuses on
the scaling function $g_\rii$ used in Eq. (\ref{alternatives_1}). For
that case one finds (see Appendix \ref{amplitudes}) the {\em
universal} ratio
\beq
\label{G_e_G_b}
{\cal G}_\rii/{\cal G}_\rv & = & 2 \Big( 1 + \epsilon \frac{n+2}{n+8} +
{\cal O} (\epsilon^2) \Big) .
\enq
With these results we finally obtain
\beq
\label{G_final}
G(p,z_1,z_2,L,t) & = & B^2 (\xi_0^+)^{d-1} {\cal R}
(z_1/\xi_0^+)^{1-\eta} g_\rii (p z_2, z_1/\xi, z_2/\xi, z_2/L)
\enq
where ${\cal R} = 2 {\cal U} \Big( 1 + \epsilon \frac{n+2}{n+8} + {\cal O}
(\epsilon^2) \Big) \simeq 4.21$ is a universal number. Thus in all our
subsequent formulae for {\em film} properties their {\em absolute}
values are determined and fixed by the two nonuniversal {\em bulk}
amplitudes $B$ and $\xi_0^+$.

The actual order parameter $O\!P$ for a particular second order phase
transition is proportional to the field $\phi$ introduced in
Eq. (\ref{free_energy}), i.e., $O\!P({\bf x}) = {\it b} \phi ({\bf x})$.
The value of ${\it b}$ depends on the particular system (binary alloy,
liquid, ferromagnet etc.). Moreover, any rescaling of ${\it b}$ by a
dimensionless number renders another order parameter $O\!P$ which is
equally valid for describing the singular behavior of the phase
transition. We emphasize that Eqs. (\ref{alternatives_1}),
(\ref{G_e_B_xi0}), and (\ref{G_e_G_b}) remain valid if $G$ is replaced
by $\langle O\!P ({\bf x}) O\!P ({\bf x}') \rangle$, $\langle \phi
({\bf x}) \rangle$ by $\langle O\!P ({\bf x}) \rangle$, and $B$ by
$B'= {\it b} B$; these replacements have to be carried out if the
present fieldtheoretic results are used to interpret, e.g., the
intensity of scattered X-rays or neutrons (see, c.f.,
Sec. \ref{crosssection}). The actual choice of the $O\!P$, as it
enters into the expression for the scattering cross section, is borne
out and tight to the relation $\langle O\!P ({\bf x}) \rangle = B'
(-t)^\beta $.


\section{Explicit properties of the two-point correlation function}
\label{correlation}
The discussion of the correlation function consists of three
parts. First, we set $z_1=z_2$ and analyze its nonanalytic behavior
in certain limits. Then, we take into account the case $z_1 \not=
z_2$, which serves to understand the correlations perpendicular to the
surfaces. Moreover, the discussion of this latter case turns out to be
very useful for carrying out the integrations appearing in the
scattering cross section to be analyzed in
Sec. \ref{crosssection}. The film excess susceptibility is discussed
in the last part.


\subsection{Lateral two-point correlation function for $z_1=z_2$}
\label{correlation_zz}
In order to investigate various asymptotic properties of the lateral
behavior of the two-point correlation function we resort to short
distance expansions (SDE) \cite{Diehl:81:a}, distant wall corrections 
(DWC) \cite{Eisenriegler:96}, and results of the perturbation theory
supported by appropriate exponentiations of the explicit
$\epsilon$-expansion results. With $z_1=z_2=z$, in the present context
a representation of the form 
\beq
\label{corr_fctn_z1_eq_z2}
G(p,z,L,t) & = & {\cal G}_\rii z^{1-\eta} g(pz,z/\xi,z/L) 
\enq
is useful. According to Eq. (\ref{alternatives_1}) one has
$g(u,v,w)=g_\rii(u,v,v,w)$ with $g(0,0,0)=1$ (Eq. (\ref{norm_rii})). For
semi-infinite systems, i.e., $L=\infty$ the SDE in the cases $t=0$, $p
\to 0$ and $p=0$, $t \to 0$ \cite{Gompper:84,Gompper:86:a} leads to
the asymptotic behaviors   
\beq
\label{corr_p}
G(p,z,L=\infty,t=0) & = & {\cal G}_\rii z^{1-\eta} g_1(u=pz) \\
& _{\longrightarrow \atop p \to 0} & {\cal G}_\rii z^{1-\eta}[1 + A_1
(pz)^{-1+\eta_\|} + \dots] 
\nonumber 
\enq
and
\beq
\label{corr_t}
G(p=0,z,L=\infty,t) & = & {\cal G}_\rii z^{1-\eta}g_2(v=z/\xi) \\
& _{\longrightarrow \atop t \to 0} & {\cal G}_\rii
z^{1-\eta} [1 + B_1 (z/\xi)^{-1+\eta_\|} + \dots] \nonumber \\
& = & {\cal G}_\rii z^{1-\eta} [1 + B_1 (z/\xi^+_0)^{-1+\eta_\|}
t^{-\gamma_{11}} + \dots], \nonumber 
\enq
respectively, with $\gamma_{11}=\nu (\eta_\|-1)$, $g_1(u) =
g(u,v=0,w=0)$, $g_1(0)=1$, $g_2(v) = g(u=0,v,w=0)$, and $g_2(0)=1$. In
the case $p=0$, $t=0$ one has
\beq
\label{corr_L_1}
G(p=0,z,L,t=0) & = & {\cal G}_\rii z^{1-\eta} g_3(w=z/L)
\enq
with $g_3(w)=g(u=0,v=0,w)$ so that $g_3(0)=1$. In order to infer the first
nontrivial dependence on $L$ for $L \to \infty$, according to
Eq. (\ref{corr_L_1}) one can equally consider the limit $z \to 0$
for $L$ fixed. To this end we consider the SDE of the renormalized
film correlation function in real space:
\beq
\label{sde_dwc}
\langle \phi({\bf x}_\|,z) \phi(0,z) \rangle & _{\longrightarrow
 \atop z \to 0} & \mu^{-2} (\mu z)^{2(x_s-x)} 
 \langle \phi_\perp({\bf x}_\|) \phi_\perp(0) \rangle \\
& = & \mu^{d-2} (\mu z)^{2(x_s-x)} (\mu x_\|)^{-2x_s} Y(x_\|/L) \nonumber.
\enq
Here $\phi_\perp$ denotes the normal derivate of $\phi$ taken at one
of the surfaces and $Y(y)$ is a dimensionless scaling function for the
film which is universal up to nonuniversal prefactor. The scaling 
dimensions of $\phi$ and $\phi_\perp$ are $x=\frac{1}{2}(d-2+\eta)$
and $x_s=\frac{1}{2}(d-2+\eta_\|)$, respectively. 
The scaling function $Y(x_\|/L)$ describes the influence of the
distant wall at $z=L$ on the lateral correlations close to the near
wall at $z=0$. In order to obtain its leading asymptotic behavior for
$x_\|/L \to 0$ we use the identity
\beq
\label{G_L_identity}
G(p=0,z,L,t=0) & = & G(p=0,z,L=\infty,t=0) \\
&& - \int_L^\infty \frac{\partial G(p=0,z,L',t=0)}{\partial L'} d L'
. \nonumber 
\enq
The first term on the rhs is equal to ${\cal G}_\rii z^{1-\eta}$
(compare Eqs. (\ref{corr_p}) and (\ref{corr_t})). The leading
correction is given by using the SDE in Eq. (\ref{sde_dwc}) for the
second term:
\beq
\label{leading_L_correction}
-\int\limits_L^\infty \frac{\partial G(p=0,z,L',t=0)}{\partial L'} d L'&=&-
\int d^{d-1} x_\| \int\limits_L^\infty d L' \frac{\partial}{\partial L'} 
\langle \phi({\bf x}_\|,z) \phi(0,z) \rangle \\
& _{\longrightarrow \atop L \to \infty} & - \int d^{d-1} x_\|
\int\limits_L^\infty d L' \frac{\partial}{\partial L'} \mu^{d-2} (\mu
z)^{2(x_s-x)} (\mu x_\|)^{-2x_s} Y(x_\|/L') \nonumber \\
& = & \mu^{-1} (\mu z)^{1-\eta} \Big( \frac{z}{L}
\Big)^{-1+\eta_\|}\tilde C 
\nonumber
\enq
with $\tilde C = \frac{1}{\eta_\|-1} \int d^{d-1}y \esp
y^{-(d-3+\eta_\|)} Y'(y)$. Thus we find $g_3(w \to 0) = 1 + C_1
w^{-1+\eta_\|}$ where $C_1 = \tilde C \mu^{-\eta}/{\cal G}_\rii$ is a
universal number, i.e.,
\beq
\label{corr_L}
G(p=0,z,L \to \infty,t=0) & = & {\cal G}_\rii z^{1-\eta}[1 + C_1
(z/L)^{-1+\eta_\|} + \dots] .
\enq
Finally we note that due to the normalization $g(0,0,0)=1$ the scaling
function $g(u,v,w)$ is given by the ratio
$g(pz,z/\xi,z/L)=G(p,z,L,t)/G(p=0,z,L=\infty,t=0)$ from which the
prefactors ${\cal G}_\rii z^{1-\eta}$ appearing in
Eq. (\ref{corr_fctn_z1_eq_z2}) drop out. The $\epsilon$-expansions of
the amplitudes of the leading asymptotic terms follow from
Eqs. (\ref{g_p_to_0}), (\ref{g_t_to_0}), and (\ref{g_L_to_oo})) in
Appendix \ref{zz_appendix}:
\beq
\label{A1_B1_C1}
A_1 & = & - \Big[ 1 + \epsilon \frac{n+2}{n+8} \left( 1 - C_E - \ln 2
\right) + {\cal O} (\epsilon^2) \Big] , \\
B_1 & = & - \Big[ 1 + \epsilon \frac{n+2}{n+8} \left( 1 - C_E \right)
+ {\cal O} (\epsilon^2) \Big] , \nonumber \\
C_1 & = & - \Big[ 1 + \epsilon \frac{n+2}{n+8} \left( \frac{\pi^2}{18} - C_E
+2(S_2+I_1) - 1\right)
+ {\cal O} (\epsilon^2) \Big] . \nonumber
\enq
$C_E \approx 0.5772$ is Euler's constant, $S_2 \simeq 0.083$ and $I_1
\simeq 0.287$ are given by Eq. (\ref{S2_I1}) in Appendix
\ref{zz_appendix}. Within the $\epsilon$-expansion the full forms of
the scaling functions $g_1(u)$, $g_2(v)$, and $g_3(w)$ can be found in
Appendix \ref{zz_appendix} (see Eqs. (\ref{g_p}) - (\ref{g_L})).

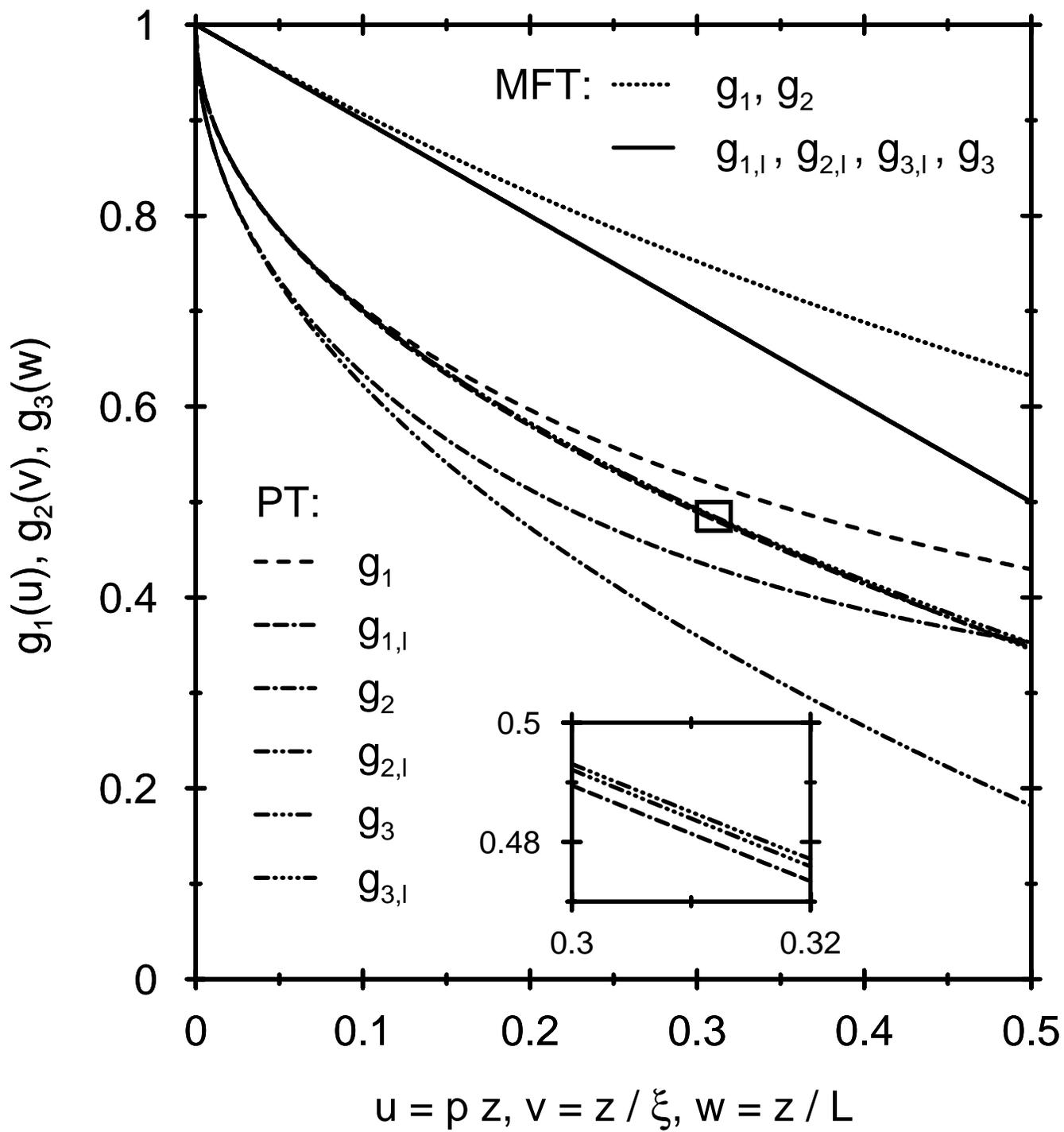
\begin{figure}
\caption{
The three scaling functions describing the lateral
correlations $g(pz, z/\xi, z/L)$ in the film
(Eq. (\ref{corr_fctn_z1_eq_z2})) for the limiting cases $p=0$, $\xi =
\infty$, or $L=\infty$:  
$g_1(u=pz) \equiv g(pz,0,0)$ ($T=T_c, L= \infty$, Eq. (\ref{corr_p})), 
$g_2(v=z/\xi) \equiv g(0,z/\xi,0)$ ($p=0, L=\infty$,
Eq. (\ref{corr_t})), and $g_3(w=z/L) \equiv g(0,0,z/L)$ ($p=0,
T=T_c$, Eq. (\ref{corr_L_1})). The two uppermost curves correspond to
the mean-field results for $g_i$, $i=1,2,3$, and to their 
{\em l}eading behavior $g_i(x_i \to 0) = g_{i,l} (x_i)$, respectively,
with $x_1=u$, $x_2=v$, and $x_3=w$; within MFT $g_1=g_2$,
$g_{1,l}=g_{2,l}=g_{3,l}$, and $g_3=g_{3,l}$. The lower six curves
correspond to $g_i(x_i)$ (Eqs. (\ref{g_p}), (\ref{g_t}), and
(\ref{g_L})) and $g_{i,l} (x_i)$ 
as obtained by {\em p}erturbation {\em t}heory for $d=3$. The
difference between $g_3$ and $g_{3,l}$ is revealed only in the
inset: $g_{1,l} (u) = 1 + A_1 u^{-1+\eta_\|}$, $g_{2,l}(v) = 1 + B_1
v^{-1+\eta_\|}$, and $g_{3,l}(w) = 1+ C_1 w^{-1+\eta_\|}$. Within MFT
one has $A_1=B_1=C_1=-1$ (Eq. (\ref{A1_B1_C1})) and $\eta_\|=2$
whereas for $(n,d)=(1,3)$ PT yields $A_1 \simeq -0.9099$, $B_1 \simeq
-1.1409$, $C_1 \simeq -0.9035$, and $\eta_\| \simeq 1.48$. For
vanishing scaling arguments all scaling functions attain 1.
}
\label{fig_p_t_L}
\end{figure}

In Fig. \ref{fig_p_t_L} we display the three scaling functions $g_i$,
$i=1,2,3$, (Eqs. (\ref{g_p}) - (\ref{g_L})) corresponding to
Eqs. (\ref{corr_p}), (\ref{corr_t}), and (\ref{corr_L_1}) as obtained
within mean-field theory (MFT), i.e., for $\epsilon=0$ and from
renormalization group guided perturbation theory (PT) as well as their
leading behavior $g_i (x_i \to 0) = g_{i,l}(x_i)$, $x_1=u, x_2=v, x_3
=w$. Within MFT the three scaling functions have
the same limiting form for small scaling variables with
$A_1=B_1=C_1=-1$ and the critical exponent 
$\eta_\|=2$. Beyond MFT, in Fig. \ref{fig_p_t_L} we use
$\eta_\|=1.48$ as the best available 
estimate \cite{Diehl:86} whereas the amplitudes are evaluated in first
order in $\epsilon$ (Eq. (\ref{A1_B1_C1}) for $(n,\epsilon)=(1,1)$)
so that $A_1 \simeq -0.9099$, $B_1 \simeq -1.1409$, and $C_1 \simeq
-0.9035$. Within mean-field theory $g_1=g_2$ and the leading
asymptotic behavior $g_{3,l}$ provides already the full scaling
function $g_3$. Beyond MFT there is a small difference between $g_3$
and $g_{3,l}$. This difference is much bigger for the scaling
functions $g_1$ and $g_2$ describing the semi-infinite system.  

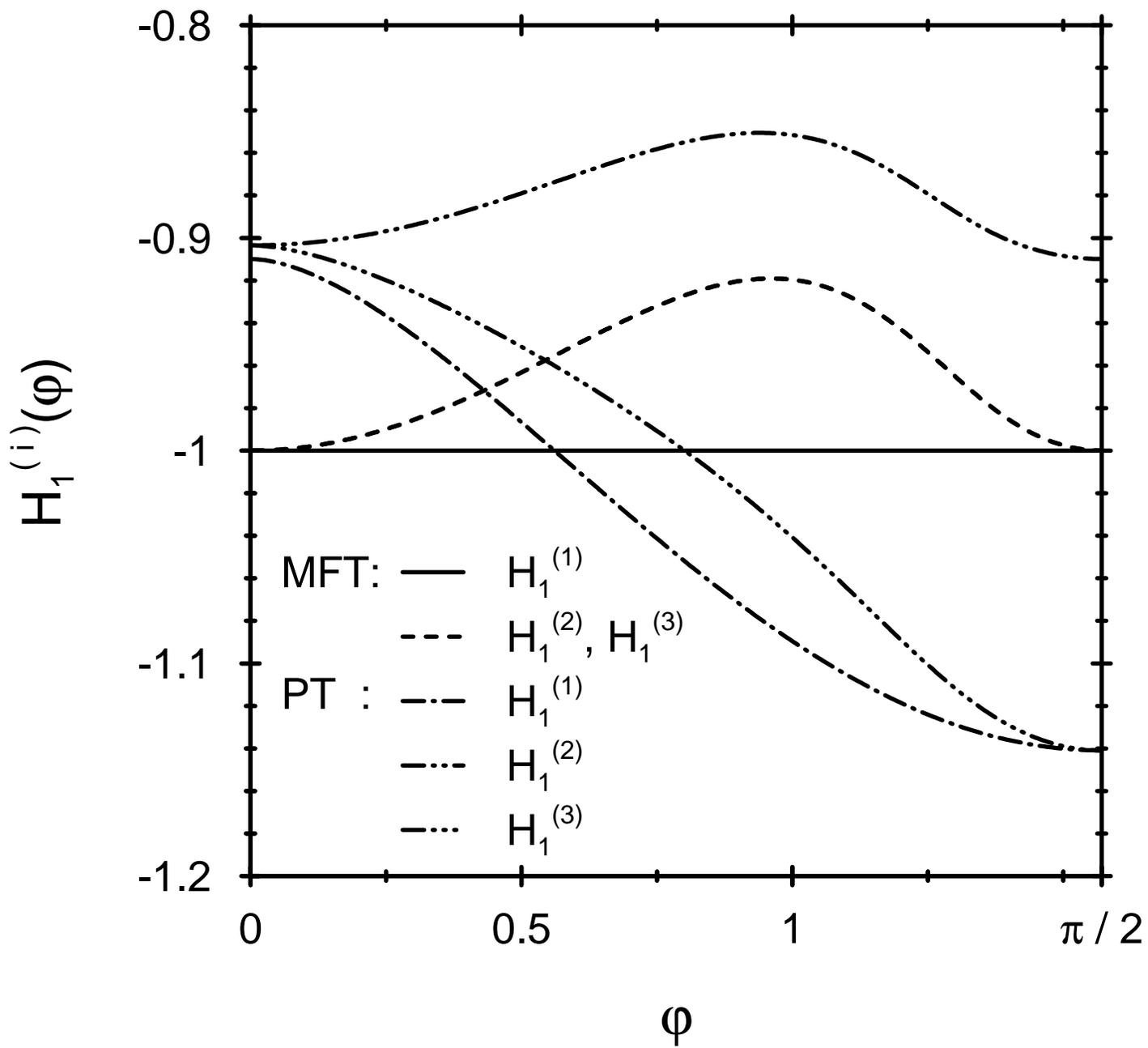
\begin{figure}
\caption{
$G(p \to 0,z,L = \infty, t \to 0)$, $G(p \to 0,z,L \to \infty, t =
0)$, and $G(p = 0,z,L \to \infty, t \to 0)$ attain their maximum value 
$G(p = 0,z,L = \infty, t = 0)={\cal G}_\rii z^{1-\eta}$ via cusplike
singularities 
$H^{(1)}_1(\varphi) [z (p^2+\xi^{-2})^{1/2}]^{-1+\eta_\|}$
with $\varphi = \arctan ((p \xi)^{-1})$,
$H^{(2)}_1(\varphi) [z (p^2+L^{-2})^{1/2}]^{-1+\eta_\|}$
with $\varphi = \arctan (p L)$, and
$H^{(3)}_1(\varphi) [z (\xi^{-2}+L^{-2})^{1/2}]^{-1+\eta_\|}$
with $\varphi = \arctan (L / \xi)$,
respectively, interpolating smoothly between the singularity
$A_1 (pz)^{-1+\eta_\|}$ for $(t=0, L=\infty)$ and
$B_1 (z/\xi)^{-1+\eta_\|}$ for $(p=0, L=\infty)$,
$C_1 (z/L)^{-1+\eta_\|}$ for $(p=0, t=0)$ and
$A_1 (pz)^{-1+\eta_\|}$ for $(L=\infty, t=0)$, and
$C_1 (z/L)^{-1+\eta_\|}$ for $(t=0, p=0)$ and
$B_1 (z/\xi)^{-1+\eta_\|}$ for $(L=\infty, p=0)$, respectively. In ${\cal O}
(\epsilon)$ of perturbation theory (PT) the amplitude functions
$H^{(i)}_1(\varphi)$, $i=1,2,3$, are given by Eqs. (\ref{H1_1}),
(\ref{H1_2}), and (\ref{H1_3}). In ${\cal O} (\epsilon)$ one has 
$H^{(1)}_1(0)=H^{(2)}_1(\pi/2) = A_1 \simeq -0.9099$,
$H^{(1)}_1(\pi/2)=H^{(3)}_1(\pi/2)=B_1 \simeq -1.1409$, and
$H^{(2)}_1(0)=H^{(3)}_1(0)=C_1 \simeq -0.9035$. Within MFT
$H^{(i)}_1(\varphi=0)= H^{(i)}_1(\varphi=\pi/2)=-1$ and
$H^{(1)}_1(\varphi)$ is constant; moreover $H^{(2)}_1(\varphi) =
H^{(3)}_1(\varphi)$ but not constant.
} 
\label{fig_H1H2H3_phi}
\end{figure}

The above discussion demonstrates that, for $z$ fixed, the two-point
correlation function $G(p,z,L,t)$ has a finite value
$G(p=0,z,L=\infty,t=0)$ which is attained via cusplike singularities:
$\sim p^{-1+\eta_\|} \esp (p \to 0, 1/L =0, t=0)$, $\sim
(1/L)^{-1+\eta_\|} \esp (1/L \to 0, p=0, t=0)$, $\sim
(1/\xi)^{-1+\eta_\|} \esp (t \to 0, p=0, 1/L =0)$. In terms of these
variables the critical exponent is the same for all three cases and
only the amplitudes differ. These singularities remain if only one out of
the above three variables is zero and the remaining two both vanish. This
behavior, which includes the smooth interpolation between the
corresponding amplitudes, is described by the scaling functions $h_1$,
$h_2$, and $h_3$ of two variables instead of the scaling functions with
one variable as $g_1(u)$, $g_2(v)$, and $g_3(w)$:  
\beq
\label{g_pt}
G(p,z,L=\infty,t) & = & {\cal G}_\rii z^{1-\eta} h_1(u,v), \esp
h_1(u,v) = g(u,v,w=0) , 
\enq
\beq
\label{g_pL}
G(p,z,L,t=0) & = & {\cal G}_\rii z^{1-\eta} h_2(u,w), \esp h_2(u,w) =
g(u,v=0,w) , 
\enq
and
\beq
\label{g_tL}
G(p=0,z,L,t) & = & {\cal G}_\rii z^{1-\eta} h_3(v,w), \esp h_3(v,w) =
g(u=0,v,w) 
\enq
with $u=pz$, $v=z/\xi$, and $w=z/L$. All three scaling function can be
obtained from Eq. (\ref{g_ptL_z1z2}). Since the discussion of all three
scaling functions is analogous we demonstrate our analysis only for
$h_3(v,w)$. We introduce polar coordinates $\omega$ and $\varphi$
\beq
\label{polar_coordinates}
\omega & = & \sqrt{v^2+w^2} = z
\sqrt{\xi^{-2}+L^{-2}}, \esp \varphi = \arctan(v/w) = \arctan (L/\xi)
\esp , \\
v & = & \omega \sin \varphi, \esp w = \omega \cos \varphi
\nonumber
\enq
which leads to
\beq
\label{h3_transf}
h_3(v,w) & = & h_3 (\omega \sin \varphi, \omega \cos \varphi) =
h^{(3)}_{polar} (\omega,\varphi) . 
\enq
Since the limit $\omega \to 0$, i.e., $1/\xi \to 0$ and $1/L \to 0$,
is equivalent to the limit $z \to 0$ for $\xi$ and $L$ fixed the
resulting singularity is compatible with the SDE so that
\beq
\label{h3_polar_0}
h^{(3)}_{polar} (\omega \to 0, \varphi) & = & H^{(3)}_0(\varphi) +
H^{(3)}_1(\varphi) \omega^{-1+\eta_\|} + \dots \esp . 
\enq
The explicit form of the scaling function $h_3(v,w)$ as obtained from
perturbation theory in ${\cal O}(\epsilon)$ is in accordance with
Eq. (\ref{h3_polar_0}) and renders explicit results for the
coefficients $H^{(3)}_0(\varphi)$ and $H^{(3)}_1(\varphi)$:
\beq
\label{H_0_phi}
H_0(\varphi) & = & h^{(3)}_{polar} (\omega=0,\varphi) = 
h_3(v=0,w=0) = 1
\enq
is independent of $\varphi$ and equal to 1 due to the normalization
$g(u=0,v=0,w=0)=1$. With this result the $\epsilon$-expansion of
$H^{(3)}_1(\varphi)$ follows by comparing the $\epsilon$-expansion of the
rhs of Eq. (\ref{h3_polar_0}) with the limit $\omega \to 0$ of the
$\epsilon$-expansion of $h^{(3)}_{polar}(\omega,\varphi)$. As expected
one finds that $H^{(3)}_1(\varphi)$ interpolates smoothly between the value
$H^{(3)}_1(\varphi=0)=C_1$ (see Eq. (\ref{A1_B1_C1})) corresponding to the
amplitude of the singularity $\sim (1/L)^{-1+\eta_\|}$ for $u=0$ and
$v=0$ and the value $H^{(3)}_1(\varphi=\pi/2)=B_1$ (see
Eq. (\ref{A1_B1_C1})) corresponding to the amplitude of the
singularity $\sim (1/\xi)^{-1+\eta_\|}$ for $u=0$ and $w=0$. 
In Fig. \ref{fig_H1H2H3_phi} all three amplitude functions
$H^{(1)}_1(\varphi)$, $H^{(2)}_1(\varphi)$, and $H^{(3)}_1(\varphi)$
(see Eqs. (\ref{H1_1}) - (\ref{H1_3})) are shown in mean-field theory
(MFT) and in first order in $\epsilon$ (PT). Within MFT
$H^{(1)}_1(\varphi)$ of the semi-infinite system is constant and 
$H^{(2)}_1(\varphi) = H^{(3)}_1(\varphi)$ exhibit a nontrivial
dependence on $\varphi$. Beyond MFT all three functions
interpolate between the amplitudes $A_1$, $B_1$, and $C_1$ (see
Eq. (\ref{A1_B1_C1})) in a nontrivial way.

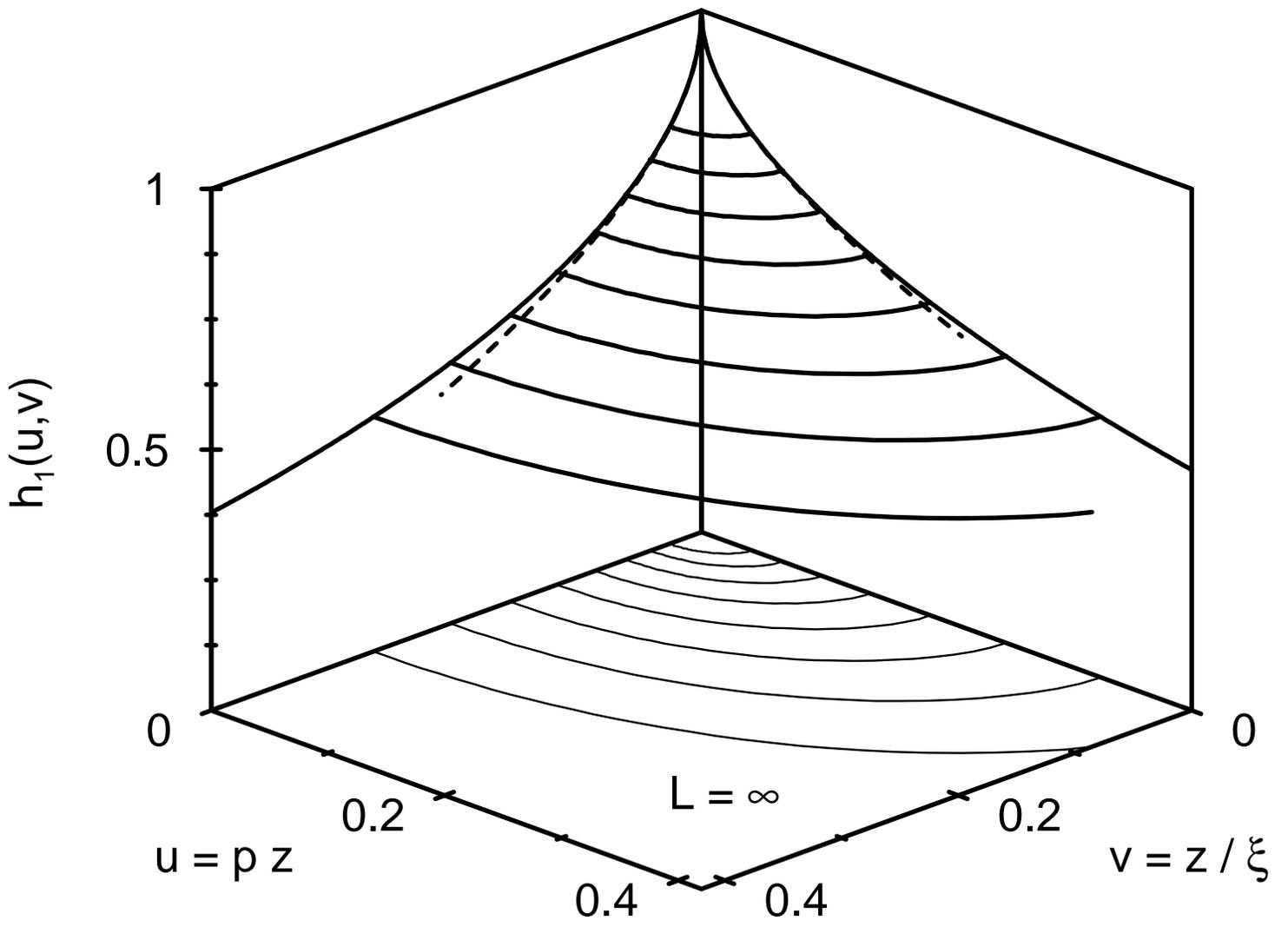
\begin{figure}
\caption{
The exponentiated scaling function $h_1(u=pz,v=z/\xi)$
(Eq. (\ref{g_pt})) corresponding to the case $L=\infty$. We show the
contour lines $h_1(u,v)=h^{(1)}_{polar}
(\omega=(u^2+v^2)^{1/2},\varphi=\arctan (v/u))$ for
$h_1 \esp = \esp 0.8,0.75,0.7,0.65,0.6,0.55,0.5,0.45$ \esp with their 
projections onto the $u$$v$ plane as well as
$h_1(u,v=0)= g_1(u)$ (Eq. (\ref{corr_p})) and $h_1(u=0,v) = g_2(v)$
(Eq. (\ref{corr_t})) which are discussed in Fig. \ref{fig_p_t_L}. The
dashed lines correspond to the leading singularities $g_1(u \to 0) = 1
+ A_1 u^{-1+\eta_\|}$ and $g_2(v \to 0) = 1 + B_1 v^{-1+\eta_\|}$,
respectively.  
}
\label{fig_cusp_pt}
\end{figure}

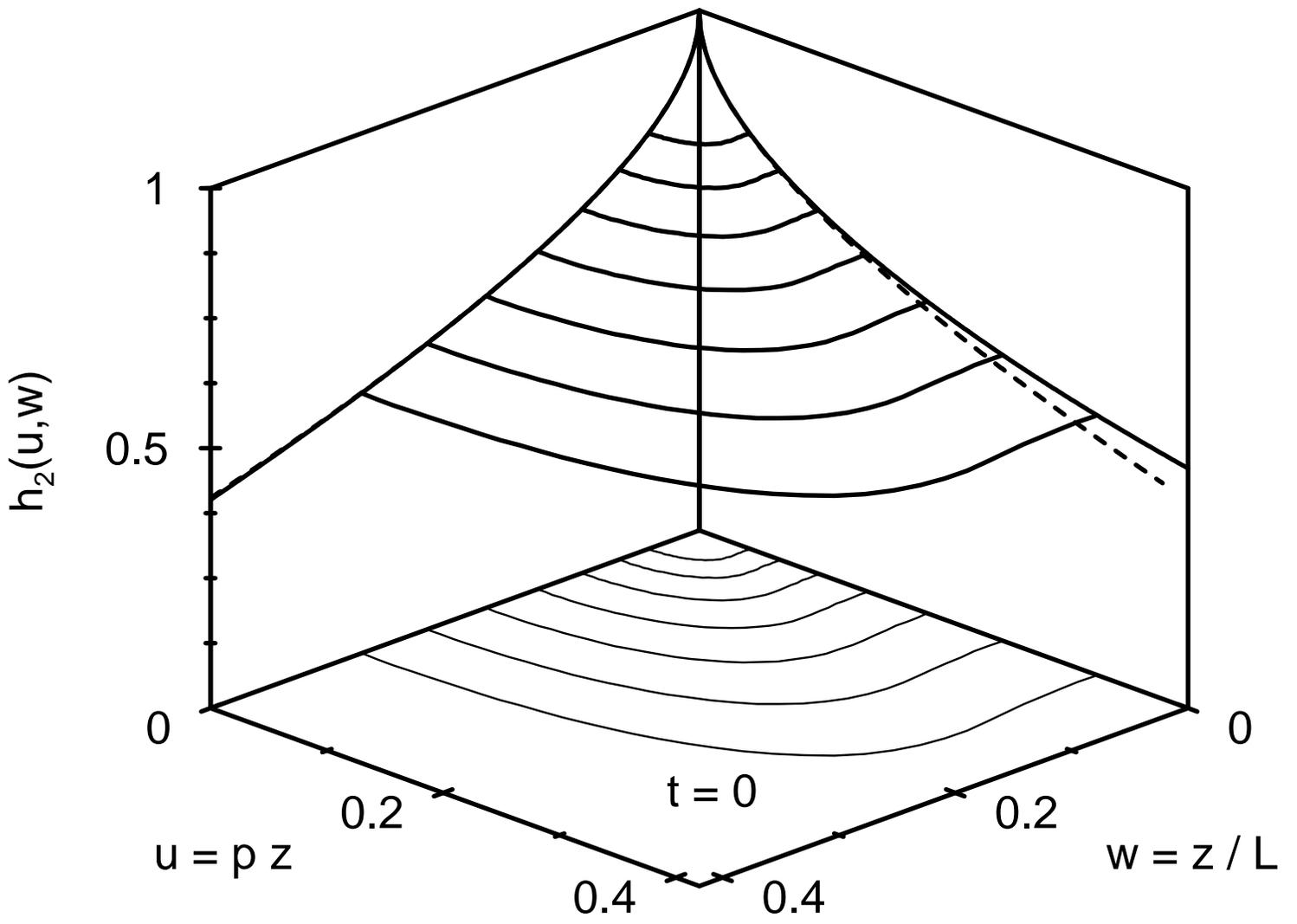
\begin{figure}
\caption{
The exponentiated scaling function $h_2(u=pz,w=z/L)$
(Eq. (\ref{g_pL})) at bulk criticality $t=0$. We show the
contour lines $h_2(u,w)=h^{(2)}_{polar}
(\omega=(u^2+w^2)^{1/2},\varphi=\arctan (u/w))$ for
$h_2=0.8,0.75,0.7,0.65,0.6,0.55,0.5$ with their 
projections onto the $u$$w$ plane as well as
$h_2(u,w=0)= g_1(u)$ (Eq. (\ref{corr_p})) and $h_2(u=0,w) = g_3(w)$
(Eq. (\ref{corr_L_1})). The dashed lines correspond to the leading
singularities $g_1(u \to 0) = 1 + A_1 u^{-1+\eta_\|}$ and $g_3(w \to
0) = 1 + C_1 w^{-1+\eta_\|}$, respectively. In the latter case the
difference between the leading behavior and the full scaling function
$g_3(w)$ is hardly visible. Thus the leading dependence on $z/L$ for
$p=0$, $t=0$ remains valid nearly up to the middle of the film at
$z/L=0.5$. 
}
\label{fig_cusp_pL}
\end{figure}

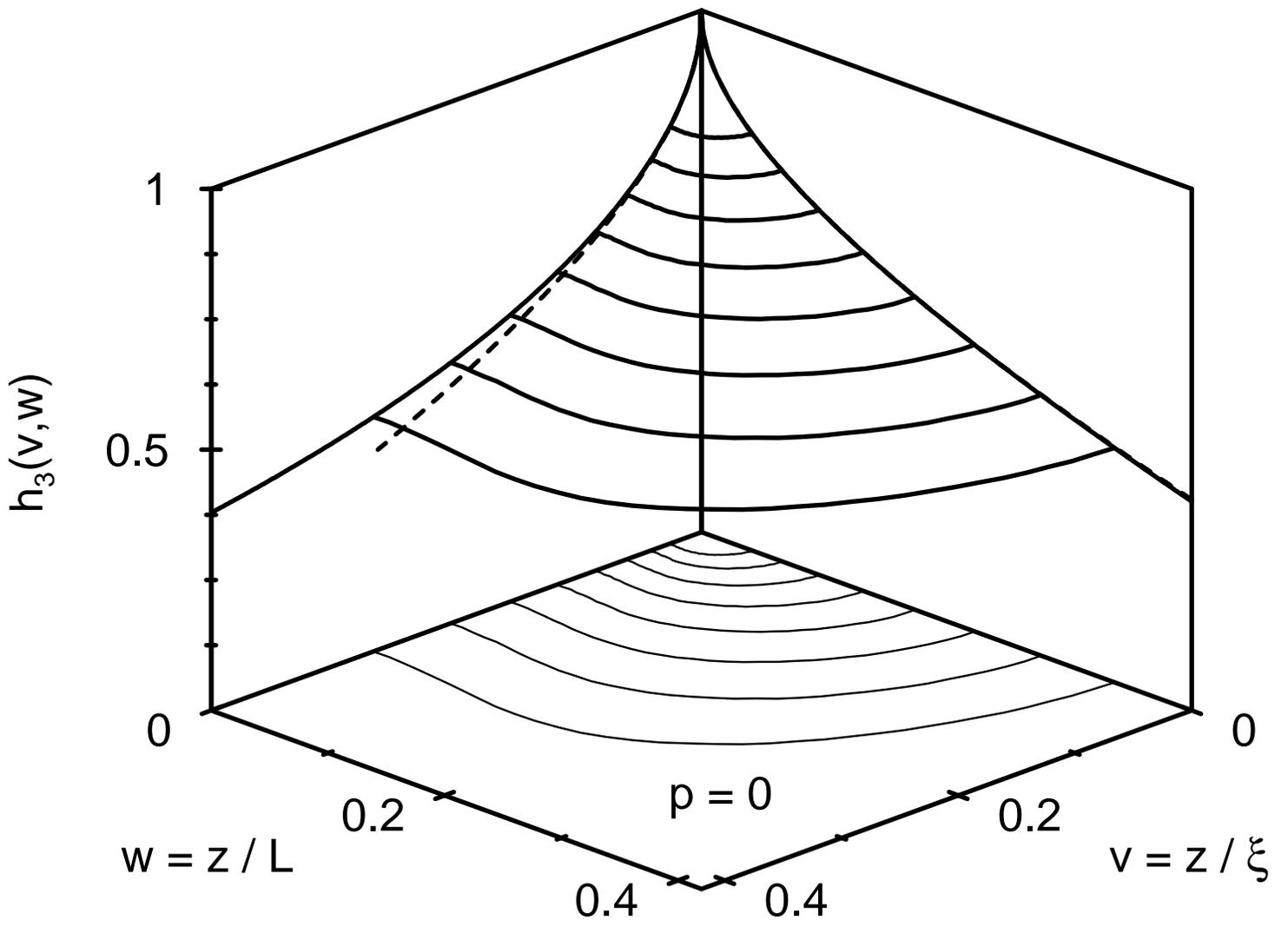
\begin{figure}
\caption{
The exponentiated scaling function $h_3(v=z/\xi,w=z/L)$
(Eq. (\ref{g_tL})) for lateral momentum $p=0$. We show the
contour lines $h_3(v,w)=h^{(3)}_{polar}
(\omega=(v^2+w^2)^{1/2},\varphi=\arctan (v/w))$ for
$h_3=0.8,0.75,0.7,0.65,0.6,0.55,0.5,0.45$ with their 
projections onto the $v$$w$ plane as well as
$h_3(v,w=0)= g_2(v)$ (Eq. (\ref{corr_t})) and $h_3(v=0,w) = g_3(w)$
(Eq. (\ref{corr_L_1})). The dashed lines correspond to the leading
singularities $g_2(v \to 0) = 1 + B_1 v^{-1+\eta_\|}$ and $g_3(w \to
0) = 1 + C_1 w^{-1+\eta_\|}$, respectively. In the latter case the
difference between the leading behavior and the full scaling function
$g_3(w)$ is hardly visible. Thus the leading dependence on $z/L$ for
$p=0$, $t=0$ remains valid nearly up to the middle of the film at
$z/L=0.5$. 
}
\label{fig_cusp_tL}
\end{figure}

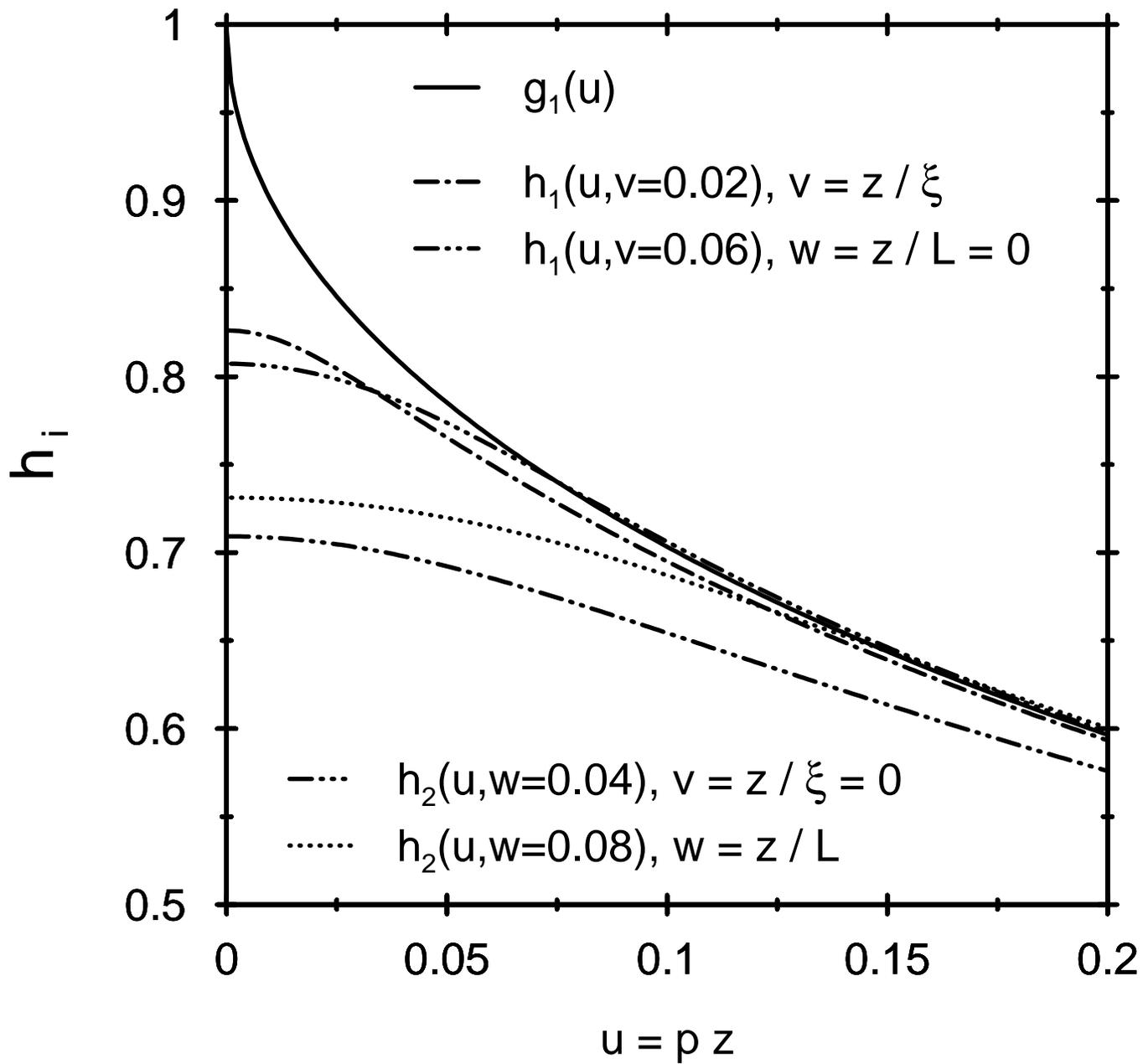
\begin{figure}
\caption{
The scaling function $g_1(u=pz)$ (Eq. (\ref{corr_p})) with the
cusplike singularity $g_1(u \to 0) =$ $ 1 + A_1 u^{-1+\eta_\|}$ evolves
out of the scaling functions $h_1 (u,v=z/\xi)$ (Eq. (\ref{g_pt})) and
$h_2 (u,w=z/L)$ (Eq. (\ref{g_pL})) in the limits $v \to 0$ and $w \to
0$, respectively, which are analytic functions of $u$, with a maximum
at $u=0$, for $v \not= 0$ or $w \not= 0$. The various curves
correspond to vertical cuts of the surface shown in
Fig. \ref{fig_cusp_pt} for $v=const.$ with $w=z/L=0$ and in
Fig. \ref{fig_cusp_pL} for $w=const.$ with $v=z/\xi=0$, respectively. 
}
\label{fig_p_cusp_tL}
\end{figure}

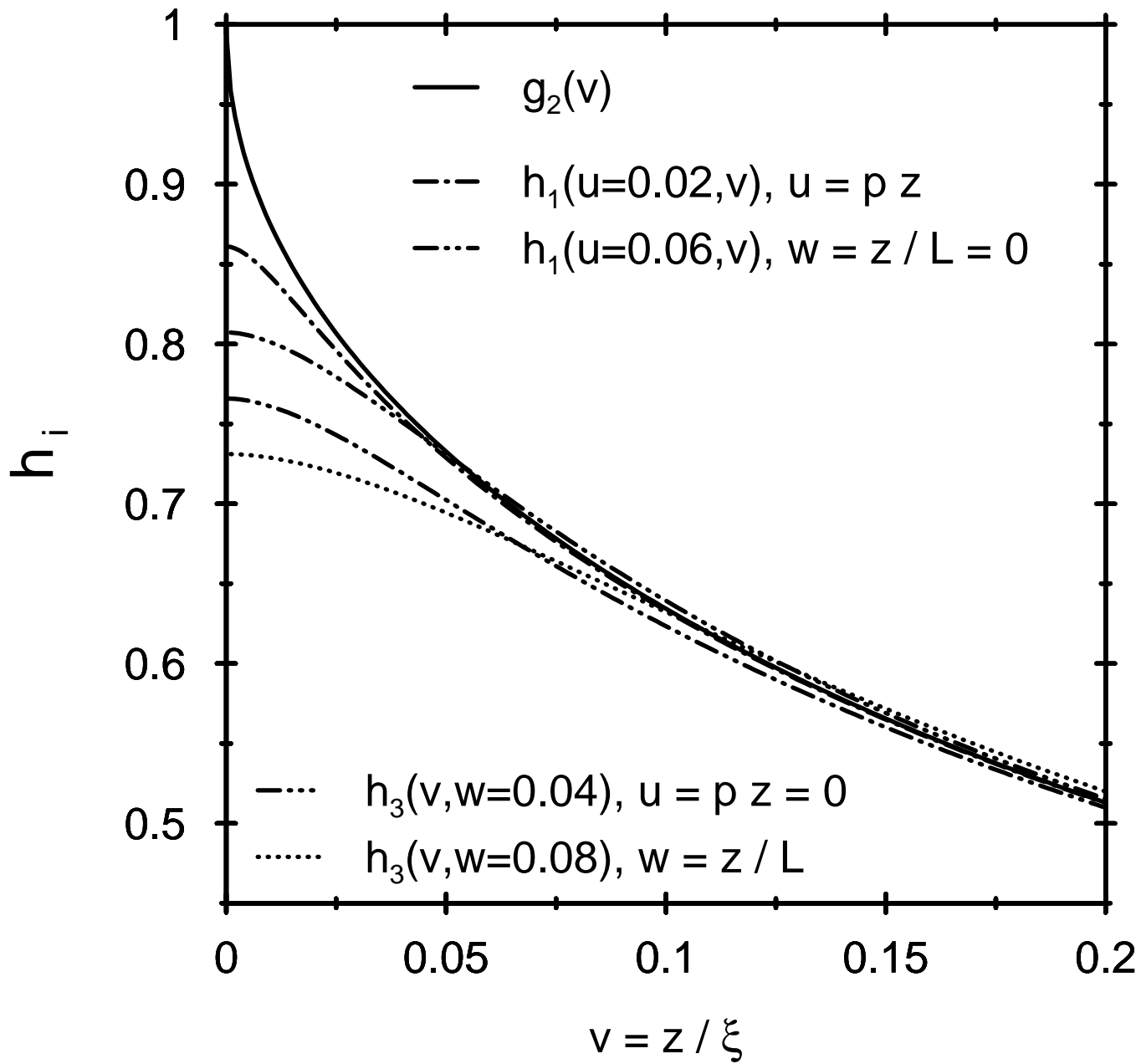
\begin{figure}[b!]
\caption{
The scaling function $g_2(v=z/\xi)$ (Eq. (\ref{corr_t})) with the
cusplike singularity $g_2(v \to 0) =$$ 1 + B_1 v^{-1+\eta_\|}$ evolves
out of the scaling functions $h_1 (u=pz,v)$ (Eq. (\ref{g_pt})) and
$h_3 (v,w=z/L)$ 
(Eq. (\ref{g_tL})) in the limits $u \to 0$ and $w \to
0$, respectively, which are analytic functions of $v$, with a maximum
at $v=0$, for $u \not= 0$ or $w \not= 0$. The various curves
correspond to vertical cuts of the surface shown in
Fig. \ref{fig_cusp_pt} for $u=const.$ with $w=z/L=0$ and in
Fig. \ref{fig_cusp_tL} for $w=const.$ with $u=pz=0$, respectively. 
}
\label{fig_t_cusp_pL}
\end{figure}

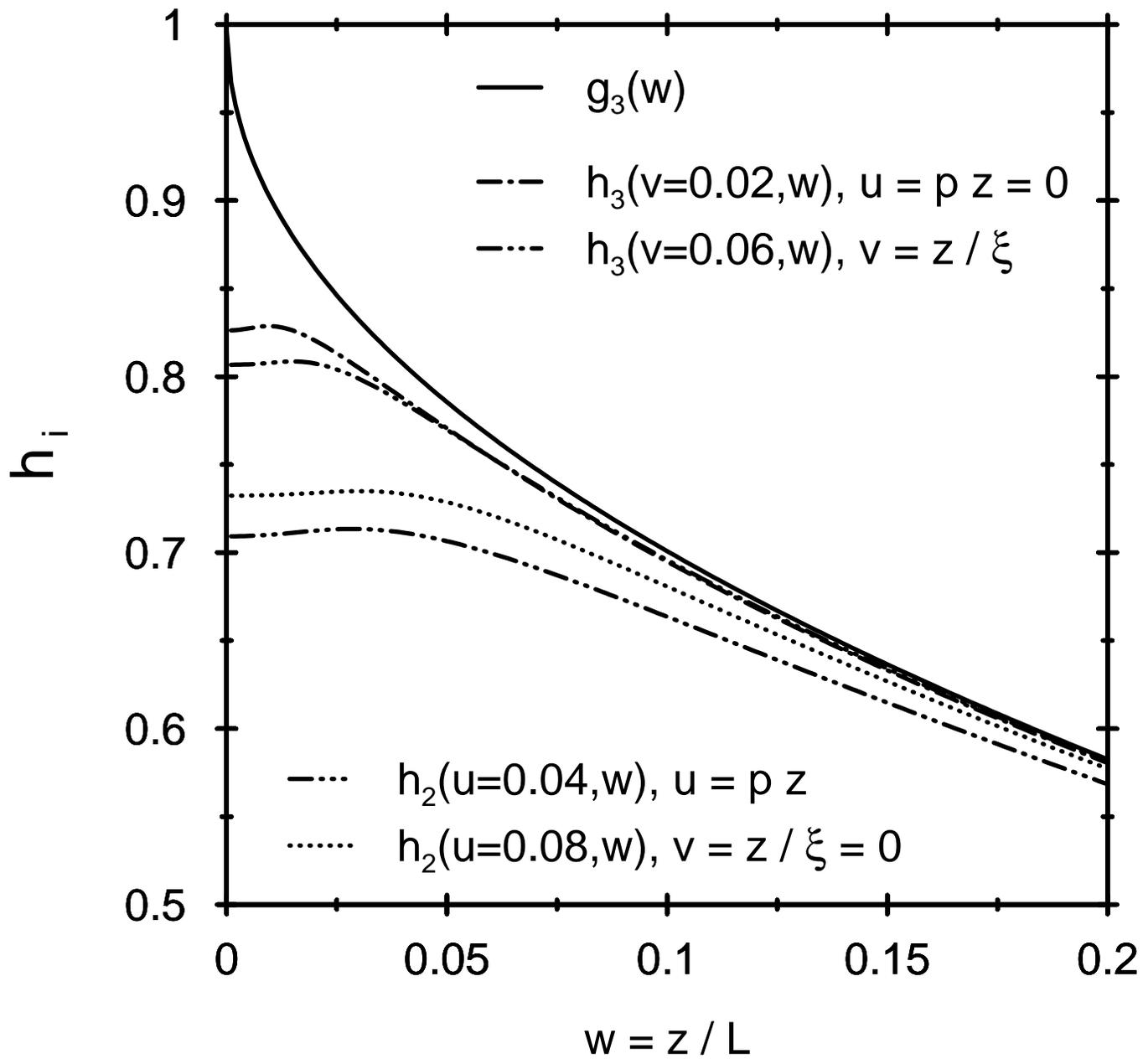
\begin{figure}
\caption{
The scaling function $g_3(w=z/L)$ (Eq. (\ref{corr_L_1})) with the
cusplike singularity $g_3(w \to 0) =$ $ 1 + C_1 w^{-1+\eta_\|}$ evolves
out of the scaling functions $h_3 (v=z/\xi,w)$ (Eq. (\ref{g_tL})) and
$h_2 (u=pz,w)$ (Eq. (\ref{g_pL})) in the limits $v \to 0$ and $u \to
0$, respectively, which are analytic functions of $w$ for $u \not= 0$
or $w \not= 0$. The various curves correspond to vertical cuts of the
surface shown in 
Fig. \ref{fig_cusp_tL} for $v=const.$ with $pz=0$ and in
Fig. \ref{fig_cusp_pL} for $u=const.$ with $z/\xi=0$. We note that,
different than in Figs. \ref{fig_p_cusp_tL} and \ref{fig_t_cusp_pL},
the scaling functions $h_3(v \not= 0,w)$ and $h_2(u \not= 0,w)$ are
nonmonotonous functions and exhibit a maximum at $w \not=0$ and a
local minimum at $w=0$. 
}
\label{fig_L_cusp_pt}
\end{figure}

In Figs. \ref{fig_cusp_pt}, \ref{fig_cusp_pL}, and \ref{fig_cusp_tL} we
display the full scaling functions $h_1(u,v)$, $h_2(u,w)$, and $h_3(v,w)$,
respectively. In order to obtain such a scaling function beyond the
leading asymptotic form we first subtract its leading contribution in
its $\epsilon$-expanded form in ${\cal O} (\epsilon)$ from the full
expression of the scaling function and add the leading exponentiated
contribution afterwards. This exponentiation scheme is consistent with
the explicit expanded form up to and including ${\cal O} (\epsilon)$.
In Figs. \ref{fig_p_cusp_tL}, \ref{fig_t_cusp_pL}, and
\ref{fig_L_cusp_pt} we show cross sections of the three-dimensional
plots in order to illustrate the emergence of the $p^{-1+\eta_\|}$ cusplike
singularity upon varying $t$ or $L$, the $(1/\xi)^{-1+\eta_\|}$ cusplike
singularity upon varying $p$ or $L$, and the $(1/L)^{-1+\eta_\|}$
cusplike singularity upon varying $t$ or $p$, respectively. 


\subsection{Perpendicular correlations}
\label{correlation_z1z2}
In a semi-infinite system the perpendicular correlations in real space
define the exponent $\eta_\perp = (\eta + \eta_\|)/2$ through the
limit $G(x_\|,z_1 \to \infty,z_2,L=\infty,t) \sim
z_1^{-(d-2+\eta_\perp)}$ with $x_\|$ and $z_2$ fixed.
A Fourier transformation leads to the relation
$G(p=0,z_1,z_2,L=\infty,t) \sim z_1^{1-\eta_\perp}$ with $z_2$ fixed
and $z_1 \to \infty$. Note that in real space
$G(x_\|,z_1,z_2,L=\infty,t=0)$ {\em increases} as function of $z_1$ for
$z_2$ and $x_\|$ fixed, reaches a maximum at a certain value $z_1^\ast
=z_2 f(x_\|/z_2)$ and finally vanishes for $z_1 \to \infty$. This
increase for $0 < z_1 < z_1^\ast$ leads to the divergence $\sim
z_1^{1-\eta_\perp}$, $1-\eta_\perp \simeq 0.25$, of
$G(p=0,z_1,z_2,L=\infty,t=0) = \int d x_\| \langle \phi (0,z_1) \phi
(x_\|,z_2) \rangle$. The coordinates $z_1$ and $z_2$ can be
interchanged. Actually conformal invariance fixes completely the
functional form of $G(p=0,z_1,z_2,L=\infty,t=0)$ (see
Ref. \cite{Cardy:84}). SDE leads up to a constant amplitude to the
expression 
\beq
\label{gompper_cft_z1z2}
G(p=0,z_1,z_2,L=\infty,t=0) & \sim & (z_1 z_2)^\frac{1-\eta}{2} \Big
( \Theta(z_1-z_2) \frac{z_2}{z_1} + \Theta(z_2-z_1) \frac{z_1}{z_2} 
\Big)^\frac{\eta_\|-1}{2} + \dots
\enq
(see Eq. (4.68) in Ref. \cite{Gompper:86:b}). The explicit calculation
to first order in $\epsilon$ gives
\beq
\label{gompper_firstorder_z1z2}
G(p=0,z_1,z_2,L=\infty,t=0) & = & {\cal G}_\rii (z_1 z_2)^\frac{1-\eta}{2} \Big
( \Theta(z_1-z_2) \frac{z_2}{z_1} + \Theta(z_2-z_1) \frac{z_1}{z_2} 
\Big)^\frac{\eta_\|-1}{2}
\enq
for arbitrary $z_1$ and $z_2$. 
This perturbation theory guided result for $d=3$ has a structure
similar to the exact result from conformal theory in $d=2$ (see  
Refs. \cite{Cardy:84} and \cite{Gompper:86:b}). Therefore one is led
to the 
conclusion that Eq. (\ref{gompper_firstorder_z1z2}) is a good
approximation for the exact correlation function in $d=3$. Guided by
these considerations we find that in the case that the variables $p$,
$t$, and $1/L$ are small but nonzero the explicit results for $G$
obtained from the $\epsilon$-expansion can be cast into the following
forms:
\beq
\label{corr_p_z1z2}
&&G(p \to 0,z_1,z_2,L=\infty,t=0) \\
&=& {\cal G}_\rii \Big( \Theta(z_2-z_1)
z_1^{1-\eta} \Big( \frac{z_2}{z_1} \Big)^{1-\eta_\perp} \Big( 1 +
A_1 (pz_2)^{-1+\eta_\|} + \dots \Big) \nonumber \\
&& + \esp \Theta(z_1-z_2)
z_2^{1-\eta} \Big( \frac{z_1}{z_2} \Big)^{1-\eta_\perp} \Big( 1 +
A_1 (pz_1)^{-1+\eta_\|} + \dots \Big) \Big) , \nonumber 
\enq
\beq
\label{corr_t_z1z2}
&&G(p=0,z_1,z_2,L=\infty,t \to 0) \\
& = & {\cal G}_\rii \Big( \Theta(z_2-z_1)
z_1^{1-\eta} \Big( \frac{z_2}{z_1} \Big)^{1-\eta_\perp} \Big( 1 +
B_1 \Big( \frac{z_2}{\xi} \Big)^{-1+\eta_\|} + \dots \Big) \nonumber \\ 
&& + \esp \Theta(z_1-z_2)
z_2^{1-\eta} \Big( \frac{z_1}{z_2} \Big)^{1-\eta_\perp} \Big( 1 +
B_1 \Big( \frac{z_1}{\xi} \Big)^{-1+\eta_\|} + \dots \Big) \Big) , \nonumber
\enq
and
\beq
\label{corr_L_z1z2}
&&G(p=0,z_1,z_2,L \to \infty,t=0) \\
& = & {\cal G}_\rii \Big( \Theta(z_2-z_1)
z_1^{1-\eta} \Big( \frac{z_2}{z_1} \Big)^{1-\eta_\perp} \Big( 1 +
C_1 \Big( \frac{z_2}{L} \Big)^{-1+\eta_\|} + \dots \Big) \nonumber \\ 
&& \esp + \Theta(z_1-z_2)
z_2^{1-\eta} \Big( \frac{z_1}{z_2} \Big)^{1-\eta_\perp} \Big( 1 +
C_1 \Big( \frac{z_1}{L} \Big)^{-1+\eta_\|} + \dots \Big) \Big) . \nonumber
\enq
These expressions are valid for arbitrary $z_1$ and $z_2$ as long as the
scaling variables $pz_{1,2}$, $z_{1,2}/\xi$, and $z_{1,2}/L$ are
small. The explicit $\epsilon$-expansion provides the amplitudes
$A_1$, $B_1$, and $C_1$ given by Eq. (\ref{A1_B1_C1}). For the special case
$z_1=z_2$ Eqs. (\ref{corr_p_z1z2}) - (\ref{corr_L_z1z2}) reduce to
Eqs. (\ref{corr_p}), (\ref{corr_t}), and (\ref{corr_L}).
In the limits $p=0$, $t=0$, and $L=\infty$ Eqs. (\ref{corr_p_z1z2}) -
(\ref{corr_L_z1z2}) reduce to Eq. (\ref{gompper_cft_z1z2}) (recall
$\eta_\perp = (\eta_\|+\eta)/2$).

Finally we note that Eqs. (\ref{corr_p_z1z2}), (\ref{corr_t_z1z2}),
(\ref{corr_L_z1z2}), and the full film correlation function
$G(p,z_1,z_2,L,t)$ up to first order in $\epsilon$ (see
Eq. (\ref{g_ptL_z1z2}) in Appendix \ref{corr_for_z1z2}) 
satisfy the so-called product rule derived by Parry and Swain for the
correlation function algebra of inhomogeneous fluids (see Eq. (2.20)
in Ref. \cite{Parry:97}):
\beq
\label{parry_product}
G(p,z_1,z_2,L,t) G(p,z_2,z_3,L,t) & = & G(p,z_2,z_2,L,t)
G(p,z_1,z_3,L,t) 
\enq
for all $0 < z_1 \le z_2 \le z_3 < L$. The second identity derived by
Parry and Swain (see Eq. (2.21) in Ref. \cite{Parry:97}) is trivially
fulfilled in the disordered phase considered here because it involves
the derivative of the order-parameter profile which vanishes above
$T_c$. A nontrivial test of this relation would require results for
the ordered phase below $T_c$. 


\subsection{The susceptibility}
\label{excess_sus}

As it has become apparent in the previous subsection, the full
dependence of the correlation function $G$ on all its variables $p$,
$z_1$, $z_2$, $L$, and $t$ is rather complicated. Therefore it
increases the transparency to consider a spatially averaged quantity
which still displays interesting specific properties of the critical
behavior in a film geometry. The singular part of the total
susceptibility per area defined as
\beq
\label{def_exess_sus}
\chi (L,t) & = & \int_0^L d z_1 \int_0^L d z_2 \esp G(p=0,z_1,z_2,L,t)
\enq
provides such a reduced but still interesting quantity in that it
depends only on two variables $L$ and $t$. In addition this
susceptibility is directly accessible in an experiment which probes
the response of a thin magnetic film on the applied external field in
the limit of vanishing field strength. 

From the scaling properties for $G$ one obtains the following scaling
property for $\chi$ (see Eqs. (\ref{G_final}) and (\ref{G_b_num})):
\beq
\label{scaling_excess_sus}
\chi (L,t) & = & B^2 (\xi_0^+)^{d+1} \Big( \frac{L}{\xi_0^+}
\Big)^{3-\eta} {\cal R} f(y=L/\xi) = L^{3-\eta} {\cal G}_\rii f(y)
\enq
where
\beq
\label{fy_def}
f(y) & = & \int_0^1 d x_1 \int_0^1 d x_2 x_1^{1-\eta} g_\rii (0,x_1
y,x_2 y, x_2)
\enq
is a universal scaling function. For $y \to \infty$, i.e., $L \to
\infty$ and $t$ fixed the scaling function $f(y)$ vanishes as follows:
\beq
\label{f_large_y}
f(y \to \infty) & = & {\cal A} y^{-2+\eta} + {\cal B} y^{-3+\eta} +
{\cal C} y^{-3+\eta} e^{-y} + {\cal O} (e^{-2 y}) .
\enq
with
\beq
\label{large_y_abc}
{\cal A} & = & 1 - \te + {\cal O} (\epsilon^2) , \nonumber \\
{\cal B} & = & -2 \Big( 1 + \te \Big[ \pi \Big( \frac{1}{2} -
\frac{1}{\sqrt{3}} \Big) -1 \Big] \Big) + {\cal O} (\epsilon^2) , \\ 
{\cal C} & = & 4 \Big( 1 - \te \Big[ \frac{\pi}{2} \Big( 1 -
\frac{1}{\sqrt{3}} \Big) +1 \Big] \Big) + {\cal O} (\epsilon^2) , \nonumber
\enq
so that with $\gamma = \nu (2-\eta)$ and $\gamma_s = \gamma + \nu$
\beq
\label{chi_large_y}
\chi(L \to \infty,t) & = & B^2 (\xi_0^+)^{d+1} {\cal R} 
\Big\{ \frac{L}{\xi_0^+} {\cal A} t^{-\gamma} + {\cal B} t^{-\gamma_s}
\Big[ 1 + \frac{{\cal C}}{{\cal B}} e^{-L/\xi} + {\cal O} (e^{-2L/\xi})
\Big] \Big\} .
\enq
The first term $(\sim t^{-\gamma})$ corresponds to the bulk
contribution of the total susceptibility. (We recall that $\chi$ is
the total susceptibility per area $A_\|$ of one surface and that the total
volume of the system is $A_\| L$.) The universal amplitude ${\cal A}$
(Eq. (\ref{large_y_abc})) is in accordance with the corresponding
known universal amplitude ratios \cite{Bervillier:76,Privman:91}. 
The second term $(\sim t^{-\gamma_s})$ corresponds to the sum of the
excess susceptibilities of two semi-infinite systems within the
surface universality class of the ordinary transition resembling the
two bounding surfaces of the film. The corresponding universal
amplitude ${\cal B}$ (Eq. (\ref{large_y_abc})) of the semi-infinite
systems is in accordance with the corresponding result in
Ref. \cite{Diehl:85:a}. Finally, the last term $\sim e^{-L/\xi}$ in
Eq. (\ref{chi_large_y}) is the actual finite size contribution induced
by the finite distance $L$ between the two surfaces confining the
film. It is interesting to note that the structure of this finite size
term ${\cal C} t^{-\gamma_s} \exp (-L/\xi)$ differs from its
counterparts for the free energy and specific heat in two respects
(see Eqs. (4.8) and (6.14) in Ref.
[30(a)]): (i) For
ordinary - ordinary boundary conditions the finite size terms of the
latter two both vanish $\sim \exp (-2y)$ for large $y=L/\xi$. (ii) The
prefactor ${\cal C} t^{-\gamma_s}$ is replaced by ${\cal C}'
t^{-\kappa} y^{\kappa/(2\nu)}$ with $\kappa = \alpha_s -2$ (free
energy) and $\kappa = \alpha_s = \alpha + \nu$ (specific heat),
respectively. From the explicit result in ${\cal O}(\epsilon)$ we
infer that in the case of the excess susceptibility this power law in
front of the exponential is either missing or has an exponent of ${\cal
O}(\epsilon^2)$. In order to render the comparison between the finite
size scaling of the free energy and specific heat on one hand and of
the susceptibility on the other hand more transparent we rewrite the
susceptibility as
\beq
\label{chi_SD_transparent}
\chi (L,t) & = & B^2 (\xi_0^+)^{d+1} {\cal R} \Big( \frac{L}{\xi_0^+}
\Big)^{\gamma_s/\nu} 
\Big\{ {\cal A} y^{-\gamma/\nu} + {\cal B} y^{-\gamma_s/\nu} + g(y) 
\Big\}
\enq
where $(2-\eta = \gamma/\nu$, $3-\eta=\gamma_s/\nu)$
\beq
\label{chi_SD_f_g}
g(y) = f(y) - {\cal A} y^{-2+\eta} - {\cal B} y^{-3+\eta}.
\enq
The finite size scaling for the singular part ${\cal F}_{sing}$ of the
free energy of a film has the similar form $(d=(2-\alpha)/\nu$,
$d-1=(2-\alpha_s)/\nu)$ (see Eq. (4.11) in Ref. 
[30(a)])
\beq
\label{krech_free_energy}
\frac{{\cal F}_{sing}}{k_b T_c(\infty)} & = &
\frac{A_\|}{(\xi_0^+)^{d-1}} \Big( \frac{L}{\xi_0^+}
\Big)^{(\alpha_s-2)/\nu} \Big\{ A_b y^{-(\alpha-2)/\nu} + A_s
y^{-(\alpha_s-2)/\nu} + \Theta(y) 
\Big\} .
\enq
$A_\|$ is the area of the cross section of the film. Both in
Eq. (\ref{chi_SD_transparent}) and Eq. (\ref{krech_free_energy}) the first two
terms correspond to the bulk and surface contribution,
respectively. In both cases the curly bracket represents a universal
scaling function. For the susceptibility the finite size part vanishes
as
\beq
\label{chi_vanish}
g(y \to \infty) & = & {\cal C} y^{-\gamma_s/\nu} e^{-y} + {\cal O}
(e^{-2 y})
\enq
whereas for the free energy one has
\beq
\label{krech_free_energy_vanish}
\Theta (y \to \infty) & = & {\cal C}' y^{-(\alpha_s-2)/(2 \nu)} e^{-2
y} + {\cal O} (e^{-3 y}) .
\enq
At this point we note that the film susceptibility has been
also discussed by Nemirovsky and Freed (see Eqs. (3.14d) and (3.16d)
in Ref. \cite{Nemirovsky:86:a}). Instead of the $(p, z_1,
z_2)$-representation of the propagator employed here they used a
discrete spectral $(p$-$\kappa_j)$-representation. In the discrete
representation the propagator for Dirichlet boundary conditions is
given by
\beq
\label{G_d_pk}
G_{D,j}(p,\tau) & = & \frac{1}{p^2+\tau+\kappa_j^2}, \esp \kappa_j = \pi
(j+1)/L , \esp j=0,1,2, \dots \esp .
\enq
The $(p, z_1,z_2)$- and $(p$-$\kappa_j)$-representation are related by
the formula
\beq
\label{G_d_pz_and_pk}
G_D(p,z_1,z_2,L,\tau) & = & \frac{2}{L} \sum_{j=0}^\infty \sin
(\kappa_j z_1) \sin (\kappa_j z_2) G_{D,j}(p,\tau) .
\enq
The one-loop contribution to the total susceptibility is given by
\beq
\label{one_loop_total_sus}
&&- \frac{g}{2} \frac{n+2}{3}
\int\limits_0^L d z_1 \int\limits_0^L d z_2
\int \frac{d^{d-1}q}{(2\pi)^{d-1}} \int\limits_0^L dz \\
&& G_D(p=0,z_1,z,L,\tau) \esp G_D(q,z,z,L,\tau) \esp
G_D(p=0,z,z_2,L,\tau) \nonumber \\
&=&- \frac{g}{2} \frac{n+2}{3}
\int\limits_0^L d z_1 \int\limits_0^L d z_2
\int \frac{d^{d-1}q}{(2\pi)^{d-1}} \int\limits_0^L dz 
\left( \frac{2}{L} \right)^3 \sum_{m_1,m_2,m_3=0}^\infty \sin
(\kappa_{m_1} z_1) \sin (\kappa_{m_1} z) \nonumber \\ 
&&G_{D,m_1}(p=0,\tau) 
\sin^2(\kappa_{m_2} z) G_{D,m_2}(q,\tau)
\sin (\kappa_{m_3} z) \sin (\kappa_{m_3} z_2) G_{D,m_3}(p=0,\tau)
\nonumber .
\enq
After performing the integrations one has to evaluate the triple
sum. In their 
calculation of the susceptibility Nemirovsky and Freed omitted the
terms $m_1 \not= m_3$ in the above sum which leads to an erroneous
expression for the scaling function $f(y)$. If, however, all terms in
the triple sum are properly taken into account, one obtains, as
expected, the same correct result for $f(y)$ as via the $(p, z_1,
z_2)$-representation.

The above discussion is focused on the limit $y=L/\xi \to \infty$,
i.e., on increasing the film thickness at a fixed temperature. In the
opposite limit $y \to 0$ the film thickness is kept fixed and one
approaches the bulk critical temperature $T_c(L=\infty)$ where $\xi$
diverges as $\xi_0^+ ((T-T_c(L=\infty))/T_c(L=\infty))^{-\nu}$. For
Dirichlet boundary conditions as considered here the critical
temperature of the film occurs at a lower temperature $T_c(L) <
T_c(L=\infty)$. Therefore the film is not critical at $T_c(L=\infty)$
and thus the susceptibility is an analytic function of $t$ around $t=
(T-T_c(L=\infty))/T_c(L=\infty) =0$. Therefore the finite size scaling
function $g(y \to 0)$ has the following form:
\beq
\label{g_for_small_y}
g(y \to 0) & = & - {\cal A} y^{-\gamma/\nu} - {\cal B}
y^{-\gamma_s/\nu} + {\cal D} + {\cal E} y^{1/\nu} + {\cal O}
(y^{2/\nu}) 
\enq
with
\beq
\label{D_sus}
{\cal D} & = & \frac{1}{12} \Big( 1 - \tilde \epsilon \Big(
\frac{\pi^2}{60} + 12 b_2 +1\Big) \Big) + {\cal O} (\epsilon^2) 
\enq
and
\beq
\label{E_sus}
{\cal E} & = & - \frac{1}{120} \Big( 1 + \tilde \epsilon \Big( 4 a_2 -
\frac{17 \pi^2}{504} + 102 b_4 - 10 b_2 -1 \Big) \Big) + {\cal O}
(\epsilon^2)
\enq
so that
\beq
\label{f_small_y}
f(y \to 0) & = & {\cal D} + {\cal E} y^{1/\nu} + {\cal O} (y^{2/\nu}) .  
\enq
The numbers $a_2$, $b_2$, and $b_4$ are given in Eq. (\ref{a2b2b4}) in
Appendix \ref{excess_app}. For $(n,d)=(1,3)$ the values of the
amplitudes to first order in $\epsilon$ are ${\cal D} \simeq 0.08142$
and ${\cal E} \simeq -0.01375$; for ${\cal A}$ and ${\cal B}$ see
Eq. (\ref{large_y_abc}). The explicit form of the scaling function and
its limiting behaviors are given in Appendix \ref{excess_app}.
Figure \ref{fig_f_d} (a) shows $f(y)$ within mean-field theory and within
perturbation theory in first order $\epsilon$ as well as its
corresponding asymptotic behaviors for large and small values of $y$,
and Fig. \ref{fig_f_d} (b) displays $g(y)$ for large values of $y$.

Our investigations are restricted to temperatures $T \ge
T_c$. Recently Leite, Sardelich, and Coutinho-Filho (LSC)
\cite{Leite:99} have analyzed amplitude ratios of the specific heat 
and the susceptibility above $(T>T_c)$ and below $(T<T_c)$ the bulk
critical temperature in the parallel plate geometry for various
boundary conditions. These  
amplitude ratios are functions of the scaling variable $L/\xi_\pm$
(where $\xi_\pm$ is the correlation length above $(+)$ and below $(-)$
the bulk critical temperature) and describe the surface excess and
finite-size contributions of the system. 
Their result for the amplitude function of the susceptibility above $T_c$
(see the expression for $C_+$ in Eq. (22) in Ref. \cite{Leite:99}) can
be expressed in terms of the
scaling function $f(y)$ as introduced in
Eq. (\ref{scaling_excess_sus}). Within this framework the results of
LSC to first order in $\epsilon$ for Dirichlet boundary conditions are
equivalent to the following version of the scaling function $f(y)$: 
\beq
\label{f_d_LCS}
f_{LSC} (y) & = & y^{-2} \Big[ 1 - \frac{\epsilon}{3} +
\frac{\epsilon}{3} \int_0^1 d s f_{1/2} \left( \sqrt{s} \frac{y}{\pi}
\right) - \frac{\epsilon \pi}{6 y} \Big] + {\cal O} (\epsilon^2) 
\enq
with 
\beq
\label{f12}
f_{1/2} (a) & = & \int_a^\infty 
\frac{(u^2-a^2)^{-1/2} d u}{\exp (2 \pi u) -1} \esp .
\enq
For small values of the scaling variable $y$
this scaling function $f_{LSC}(y)$ deviates even within mean-field
theory {\em qualitatively} from the actual correct form $f(y)$ given
in Eqs. (\ref{fyd_2}) and (\ref{f_small_y}). Moreover, already for $y
\lesssim 10$ the difference between $f_{LSC}$ and $f$ becomes larger
than 10\% in ${\cal O}(\epsilon)$ and larger then 25\% within
mean-field theory. These discrepancies are due to the fact that even
within mean-field theory the results of LSC do not reproduce the
correct surface excess contributions \cite{Diehl:85:a} and finite-size 
contribution (Eq. (\ref{fyd_2})).


\section{Scattering cross section}
\label{crosssection}
%
%
\subsection{Scattering theory}
As pointed out in the Introduction the diffuse scattering of X-rays
and neutrons under grazing incidence allows one to probe the local
structure factor near interfaces and in thin films. In this section we
discuss how the singularities of the two-point correlation function
near criticality in a film, as calculated above, translate into
singularities of the diffuse scattering intensity under the
aforementioned experimental conditions.

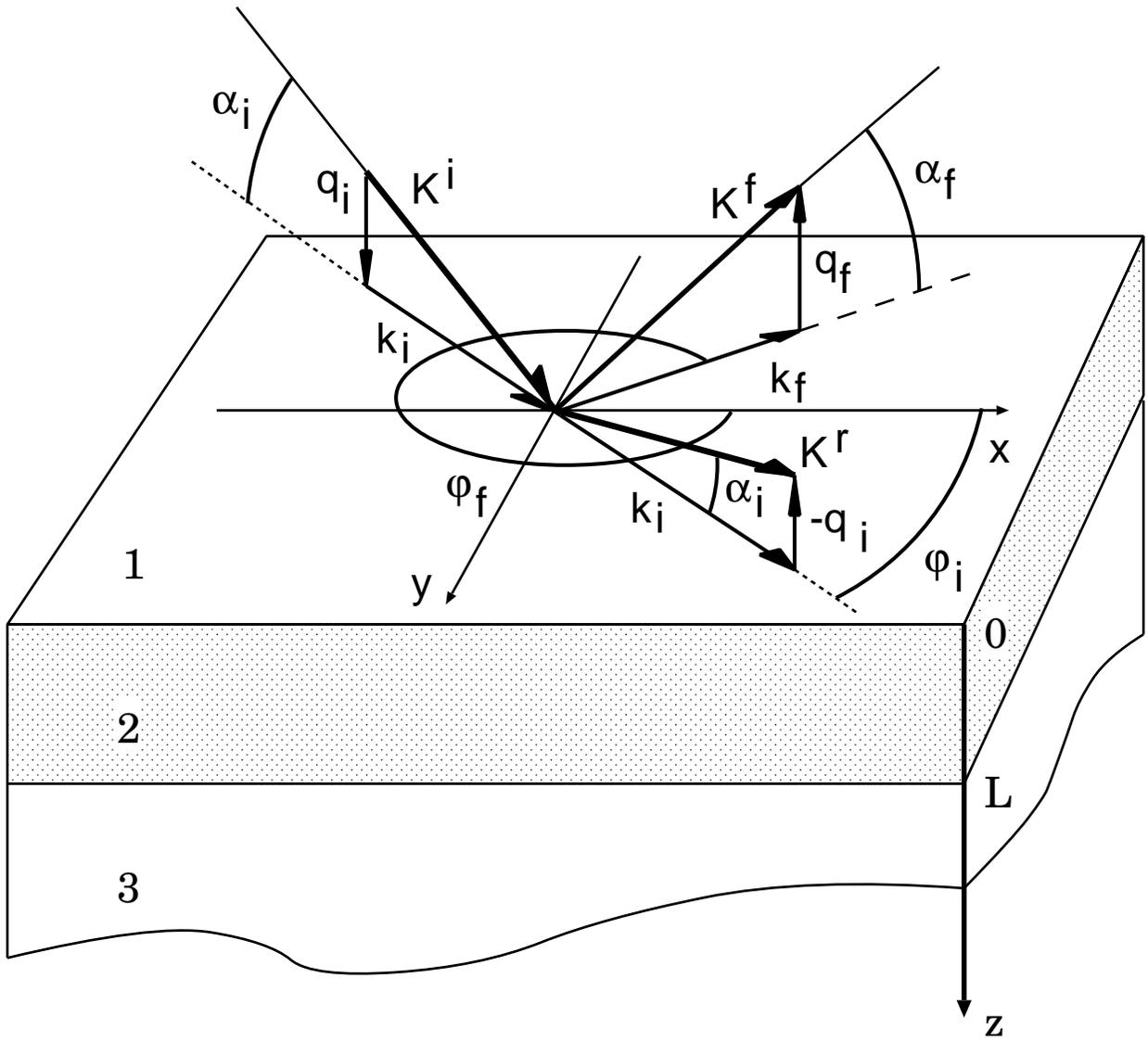
\begin{figure}
\hyphenation{halfspace} 
\caption{
A film $(0 < z < L)$ filled with material 2 is sandwiched in between a
halfspace $z<0$ filled with material 1 (typically vacuum) and a 
halfspace $z>L$ filled with material 3 acting as a supporting
substrate for the film. A plane wave with wave vector ${\bf K}^i =
({\bf k}_i,q_i)=K^i(\cos \alpha_i \cos \varphi_i,\cos \alpha_i \sin
\varphi_i, \sin \alpha_i)$ impinges on the 1-2 interface at $z=0$. The
reflected beam has the wave vector ${\bf K}^r=({\bf k}^i,-q_i)$; the
transmitted beam is not shown. Fluctuations in the film give rise to an
off specular elastic diffuse scattering with ${\bf K}^f$ 
$=({\bf k}_f,q_f)$ 
$=K^f(\cos \alpha_f \cos \varphi_f,\cos \alpha_f \sin \varphi_f, -\sin 
\alpha_f)$, $K^f=K^i=K$.
}
\label{fig_geom}
\end{figure}

We consider a film $(0 \le z \le L)$ composed of a material 2
sandwiched in between two halfspaces filled with material 1 $(z < 0)$ and
3 $(z >L)$, respectively (see Fig. \ref{fig_geom}). An incoming plane
wave of X-rays or neutrons with momentum ${\bf K}^i = ({\bf k}_i,q_i)$
impinges on the 1-2 interface at an angle of incidence $\alpha_i$ so
that $q_i = K^i \sin \alpha$ and ${\bf k}_i = K^i \cos \alpha_i (\cos
\varphi_i, \sin \varphi_i, 0)$. $\lambda = 2 \pi / K^i$ is the
wavelength of the X-rays or neutrons. We assume that the media 1 and 3
are homogeneous and that the 1-2 and 2-3 interfaces are laterally flat
so that their contributions to diffuse scattering can be
ignored. Within the plane of incidence there is a specularly reflected
wave with ${\bf K}^r = ({\bf k}_i,-q_i)$. The mean value of the
electron density in the case of X-rays and of the scattering length in
the case of neutrons determine the intensity of the reflected beam
whereas fluctuations around the mean value give rise to scattered
intensity in off-specular directions ${\bf K}^f = ({\bf k}_f, q_f <
0)$ with $q_f = -K^f \sin \alpha_f$ and ${\bf k}_f = K^f \cos \alpha_f
(\cos \varphi_f, \sin \varphi_f, 0)$. We consider only elastic
scattering, i.e., $K^i = K^f = K^r \equiv K$. (For the more complex
case of neutron scattering under grazing incidence from magnetic
systems see Ref. \cite{Dietrich:85}.)

In order to proceed we assume that the mean values of the electron
density or of the scattering length density in each medium is constant
and varies steplike across the two interfaces 1-2 and 2-3. This gives
rise to the following indices of refraction \cite{Dosch:92}:
\beq
\label{n1n2n3}
z < 0 & : & n = n_1 = 1 \\
0 < z < L & : & n = n_2 = 1-\delta_2 + i \beta_2 \nonumber \\
z > L & : & n = n_3 = 1- \delta_3 + i \beta_3 \nonumber .
\enq
In Eq. (\ref{n1n2n3}) we consider the case that medium 1 is vacuum and
the generic case 
for hard X-rays that $Re \esp n < 1$ in condensed matter. Although for
neutrons one can also have $Re \esp n > 1$, in order to limit the number of
possible relative values of the indices of refraction for the
materials 1, 2, and 3 we do not analyze this latter case in more
detail. For  X-rays $\delta = \lambda^2 \frac{r_e}{2 \pi} \sum_i N_i
Z_i$ and the extinction coefficient $\beta = \frac{\lambda}{4 \pi}
\sum_i N_i \sigma_{a,i} \equiv \frac{\lambda \mu_{abs}}{4 \pi}$ where
$r_e = \frac{e^2}{4 \pi \epsilon_0 m c^2} = 2.814 \cdot 10^{-5}$ \AA
\esp is the classical electron radius, $N_i$ the number density of
atoms of species $i$ with $Z_i$ electrons and absorption cross section
$\sigma_{a,i}$. For neutrons $\delta = \frac{\lambda^2}{2 \pi} \sum_i
N_i b_i$ and $\beta = \frac{\lambda}{4 \pi} \sum_i N_i \sigma_{t,i}$
where $b_i$ is the nuclear scattering length of species
$i$. $\sigma_{t,i}$ is the cross section taking into account
incoherent scattering and nuclear reactions. Typically $\delta$ and
$\beta$ are of the order $10^{-5}$. For $Re \esp n < 1$ total external
reflection occurs for $\alpha < \alpha_c$. For $L=\infty$ one has
$\alpha_{c12} \simeq (2 \delta_2)^{1/2}$ whereas for $L=0$
$\alpha_{c13} \simeq (2 \delta_3)^{1/2}$. Since the angle of total
reflection depends only on the difference $n(z \to -\infty) - n(z \to
+\infty) > 0$, for any finite $0< L < \infty$ the incoming wave is
totally reflected for $\alpha < \alpha_{c13}$, independent of the
index of refraction within the film. Nonetheless the types of waves
propagating in the film depend on whether $\alpha \gtrless
\alpha_{c12}$ (see below). For the present setup the wave field has
the form $\Psi ({\bf r}, {\bf K}^i) = e^{i {\bf k}_i \cdot  
{\bf r}_\|} \psi (z,\alpha)$ with
\beq
\label{wavefield}
\psi(z,\alpha) & = & \left\{ \begin{array}{lcr}
e^{iq_1(\alpha)z}+r_L(\alpha)e^{-iq_1(\alpha)z}&,& z < 0 \\
s_+(\alpha) e^{iq_2(\alpha)z}+s_-(\alpha)e^{-iq_2(\alpha)z}&,& 0 \leq z \leq L \\
t_L(\alpha)e^{iq_3(\alpha)z}&,& z > L 
\end{array} \right.
\enq
where
\beq
\label{ssrt}
r_L(\alpha) & = & \Big( (q_1-q_2)(q_2+q_3) +
e^{2iq_2L}(q_1+q_2)(q_2-q_3) \Big) / \Lambda (\alpha) , \\
s_+(\alpha) & = & 2q_1(q_2+q_3)/ \Lambda (\alpha) \nonumber , \\
s_-(\alpha) & = & 2q_1(q_2-q_3)e^{2iq_2L}/ \Lambda (\alpha) \nonumber
, \\
t_L(\alpha) & = & 4q_1q_2e^{i(q_2-q_3)L}/ \Lambda (\alpha)
\nonumber , \\
\Lambda (\alpha) & = & (q_1+q_2)(q_2+q_3) +
e^{2iq_2L}(q_1-q_2)(q_2-q_3) \nonumber .
\enq
Since the scattering cross section is independent of the intensity of
the incoming beam without loss of generality we have set the amplitude
of $\Psi ({\bf r}, {\bf K}^i)$ equal to 1. The vertical components of
the momentum are given by 
\beq
\label{q1q2q3}
q_1 (\alpha) & = & K \sin \alpha , \\
q_j (\alpha) & = & K \sqrt{n_j^2-\cos^2 \alpha} \simeq K \sqrt{\sin^2
  \alpha - 2 \delta_j + 2 i \beta_j} \nonumber \\
& = & K \sqrt{\sin^2 \alpha - \sin^2 \alpha_{c1j} + 2 i \beta_j} \esp , \esp
j = 2,3 \nonumber .
\enq
In the limiting case that the film turns into a semi-infinite
substrate, i.e., $L=\infty$ one has
\beq
\label{si_eigenfunctions}
\psi_{\infty/2}(z,\alpha) & = & \left\{ \begin{array}{lcr}
e^{iq_1(\alpha)z}+r_{\infty/2}(\alpha)e^{-iq_1(\alpha)z}&, & z < 0 \\
t_{\infty/2}(\alpha)e^{iq_2(\alpha)z}&, & z \geq 0  
\end{array} \right.
\enq
with
\beq
\label{rt}
r_{\infty/2}(\alpha) & = & (q_1-q_2)/(q_1+q_2) , \\
t_{\infty/2}(\alpha) & = & 2q_1/(q_1+q_2) \nonumber . 
\enq
The vertical momentum components $q_j(\alpha)$ have a positive
imaginary part which is due to the extinction coefficient $\beta_j$
for $\alpha > \alpha_{c1j}$ and which is present for $\alpha <
\alpha_{c1j}$ even in the absence of absorption. This gives rise to an
exponentially damped evanescent wave with a penetration depth $l_j =
(Im \esp q_j(\alpha))^{-1}$ which increases steeply for $\alpha \nearrow
\alpha_{c1j}$ and would diverge if $\beta_j = 0$. Within the film
there is a superposition of two fields $s_+(\alpha) e^{i q_2(\alpha) z}$ and
$s_-(\alpha) e^{-i q_2(\alpha) z}$ (Eq. (\ref{wavefield})); in the
three cases $\alpha < \alpha_{c12}$ and $\beta_2 = 0$, $\alpha >
\alpha_{c12}$ and $\beta_2 \not= 0$, and $\alpha < \alpha_{c12}$ and
$\beta_2 \not= 0$, $q_2(\alpha)$ has a nonzero imaginary part leading
to an exponentially increasing and decreasing contribution for 
increasing $z$. The decreasing part corresponds to the damping of the
incident wave whereas the increasing part corresponds to the damping of
the reflected wave generated by the interface 2-3. 

Equation (\ref{wavefield}) describes the wave field $\Psi ({\bf r},
{\bf K}^i)$ in the absence of any fluctuations. This wave field is
scattered at the fluctuating inhomogeneities within the film giving
rise to diffuse scattering intensities in off-specular directions. 

The computation of this intensity requires one to specify the nature
of fluctuations. In the present context this amounts to specifying the
kind of system undergoing the continuous phase transition in the film
and to choose the appropriate order parameter. As described in the
Introduction the most promising candidates for these kind of phenomena
are binary alloys undergoing a continuous order-disorder phase
transition concerning the occupation of fixed lattice sites $\{ {\bf R}_l
\}$. (Magnetic films are equally well suited. However, the magnetic
scattering of neutrons \cite{Dietrich:85} or of X-rays is more
complicated and requires separate analyses. Although the details 
will differ from the analysis given below, the key features of
the singularities are expected to be born out similarly.) In these systems a
given configuration is characterized by spin-type variables $\{ S_l =
\pm 1\}$ such that $S_l = +1 (-1)$ states that the lattice site ${\bf
  R}_l$ is occupied by a $B (A)$ atom. Accordingly the number density
of electrons for such a configuration is
\beq
\label{number_density}
\rho ({\bf r}) & = & \frac{1}{2} \sum_l \left\{ 
\rho_B ({\bf r} - {\bf R}_l)
+\rho_A ({\bf r} - {\bf R}_l)
+ S_l \left[ 
\rho_B ({\bf r} - {\bf R}_l)
-\rho_A ({\bf r} - {\bf R}_l) \right]
\right\}
\enq
where $\rho_{A(B)} ({\bf r})$ is the electron number density in a single
unit cell $V_{cell}$ occupied by an $A (B)$ atom. (In the case of neutron
scattering $\rho ({\bf r})$ stands for the scattering length density
and $\rho_{A(B)} ({\bf r}) = b_{A(B)} \delta ({\bf r})$ where
$b_{A(B)}$ is the mean scattering length of the nuclei of species $A
(B)$.) The ordered state of this system corresponds to a configuration
in which the sign of $S_l$ alternates from one lattice site to any of the
neighboring ones. In this ground state the staggered ''magnetization''
$O\!P_l = S_l 
e^{i}\mbox{\boldmath${^\tau}$\unboldmath}^{ _m \cdot {\bf R}_l}$ 
is spatially constant if the reciprocal lattice vector
\boldmath$\tau$\unboldmath$_m$ of the sublattice structure is chosen such
that $e^{i}\mbox{\boldmath$^\tau$\unboldmath}^{_m \cdot ({\bf R}_l
-{\bf R}_{l'})} = -1$ for nearest neighbor sites ${\bf R}_l, 
{\bf R}_{l'}$. In the reciprocal space the positions of the reciprocal
sublattice vectors \boldmath$\tau$\unboldmath$_m$ are halfway in between the
reciprocal lattice vectors ${\bf G}_m$ with $e^{i {\bf G}_m \cdot 
{\bf R}_l}$ characterizing the underlying lattice structure of the
solid. (For the sake of simplicity as far as the scattering theory is
concerned we do not consider here explicitly the case of
systems like $Fe_3Al$ whose description requires the introduction of
several sublattices.) Upon approaching the critical temperature of the
continuous order-disorder transition the thermal average $\langle
O\!P_l \rangle$ vanishes qualifying $O\!P_l$ as an appropriate order
parameter. 
    
In the critical contribution to the {\em bulk} scattering cross
section a nonzero value of 
$\langle O\!P_l \rangle$ leads to superlattice Bragg peaks
\cite{Dietrich:95}: 
\beq
\label{bulk_super_bragg}
\left( \frac{d \sigma}{d \Omega} \right)^{Bragg}_{bulk} & = & r_e^2
\left( \frac{{\bf K}^f}{K} \times {\bf e} \right)^2 
\langle O\!P_l \rangle^2
\frac{N_V}{V_{cell}} (2 \pi)^3 \sum_m 
\left| \tilde F e^{-W} \right|^2 
\delta ({\bf K}^i-{\bf K}^f-
\mbox{\boldmath$\tau$\unboldmath}_m ) 
\enq
where
$r_e$ is the classical electron radius, ${\bf K}^f/K$ are the
directions of observation, ${\bf e}$ is the polarization vector of the
incoming electromagnetic wave, $\tilde F=(F_A-F_B)/2$ where $F_{A(B)}
({\bf K}) = \int_{V_{cell}} d^3 r \rho_{A(B)} (r) e^{i {\bf K \cdot r}}$ 
is the atomic form factor of the atom $A(B)$, $e^{-W({\bf K})}$ is the
Debye Waller factor and $N_V$ is the number of lattice sites in the
sample. With the independent knowledge of all prefactors in
Eq. (\ref{bulk_super_bragg}) the asymptotic temperature dependence of
$\left( \frac{d \sigma}{d \Omega} \right)^{Bragg}_{bulk}$ yields
$\langle O\!P_l \rangle = {\cal B}' (-t)^\beta$. As discussed in
Sec. \ref{model}  this experimental value for ${\cal B}'$ enters into
Eq. (\ref{G_final}) and there replaces $B$ if $G(p,z_1,z_2,L,t)$
corresponds to the pair correlation function $\langle O\!P_l O\!P_{l'}
\rangle$ as considered below. Similarly the singular diffuse
scattering around a superlattice Bragg peak
$\mbox{\boldmath$\tau$\unboldmath}_m$ is given by
\beq
\label{bulk_diffuse}
\left( \frac{d \sigma}{d \Omega} \right)^{diffuse}_{bulk} & = &
r_e^2 \left( \frac{{\bf K}^f}{K} \times {\bf e} \right)^2
\left| \tilde F e^{-W} \right|^2 \sum_{{\bf R}_l, {\bf R}_{l'}} \left(
  \langle O\!P_l O\!P_{l'} 
\rangle - \langle O\!P_l \rangle \langle O\!P_{l'} \rangle \right) 
e^{i {\bf q} \cdot ({\bf R}_l -{\bf R}_{l'})} \\
& \rightarrow & r_e^2 \left( \frac{{\bf K}^f}{K} \times {\bf e} \right)^2
\left| \tilde F e^{-W} \right|^2 \frac{N_V}{V_{cell}} \esp G_{bulk} 
({\bf q},t) \nonumber 
\enq
with ${\bf q} = {\bf K}^f -{\bf K}^{i} -
\mbox{\boldmath$\tau$\unboldmath}_m$. In the second part of
Eq. (\ref{bulk_diffuse}) we have performed the continuum
limit replacing the lattice sums by integrals (see
Eq. (\ref{op_corr_q}) and the last paragraph in Sec. \ref{model})
because for $\xi \to \infty$ the lattice structure becomes
irrelevant. 
From studying the temperature dependence of Eq. (\ref{bulk_diffuse})
for $T > T_c$ one can infer the correlation length $\xi$ and its amplitude
$\xi^+_0$ introduced in Sec. \ref{model}. We note that for $q$
small compared with the inverse lattice spacing $a$ Eqs. (\ref{op_corr_q}) 
and (\ref{D_B_xi_0}) can be applied to Eq. (\ref{bulk_diffuse})
provided $B$ is replaced by ${\cal B}'$ as determined from
Eq. (\ref{bulk_super_bragg}). 

Equipped with this knowledge about the critical bulk scattering (i.e.,
above the angle of total reflection and for a bulk sample) we can
now turn to the critical diffuse scattering from the film. Within the
so-called distorted wave Born approximation and for the model of the
film as described above one finds the following expression for the
singular part of the coherent scattering cross section
\cite{Dietrich:95}: 
\beq
\label{film_diffuse}
\frac{d \sigma}{d \Omega} & = & r_e^2 \frac{A_\|}{(V^\|_{cell} a)^2} 
\left| \tilde F e^{-W} \right|^2 \Sigma \esp , \\
\Sigma & = & \int\limits_0^L d z_1 \int\limits_0^L d z_2 \psi_f (z_1)
\psi_i (z_1) \psi_i^\ast (z_2) \psi_f^\ast (z_2) G(p,z_1,z_2,L,t) 
\esp , \nonumber  
\enq
where $A_\| = N_\| V^\|_{cell}$ is the illuminated surface area where
$N_\|$ is the number of lattice sites at the surface and $V^\|_{cell}$
is the two-dimensional unit cell of the surface, $a$ is the lattice
spacing of the cubic lattice, $\psi_{i,f}(z) \equiv
\psi(z,\alpha=\alpha_{i,f})$ (see Eq. (\ref{wavefield}) and
Fig. \ref{fig_geom}), and ${\bf p} = {\bf k}^f - {\bf k}^i -
\mbox{\boldmath$\tau$\unboldmath}_m$ assuming that the film surfaces
are cut such that $\mbox{\boldmath$\tau$\unboldmath}_m$ is parallel to
them. $G$ is the lateral Fourier transform of the two-point order
parameter correlation function: 
\beq
\label{g_p_discret}
G(p,z_1,z_2,L,t) & = & \frac{V^\|_{cell}}{N_\|} \sum_{{\bf r}_\|^{(m)}, 
{\bf r}_\|^{(m')}} e^{i {\bf p} \cdot ({\bf r}_\|^{(m)} -  
{\bf r}_\|^{(m')})}  \\
&& \times \left(
\langle O\!P ({\bf r}_\|^{(m)},z_1 ) 
O\!P ({\bf r}_\|^{(m')},z_2 )  \rangle -
\langle O\!P ({\bf r}_\|^{(m)},z_1 ) \rangle
\langle O\!P ({\bf r}_\|^{(m')},z_2 )  \rangle \right) \nonumber \\
& \rightarrow & \int d^2 r_\| \esp 
e^{-i {\bf p} \cdot {\bf r}_\|} \esp G({\bf r}_\|,z_1,z_2,L,t) \nonumber
\enq
on the lattice and in the continuum limit, respectively.
Thus after replacing the nonuniversal amplitude $B$ in
Eq. (\ref{G_final}) by ${\cal B}'$ as obtained from
Eq. (\ref{bulk_super_bragg}) for $\langle O\!P_l \rangle$ we can study
the scattering cross section in Eq. (\ref{film_diffuse}) by using all
the information about $G(p,z_1,z_2,L,t)$ obtained in the previous
section, provided all lengths and $1/p$ are sufficiently large
compared with the lattice spacing $a$ so that the continuum
description is applicable.

In view of the properties of the wave functions $\psi$ ($-$ only their
functional forms for $0 \le z \le L$ enter into $\Sigma$ (see
Eq. (\ref{wavefield}))$-$) and of the scaling form for
$G(p,z_1,z_2,L,t)$ (see Eq. (\ref{G_final})) one has for $\alpha_{i,f},
\alpha_{c12,c13} \ll 1$ and $\beta_{2,3}=0$:
\beq
\label{sigma}
\Sigma & = & {\cal B}' (\xi^+_0)^{d+1} {\cal R} \left
  ( \frac{L}{\xi^+_0} \right)^{3-\eta} \sigma \big( p \xi,
  \frac{L}{\xi}, \frac{l_i}{L}, \frac{l_f}{L},
  \frac{\alpha_i}{\alpha_{c12}}, \frac{\alpha_{c12}}{\alpha_{c13}}
  \big)  
\enq
where the dimensionless function $\sigma$ is given by
(Eq. (\ref{G_final})) 
\beq
\label{ss_sigma}
\sigma & = & \int\limits_0^1 d x_1 \int\limits_0^1 d x_2 \psi_f(z_1=
x_1 L) \psi_i(z_1 = x_1 L) \psi^\ast_i (z_2 = x_2 L) \psi^\ast_f (z_2
= x_2 L) \\
&& x_1^{1-\eta} g_\rii \big( p L x_2, \frac{L}{\xi} x_1, \frac{L}{\xi}
x_2, x_2 \big) \nonumber . 
\enq
The two variables $p \xi$ and $L/\xi$ of $\sigma$ stem from the
scaling function of the pair correlation function whereas the
dependences of $\sigma$ on $l_i/L$, $l_f/L$, $\alpha_i/\alpha_{c12}$,
and $\alpha_{c12}/\alpha_{c13}$ are due to the wave functions. For
$\alpha_{i,f} < \alpha_{c12}$
\beq
\label{lilf}
l_{i,f} & = & \frac{l_0^{(2)}}{\sqrt{1- \left(
      \frac{\alpha_{i,f}}{\alpha_{c12}} \right)^2}}
\enq
correspond to the {\em penetration depths of} the incoming $(i)$ and
outgoing $(f)$ evanescent wave, respectively, {\em within the film
  material 2}. $l_0^{(2)} = (K \alpha_{c12})^{-1}$ is the minimal
penetration 
depth $l_{i,f} (\alpha_{i,f}=0)$ in the film material. Typically $l_0$
is of the order of 30 \AA \cite{Dosch:92}. For $\alpha_i >
\alpha_{c12}$ and $\alpha_f > \alpha_{c12}$ the corresponding
quantities $l_i$ and $l_f$, respectively, are purely imaginary.

%
\subsection{Interplay of length scales}
\label{interplay}

The scattering cross section reflects the rich interplay of five
length scales: $1/p$, $\xi$, $l_i$, $l_f$, and $L$. Scaling reduces
that to four independent scaling variables; moreover there is a
parametric dependence on $\alpha_i/\alpha_{c12}$ and on the material
constant $\alpha_{c12}/\alpha_{c13}$. It is beyond the scope of the
present analysis to provide an exhaustive discussion of the full
dependence on all these variables. Instead we discuss some general
aspects and analyze a few specific cases in more detail in order to
highlight the key features of the diffuse scattering intensity. The
following cases have to be distinguished (for $T \ge T_c$):
\begin{itemize}
\item[I a)] $l_{i,f} \ll L$ and total reflection at 1-3 interface:
  $\frac{d \sigma}{d \Omega}$ is proportional to the scattering volume
  $A_\| min (l_i,l_f)$ 
\begin{itemize}
\item[1.] $\xi \ll l_{i,f} \ll L$: bulk behavior convoluted with
  evanescent waves 
\item[2.] $\xi \sim l_{i,f} \ll L$: crossover bulk / $\frac{\infty}{2}$
  surface behavior convoluted with evanescent waves
\item[3.] $l_{i,f} \ll \xi \ll L$: $\frac{\infty}{2}$ surface behavior
  convoluted with evanescent waves 
\item[4.] $l_{i,f} \ll \xi \sim L$: $\frac{\infty}{2}$ surface
  behavior plus distant wall correction convoluted with evanescent
  waves 
\item[5.] $\l_{i,f} \ll L \ll \xi$: film behavior near one wall
  convoluted with evanescent waves 
\end{itemize}
\item[I b)] $l_{i,f} \ll L$ and no total reflection at 1-3 interface:
  the difference to I a) is exponentially small, i.e., $\sim
  e^{-LK}$. (The volume contribution to $\frac{d \sigma}{d \Omega}$
  from material 3 is insignificant because it does not exhibit
  critical fluctuations.)
\item[II a)] $l_{i,f} \sim L$ and total reflection at 1-3 interface:
  crossover between $\frac{d \sigma}{d \Omega} \sim A_\| min(l_i,l_f)$
  to $\frac{d \sigma}{d \Omega} \sim A_\| L$ 
\begin{itemize}
\item[1.] $\xi \ll l_{i,f} \sim L$: $\frac{\infty}{2}$ surface
  behavior convoluted with film wave functions 
\item[2.] $\xi \sim l_{i,f} \sim L$: crossover bulk / $\frac{\infty}{2}$
  surface behavior convoluted with film wave functions 
\item[3.] $l_{i,f} \sim L \ll \xi \to \infty$: film behavior
  convoluted with film wave functions 
\end{itemize}
\item[II b)] $l_{i,f} \sim L$ and no total reflection at 1-3
  interface: crossover $\frac{d \sigma}{d \Omega} \sim A_\|
  min(l_i,l_f) \to A_\| L$ (Again, the volume contribution from
  material 3 is regarded to be insignificant and is not taken into
  account.)  
\item[III a)] $l_{i,f} \gg L$ and total reflection at 1-3
  interface: $\frac{d \sigma}{d \Omega} \sim A_\| L$ 
\begin{itemize}
\item[1.] $\xi \ll L \ll l_{i,f}$: bulk behavior convoluted with film
  wave functions
\item[2.] $\xi \sim L \ll l_{i,f}$: crossover between bulk and film
  behavior (including two surface contributions and distant wall
  corrections) convoluted with film wave functions
\item[3.] $L \ll \xi \ll l_{i,f}$: film behavior convoluted with film
  wave functions
\item[4.] $L \ll \xi \sim l_{i,f}$: film behavior convoluted with film
  wave functions
\item[5.] $L \ll l_{i,f} \ll \xi \to \infty$: film behavior convoluted
  with film wave functions
\end{itemize}
\item[III b)] $l_{i,f} \gg L$ and no total reflection at 1-3
  interface: $\frac{d \sigma}{d \Omega} \sim A_\| L$ (The volume
  contribution from material 3 is regarded to be insignificant.)
\item[IV a)] $l_{i,f}$ imaginary and total reflection at 1-3
  interface: $\frac{d \sigma}{d \Omega} \sim A_\| L$ 
\begin{itemize}
\item[1.] $\xi \ll L$: three-dimensional bulk behavior probed by
  undistorted plane waves
\item[2.] $L \ll \xi$: film behavior probed by undistorted plane waves
\end{itemize}
\item[IV b)] $l_{i,f}$ imaginary and no total reflection at 1-3
  interface: $\frac{d \sigma}{d \Omega} \sim A_\| L$ (in addition to
  an insignificant volume contribution from material 3).
\end{itemize}

%
\subsection{Susceptibility from the scattering cross section}
\label{sus_cross}

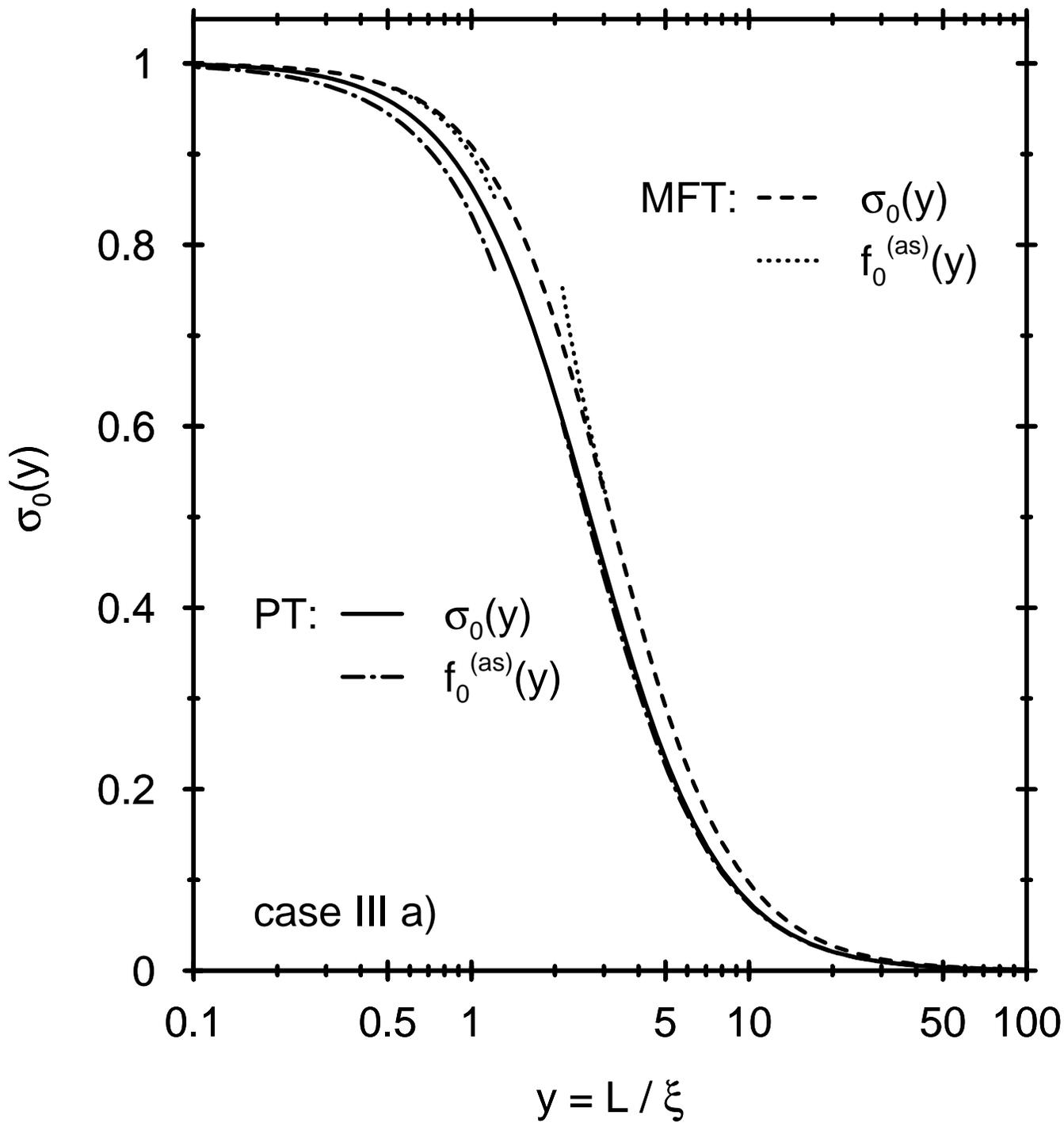
\begin{figure}
\caption{
Scaling function of the scattering cross section $\sigma$ (Eq. (\ref{ss_sigma}))
for large penetration depths $l_{i,f} \gg L$ and vanishing lateral
momentum $p=0$ as a function of the scaling variable $y=L/\xi$ within
MFT (dashed line) and perturbation theory (full line).
The dotted and dashed-dotted lines correspond to the asymptotic
behaviors $f^{(as)}_0(y)$ of the normalized scaling function $f_0(y) =
f(y)/f(0)$ of the total susceptibility $f(y)$
(Eqs. (\ref{f_large_y}) and (\ref{f_small_y})) in mean-field theory
(MFT) and in perturbation theory (PT) to first order in $\epsilon$
using in addition the best available
exponents, respectively.
}
\label{fig_sigma_suz}
\end{figure}

For large penetration depths $l_{i,f} \gg L$ the product of wave
fields in Eq. (\ref{ss_sigma}) is approximately constant. In this case
for $p=0$ the universal scaling function $\sigma$ of Eq. (\ref{ss_sigma})
reduces up to a prefactor to the scaling function $f$ of the total
susceptibility (see Eqs. (\ref{scaling_excess_sus}) and
(\ref{fy_def})), i.e.,
\beq
\label{cross_to_suz}
\sigma \left( \frac{L}{\xi} \right) = \sigma \left( p\xi=0,
  \frac{L}{\xi} , \frac{l_i}{L}=\infty, 
  \frac{l_f}{L}=\infty, \frac{\alpha_i}{\alpha_{c12}} < 1,
  \frac{\alpha_{c12}}{\alpha_{c13}} \right) & \sim & f \left(
  \frac{L}{\xi} \right) .
\enq
In this limit the dependences on $\alpha_i/\alpha_{c12}$ and on
$\alpha_{c12}/\alpha_{c13}$ drop out for $\alpha_i < \alpha_{c13}$;
for $\alpha_i > \alpha_{c13}$ there is an insignificant bulk
contribution from material 3. 
The five different cases 1. $-$ 5. in III a) are characterized by the
various contributions of asymptotic behaviors to the scaling function
$\sigma(y=L/\xi) \sim f(y)$ (see Eqs. (\ref{f_large_y}) and
(\ref{f_small_y})), i.e., bulk: ${\cal A}y^{-2+\eta}$, surface:  
${\cal B}y^{-3+\eta}$, distant wall: ${\cal C}y^{-3+\eta}e^{-y}$, and
film behavior: ${\cal D}+{\cal E}y^{1/\nu}$. In
Fig. \ref{fig_sigma_suz} we show the normalized scaling function of the
scattering cross section $\sigma_0(y) = \sigma (y) / \sigma (0)$
(Eq. (\ref{ss_sigma})) within mean-field and within first-order
perturbation theory as well as the asymptotic behaviors of the
normalized scaling function $f_0(y) = f(y)/f(0)$ of the total
susceptibility $f(y \to 0)$ 
(Eq. (\ref{f_small_y})) and $f(y \to \infty)$ (Eq. (\ref{f_large_y})) 
using mean-field exponents and amplitudes and best values for the
exponents and amplitudes to first order in $\epsilon$, respectively.
The cases III a) or b) with lateral momentum $p=0$ are the appropriate
scattering setups in order to measure the various asymptotic behaviors of
the total susceptibility by varying the temperature. 

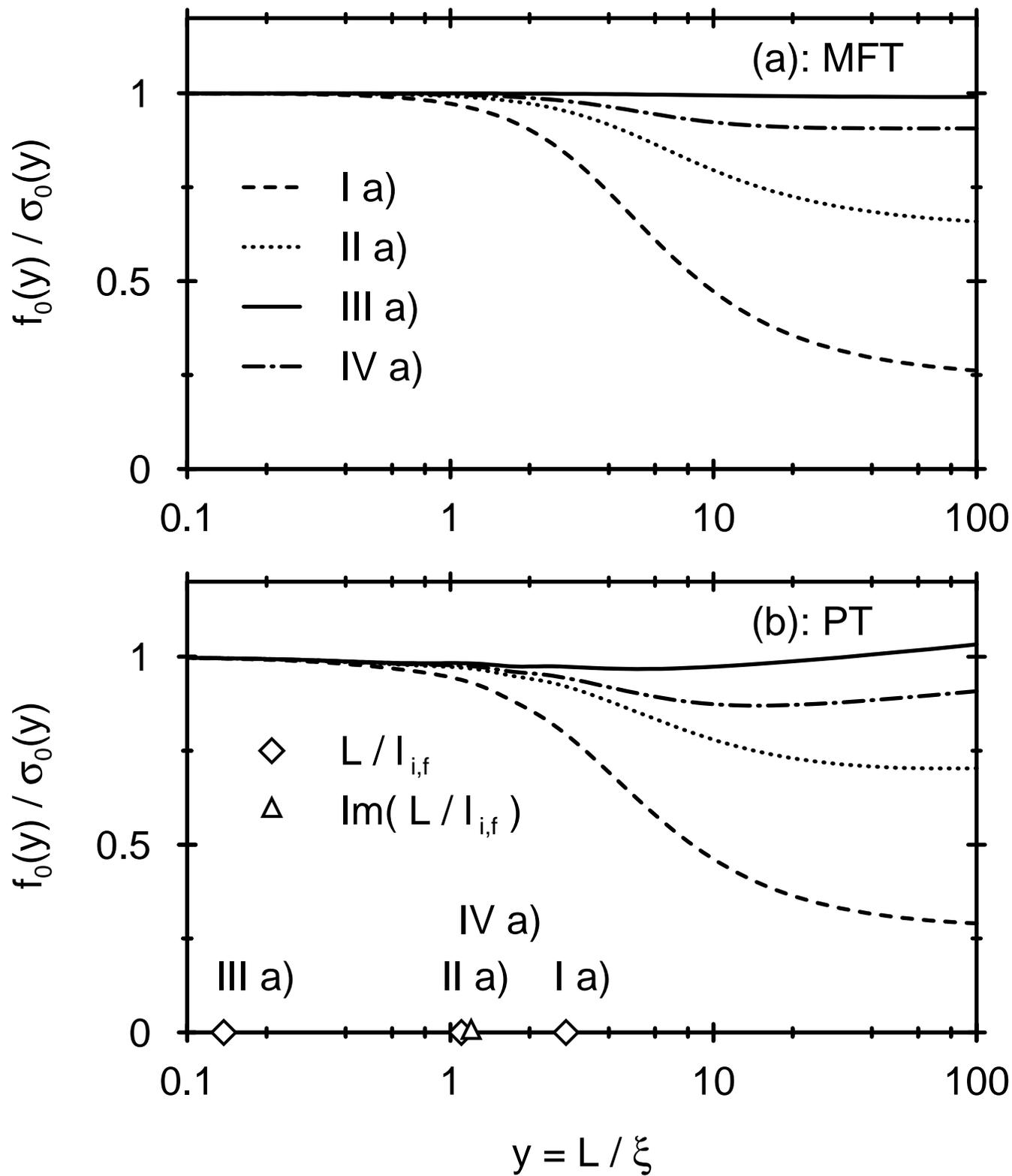
\begin{figure}
\caption{
Ratio of the normalized scaling functions of the total susceptibility
(Eq. (\ref{scaling_excess_sus})) and of the scattering cross section
(Eq. (\ref{ss_sigma})) within mean-field ((a): MFT) and within perturbation
theory ((b): PT). We use the normalization
$\sigma_0(y)=\sigma(y)/\sigma(0)$ and $f_0(y)=f(y)/f(0)$. The various
lines in (a) and (b)
correspond to different penetration depths $l_{i,f}$: I a)
$l_{i,f} \ll L$, II a) $l_{i,f} \sim L$, III a) $l_{i,f} \gg L$, and
IV a) $l_{i,f}$ imaginary (no total reflection at the interface 1-2)
as marked in (b). The curves correspond to $l_i = l_f$. In the case IV
a) the indicated value of $L/l_{i,f}$ corresponds to its imaginary
part. 
}
\label{fig_f_durch_sigma}
\end{figure}

Figure \ref{fig_f_durch_sigma} shows the ratio of the normalized
scaling functions $f_0(y)/\sigma_0(y) = \frac{f(y)}{\sigma (y)}
\frac{\sigma(0)}{f(0)}$ for all four cases I a) $-$ IV 
a) within mean-field theory (MFT) and within perturbation theory (PT),
respectively. For large penetration depths $l_{i,f} \gg L$ (see case III
a)) the deviation of the scaling 
function $\sigma_0 (y)$ of the scattering cross section from the scaling
function $f_0(y)$ of the total susceptibility is small (see the solid
lines in Fig. \ref{fig_f_durch_sigma} (a) and (b)). If the penetration depths are
of the order of the film thickness, $l_{i,f} \sim L$ (see case II a)),
the wavefields in Eq. (\ref{ss_sigma}) contribute and the deviation from
the total susceptibility becomes visible at large values of the
scaling variable $y$. For $y \to 0$ and $y \to \infty$ the dotted
lines attain constant values so that there are the same critical
exponents but different amplitudes for the leading asymptotic
behaviors of $\sigma_0$ and $f_0$. If the penetration
depths are smaller than the film thickness, $l_{i,f} \ll L$ (see case I
a)), this deviation is much more pronounced (see dashed lines). The
difference in the amplitudes is decreased if $\alpha_{i,f} >
\alpha_{c12}$, i.e., for imaginary $l_{i,f}$ (see case IV a) and
dashed-dotted lines). 

%
\subsection{Dependence on the film thickness}
\label{L_dependence}

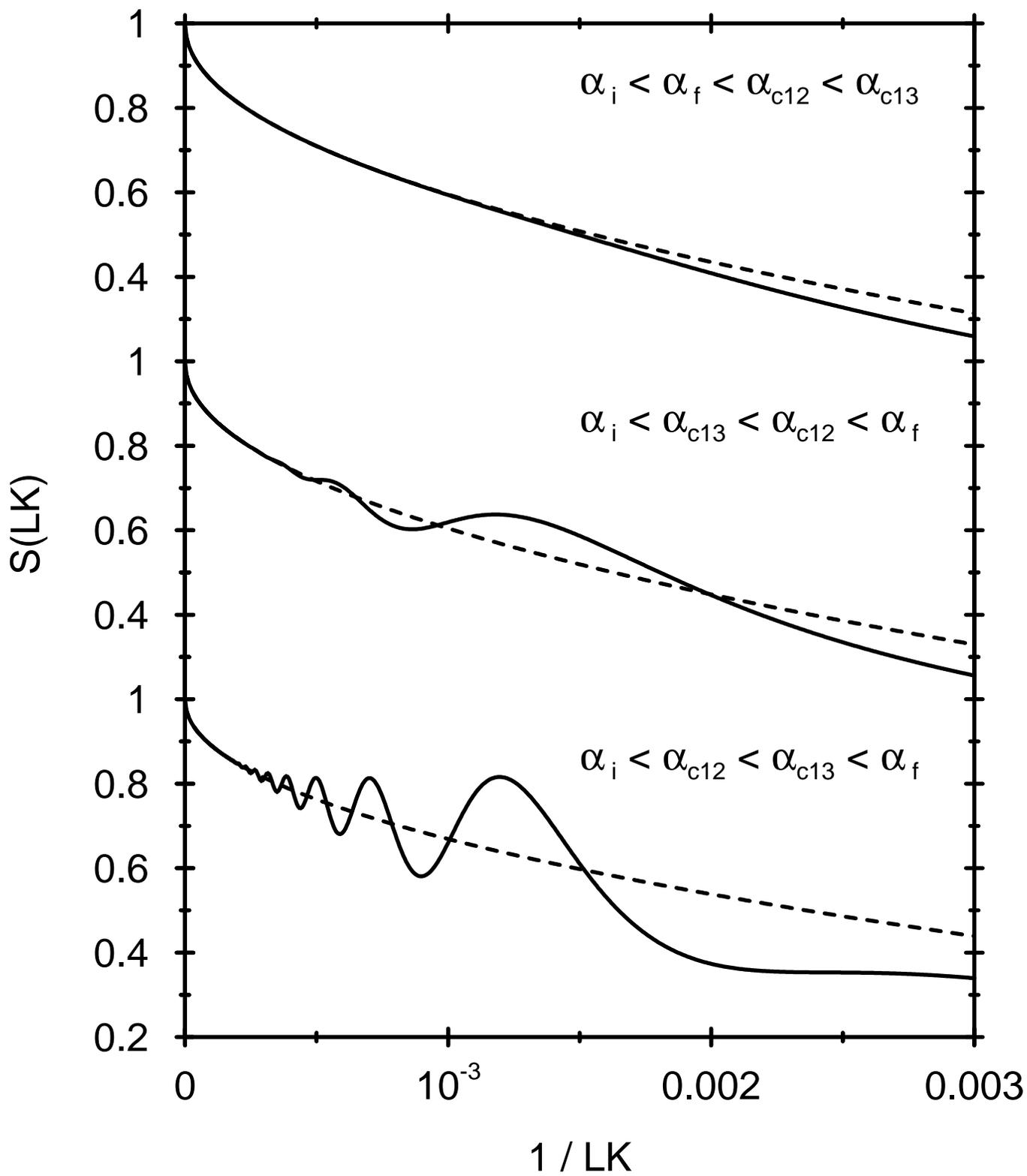
\begin{figure}
\caption{
Scattering function $S(LK;\alpha,\beta)$
(Eq. (\ref{scattering_L}), full lines) and its asymptotic form $S(LK
\to \infty;\alpha,\beta)$ (Eq. (\ref{s_LK_asymptotic_stat}), dashed
lines) for three different scattering geometries:
$\alpha_i$ $<$ $\alpha_f$ $< \alpha_{c12} < \alpha_{c13}$,  
$\alpha_i < \alpha_{c13} < \alpha_{c12} < \alpha_f$, and 
$\alpha_i < \alpha_{c12} < \alpha_{c13} < \alpha_f$. For $\alpha_{i,f}
< \alpha_{c12,c13}$ the
 scattering function decreases monotonously. If
one of the angles $\alpha_i$ or $\alpha_f$ is larger than
$\alpha_{c12}$ oscillations emerge. This effect is enhanced if
$\alpha_{c13} > \alpha_{c12}$. In the 
asymptotic form of
Eq. (\ref{s_LK_asymptotic_stat}) (dashed lines)
there are no
oscillations. In all three cases $\beta_{2,3}=0.3 \cdot 10^{-5}$ and
$(\alpha_i, \alpha_f, \alpha_{c12}, \alpha_{c13})=$ 
$(0.06^o, 0.11^o, 0.26^o, 0.36^o)$, 
$(0.06^o, 0.40^o, 0.36^o, 0.26^o)$, and
$(0.06^o, 0.40^o, 0.26^o, 0.36^o)$, respectively. 
}
\label{fig_s_LK}
\end{figure}

In order to reveal the $(1/L)^{-1+\eta_\|}$ cusp singularity in the
scattering cross section. We consider the case $p=t=0$ and introduce
the corresponding scattering function
\beq
\label{sigma_film_L}
\Sigma_L & = & \int\limits_0^L d z_1 \int\limits_0^L d z_2 \psi_f (z_1)
\psi_i (z_1) \psi_i^\ast (z_2) \psi_f^\ast (z_2) G(p=0,z_1,z_2,L,t=0),
\enq
where the wave fields are given in Eq. (\ref{wavefield}). For the
correlation function $G$ we use the asymptotic expansion given by
Eq. (\ref{corr_L_z1z2}). Furthermore we introduce the scattering
function of the semi-infinite system
\beq
\label{sigma_si_L}
\Sigma_\frac{\infty}{2} & = & \int\limits_0^\infty d z_1
\int\limits_0^\infty d z_2 
\psi^{(f)}_\frac{\infty}{2} (z_1) 
\psi^{(i)}_\frac{\infty}{2} (z_1) 
\psi^{(i)\ast}_\frac{\infty}{2} (z_2) 
\psi^{(f)\ast}_\frac{\infty}{2} (z_2) G(p=0,z_1,z_2,L=\infty,t=0),
\enq
with the wave fields and the correlation function given in
Eq. (\ref{si_eigenfunctions}) and Eq. (\ref{gompper_firstorder_z1z2}),
respectively. 
The ratio of Eqs. (\ref{sigma_film_L}) and (\ref{sigma_si_L})
defines the scattering function
\beq
\label{scattering_L}
S(LK;\alpha_i,\alpha_f,\alpha_{c12},\alpha_{c13},\beta_2,\beta_3) & =
& \frac{\Sigma_{L}}{\Sigma_\frac{\infty}{2}}
\enq
for $p=t=0$, where the film thickness $L$ and the momentum $K$ of the
scattered wave form the scaling variable, the angles
$\alpha=\{\alpha_{i,f},\alpha_{c12,c13}\}$ characterize the 
scattering geometry, and the   
extinction coefficients $\beta=\{\beta_2, \beta_3\}$ take into account photo
absorption. From Eqs. (\ref{dd_def}) - (\ref{aa_si}) in Appendix
\ref{details_crosssection} one obtains the asymptotic expansion 
\beq
\label{s_LK_asymptotic}
S(LK \to \infty;\alpha,\beta) & = & 
s_0 (LK;\alpha,\beta) + s_1(LK;\alpha,\beta) C_1 \Big( \frac{1}{LK}
\Big)^{-1+\eta_\|} + \dots ,
\enq
with 
\beq
\label{s0_s1_LK}
s_0 (LK \to \infty;\alpha,\beta) & \sim & 1 + s_0^{(1)}(LK;\alpha,\beta)
e^{-LK s_0^{(2)}(\alpha,\beta)}, \\ 
s_1 (LK \to \infty;\alpha,\beta) & \sim & s_1^{(0)}(LK=\infty;\alpha,\beta)
+ s_1^{(1)}(LK;\alpha,\beta) e^{-LK s_1^{(2)}(\alpha,\beta)} \nonumber ,
\enq
and $C_1$ given by Eq. (\ref{A1_B1_C1}). The functions $s_0$ and
$s_1$ carry the $L$-dependence of the wave functions
(see Appendix \ref{ssss}). The $L$-dependence due to the correlation
function is given by the cusp singularity $C_1 (1/LK)^{-1+\eta_\|}$. 
The range of the values of
the scaling variable $(LK)^{-1}$ is limited by the validity of the
continuum theory applied here, i.e., $L \gtrsim 30$ \AA \esp and the distorted-wave
Born approximation, i.e., $K \gtrsim 1$ \AA $^{-1}$, leading to
$(LK)^{-1} \lesssim \frac{1}{30}$. For small angles, i.e., for grazing
incidence scattering experiments Eq. (\ref{q1q2q3}) reduces
\beq
\label{q1q2q3_small}
q_1 (\alpha) & \simeq & K \alpha , \\
q_j (\alpha) & \simeq & K \sqrt{\alpha^2 - \alpha_{c1j}^2 
+ 2 i \beta_j} , \esp j = 2,3 \nonumber .
\enq
Photo absorption, $\beta_2 \not=0$, or evanescent scattering,
$\alpha_{i,f} < \alpha_{c12}$, turn $q_2$ into an imaginary quantity,
which leads to a real part of $s_0^{(2)}$ and $s_1^{(2)}$ in
Eq. (\ref{s0_s1_LK}). If at least one angle 
$\alpha_i$ or $\alpha_f$ is larger than the critical angle
$\alpha_{c12}$ the functions $s_0^{(1)}$ and $s_1^{(2)}$ have a real
and an imaginary 
part. In the latter case one expects that the scattering
function $S$ in Eq. (\ref{s_LK_asymptotic}) exhibits an oscillatory
behavior. In Fig. \ref{fig_s_LK} we show the exponentiated scattering
function and its 
asymptotic form for various scattering geometries. The exponentiated
form is obtained by subtracting the leading behavior of the one-loop
$\epsilon$-expanded result (defined by Eq. (\ref{scattering_L})) 
and by adding the leading 
behavior (see Eq. (\ref{g_as})) calculated with the best available
critical exponents ($\eta \simeq 0.031$, $\eta_\perp \simeq 0.75$, and
$\eta_\| \simeq 1.48$).
The dashed line in 
Fig. \ref{fig_s_LK} corresponds to the leading asymptotic behavior,
if the $L$ dependence of the wave fields is neglected:
\beq
\label{s_LK_asymptotic_stat}
S(LK \to \infty;\alpha,\beta) & = & 
1 + s^{(0)}_1(LK=\infty;\alpha,\beta) C_1 \Big( \frac{1}{LK}
\Big)^{-1+\eta_\|} + \dots \esp .
\enq
Thus the full lines in Fig. \ref{fig_s_LK} take into account
the whole $L$ dependence stemming from both the scattering theory and
the correlation function,
whereas the dashed lines take into account only the leading asymptotic
$L$ dependence of the correlation function.
The oscillatory behavior appearing for $\alpha_i \lessgtr \alpha_{c12}
\lessgtr \alpha_f$ stems from the scattering theory (see
Fig. \ref{fig_s_LK}). 

For the case $\alpha_{i,f} < \alpha_{c12,c13}$ in Fig. \ref{fig_s_LK}
half of the maximum value of the scattering function $S$ is reached
for $LK \simeq 1.5 \cdot 10^{-3}$. This corresponds to a film thickness
$L \simeq 600$ \AA, i.e., $200$ monolayers (with $K \simeq 1$ \AA$^{-1}$
and 1 monolayer $\simeq 3$ \AA \esp thick); 90\% of the maximum value of
$S$ is
reached for $LK \simeq 5 \cdot 10^{-5}$ which corresponds to a film
thickness $L \simeq 20000$ \AA \esp or 6700 monolayers. This
demonstrates the 
slow convergence to the semi-infinite limit. The spatial resolution is
determined by the uncertainty of the film thickness. With  $\Delta L
\simeq 3$ \AA \esp (1 monolayer) this gives $K \Delta L \simeq 3$
leading to a resolution of $\Delta (LK)^{-1} \sim 3/(LK)^2$, which is
not visible on the scale of Fig. \ref{fig_s_LK}. Based on these
considerations we conclude that the oscillations are experimentally
accessible.

%
\subsection{Emergence of cusp singularities}
\label{emergence}

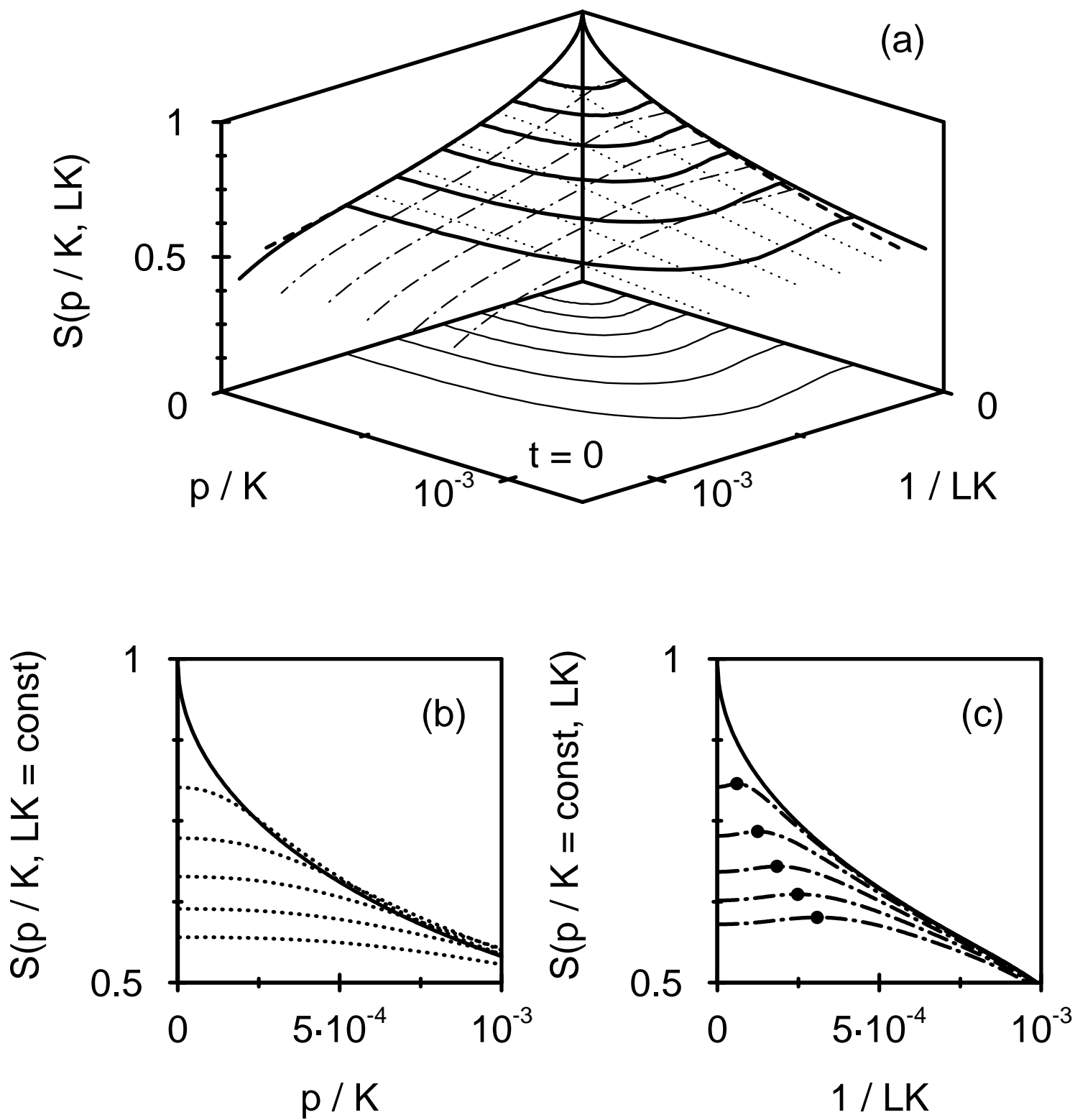
\begin{figure}
\caption{Scattering function
$S(p/K,LK;\alpha_{i,f},\alpha_{c12,c13},\beta_{2,3})$
(Eq. (\ref{scattering_pL})) for $t=0$. (a) shows the
exponentiated scaling functions $S(p/K,LK=\infty;\alpha,\beta)$
(Eq. (\ref{mft_cross_p_Loo_t})) and $S(p/K=0,LK;\alpha,\beta)$
(Eq. (\ref{mft_cross_p0_L_t0})) and their corresponding leading
asymptotic behaviors 
(Eqs. (\ref{mft_pk_cusp}) and (\ref{mft_Lk_cusp}), respectively)
(dashed lines). For the leading asymptotic behavior we use the best
available exponent $\eta_\| \simeq 1.48$ and an amplitude function
which is consistent with the mean-field expression (see
Eq. (\ref{mft_cross_pL})). The corrections to the leading asymptotic
behavior are calculated within mean-field theory.
For the scaling function $S(p/K,LK;\alpha,\beta)$ we
plot contour lines ($S=0.8, 0.75, 0.7, 0.65, 0.65, 0.6$) and their 
projections onto the $(p/K, 1/(LK))$ plane (full lines) which clearly
deviate from circular shapes, lines for $S(p/K,LK=1.5 \cdot 
10^{-4}, 3 \cdot 10^{-4}, 4.5 \cdot 10^{-4}, 6 \cdot 10^{-4}, 7.5
\cdot 10^{-4};\alpha,\beta)$ (dotted lines) and $S(p/K=1.5 \cdot
10^{-4}, 3 \cdot 10^{-4}, 4.5 \cdot 10^{-4}, 6 \cdot 10^{-4}, 7.5
\cdot 10^{-4},LK;\alpha,\beta)$ (dashed-dotted lines). In (b) and (c)
we show the aforementioned vertical cross sections. The emergence of
the $(1/(LK))^{-1+\eta_\|}$ 
cusp is not monotonous; the vertical cross sections exhibit maxima
($\bullet$) at $1/L \not= 0$. The scattering parameters are chosen
that $\alpha_i < \alpha_f < \alpha_{c12} < \alpha_{c13}$ with
$(\alpha_i, \alpha_f, \alpha_{c12}, \alpha_{c13})=$
$(0.06^o, 0.11^o, 0.26^o, 0.36^o)$ and
$\beta_2=\beta_3=0.3 \cdot 10^{-5}$. 
}
\label{fig_s_pL}
\end{figure}

In the following we analyze how the cusp singularities emerge in the
limit of vanishing scaling variables. To his end we chose as an
example a scattering function of the two scaling variables $p/K$
and $LK$. Analogous to Eq. (\ref{sigma_film_L}) we define the quantity
\beq
\label{sigma_film_pL}
\Sigma_{p,L} & = & \int\limits_0^L d z_1 \int\limits_0^L d z_2 \psi_f (z_1)
\psi_i (z_1) \psi_i^\ast (z_2) \psi_f^\ast (z_2) G(p,z_1,z_2,L,t=0) .
\enq
Together with Eq. (\ref{sigma_si_L}) this leads to the scattering
function
\beq
\label{scattering_pL}
S(p/K,LK;\alpha,\beta) & =
& \frac{\Sigma_{p,L}}{\Sigma_\frac{\infty}{2}}
\enq
where $\alpha=\{\alpha_{i,f},\alpha_{c12,c13}\}$ denotes the set of
angles and $\beta=\{\beta_{2,3}\}$ the extinction coefficients. As in
Subsec. \ref{correlation_zz} and Eq. (\ref{polar_coordinates}) we 
introduce polar coordinates
\beq
\label{polar_coord_scattering}
\omega & = & \sqrt{(p/K)^2+(L K)^{-2}} , \esp \varphi = \arctan(pL)
\esp , \\
(LK)^{-1} & = & \omega \cos \varphi, \esp p/K = \omega \sin \varphi
\esp .
\nonumber
\enq
This leads to the relation
\beq
\label{polar_scattering_function}
S(p/K,LK;\alpha,\beta) & = & S(\omega \sin \varphi, (\omega \cos
\varphi)^{-1};\alpha,\beta) = S_{polar} (\omega, \varphi;\alpha,\beta)
\enq
so that the leading asymptotic behavior is given by
\beq
\label{asympt_pol_scat_fctn}
S_{polar} (\omega \to 0, \varphi;\alpha,\beta) & = &
S_0(\varphi;\alpha,\beta) + S_1(\varphi;\alpha,\beta) \esp
\omega^{-1+\eta_\|} + \dots 
\enq
with $S_0(\varphi;\alpha,\beta) = 1$.
The amplitude $S_1$ of the leading asymptotic behavior
$\omega^{-1+\eta_\|}$ depends not only on the polar variable
$\varphi$, as it is the case for the corresponding correlation
function (see Subsec. \ref{correlation_zz}), but also on the parameters
$\alpha$ and $\beta$ characterizing the scattering process. Within
mean-field theory this amplitude is defined in Appendix
\ref{mft_crosssection} by Eq. \ref{mft_cross_pL}. In
Fig. \ref{fig_s_pL} (a) we show the exponentiated scattering function
$S(p/K,LK;\alpha,\beta)$ (Eq. (\ref{scattering_pL})), where we have
subtracted the leading asymptotic behavior from the mean-field
expression of the scattering function and added the
exponentiated form. (See Eqs. (\ref{mft_cross_pLt}) and
(\ref{mft_cross_pL}) in Subappendix \ref{mft_crosssection}); the
scattering function $S$ (Eq. (\ref{scattering_pL})) is a sum (see
Eq. (\ref{film_functions}) in Appendix \ref{ssss}) of functions of the
type ${\cal S}$ as discussed in Eq. (\ref{mft_cross_pLt}) in
Subappendix \ref{mft_crosssection}.) Figure \ref{fig_s_pL} (b)
illustrates the emergence of the $(p/K)^{-1+\eta_\|}$ cusp for
increasing film thickness, i.e., $(LK)^{-1} \to 0$. Figure
Fig. \ref{fig_s_pL} (c) shows the emergence of the 
$(1/(LK))^{-1+\eta_\|}$ cusp for vanishing lateral momentum $p/K \to
0$. In the later case the vertical cross sections of the manifold are
not monotonous; 
they exhibit a maximum ($\bullet$) at $1/L \not= 0$. Figure
\ref{fig_s_pL} corresponds to scattering angles $\alpha_{i,f} <
\alpha_{c12,c13}$ which yields a monotonous behavior of the scattering
function. Analogous considerations describe the emergence of the cusp
singularities in the $\xi$-$L$ and $\xi$-$p$ dependences (see Subappendix
\ref{mft_crosssection}). 


\section{Summary}

By using fieldtheoretic renormalization group theory we have studied
the singular part of the two-point correlation function in a film of
thickness $L$ near the critical point $T_c$ of the corresponding bulk
system. For $T \ge T_c$ and Dirichlet boundary conditions we have
obtained the following main results:
\noindent

(1) The two-point correlation function as a function of the lateral
momentum ${\bf p}$ corresponding to the $d-1$ translationally
invariant directions of the film geometry, the coordinates $z_1$ and
$z_2$ perpendicular to the parallel surfaces, the film thickness $L$, 
and temperature $t=(T-T_c)/T_c$ (or equivalently the bulk
correlation length $\xi = \xi_0^+ t^{-\nu}$) exhibits three cusp
singularities: $p^{-1+\eta_\|}$ for $t=0$ and $L=\infty$,
$(1/\xi)^{-1+\eta_\|}$ for $p=0$ and $L=\infty$, and
$(1/L)^{-1+\eta_\|}$ for $p=t=0$ (see Eqs. (\ref{corr_p_z1z2}) -
(\ref{corr_L_z1z2}) and Fig. \ref{fig_p_t_L}). The emergence of these
three cusp singularities is revealed by studying appropriate scaling
functions of two scaling variables (see Eqs. (\ref{g_pt}) -
(\ref{g_tL}) and Figs. \ref{fig_H1H2H3_phi} - \ref{fig_L_cusp_pt}).  

(2) The film correlation function calculated up to first order
perturbation theory in $\epsilon=4-d$ satisfies the so-called product
rule derived by Parry and Swain for the correlation function algebra
of inhomogeneous fluids in Ref. \cite{Parry:97} (see
Eq. (\ref{parry_product})). 

(3) By setting $p=0$ and integrating over the perpendicular
coordinates $z_1$ and $z_2$ we obtain the total susceptibility of the
film (Eq. (\ref{def_exess_sus})). Its dependence on $L$ and $\xi$ is
described by a universal scaling function $f(y=L/\xi)$ (see
Eq. (\ref{scaling_excess_sus}) and Fig. \ref{fig_f_d}) and exhibits a
typical film behavior: $f(y)$ is analytic for $y \to 0$ and $f(y \to
\infty)$ contains the bulk-, surface-, and finite-size contributions
(see Eqs. (\ref{f_small_y}) and (\ref{f_large_y}),
respectively). These properties are similar to those of the specific
heat of a critical film \cite{Krech:92:a}. Our results correct
previous findings in the literature \cite{Nemirovsky:86:a,Leite:99}
(see the discussions of Eqs. (\ref{one_loop_total_sus}) and
(\ref{f_d_LCS})). 

(4) In view of proposed experimental tests with X-rays and neutrons
under grazing incidence (see Fig. \ref{fig_geom}), as discussed in
detail in Sec. \ref{secI}, we have calculated the critical diffuse
scattering from the film within the so-called distorted wave Born
approximation . The scattering intensity is a function of the lateral
momentum transfer $p$, film thickness $L$, bulk correlation length
$\xi$, penetration depths $l_{i,f}$ of the incoming $(i)$ and outgoing
$(f)$ waves, the critical angles of total reflection $\alpha_{c12}$
and $\alpha_{c13}$ and the extinction coefficients $\beta_2$ and
$\beta_3$ of the film (2) and of the underlying substrate (3) (see
Fig. \ref{fig_geom}). 

(5) For various ratios of $L$, $\xi$, and $l_{i,f}$ the scattering
function shows the crossover between analytic, bulk, surface, and
finite-size behavior (see Figs. \ref{fig_sigma_suz} and
\ref{fig_f_durch_sigma}). By varying the temperature, a scattering
experiment for $p=0$ and $l_{i,f} \gg L$ gives access to the
aforementioned scaling function $f(y)$ of the total susceptibility
(Eq. (\ref{cross_to_suz})). 

(6) For $p=t=0$ the leading singular behavior of the scattering
function is given by the cusp singularity $(1/LK)^{-1+\eta_\|}$,
where $K$ is the momentum of the incoming wave
(Eq. (\ref{s_LK_asymptotic})). The maximal scattering intensity for $L
\to \infty$ is reached only very slowly. For certain scattering
geometries  the $L$-dependence exhibits an oscillatory behavior (see
Fig. \ref{fig_s_LK}). 

(7) The film thickness and momentum cusp singularities of the
correlation function are borne out in the scattering cross section and
are analyzed in Fig. \ref{fig_s_pL}. 


\section*{Acknowledgements}

We thank E. Eisenriegler, A. Hanke, and M. Krech for helpful
discussions.
This work has been supported by the 
German Science Foundation through
Sonderforschungsbereich 237 
{\em Unordnung und gro{\ss}e Fluktuationen}.


\appendix

\section{Amplitudes}
\label{amplitudes}
The amplitudes of the singular behavior of bulk correlation functions
are nonuniversal. There are two independent ones in the sense that any
two of them allow one to express any other in terms of these two and
universal amplitude ratios \cite{Bervillier:76,Privman:91}. As one of
these nonuniversal amplitudes in Sec. \ref{model} we have introduced
and fixed the amplitude $\xi_0^+$ of the bulk correlation length 
(see Eq.(\ref{mu_fixing})). Other nonuniversal amplitudes are given
by the  temperature dependence of the mean value of the field $\phi
({\bf x})$ below $T_c$,
\beq
\label{op_below_tc}
\langle \phi ({\bf x}) \rangle & = & B (-t)^\beta ,
\enq
by the decay of the two-point correlation function in real space at
$T_c$ for large distances $|{\bf x} - {\bf x}'|$,
\beq
\label{op_corr_r}
\langle \phi ({\bf x}) \phi ({\bf x'}) \rangle & = & D |{\bf
x}-{\bf x'}|^{-(d-2+\eta)} ,
\enq
and in momentum space for small $q$,
\beq
\label{op_corr_q}
\int d^d x e^{i {\bf q} \cdot ({\bf x}-{\bf x}')} \langle \phi 
({\bf x}) \phi ({\bf x}') \rangle = G_{bulk} (q,t=0) = \hat D
q^{-2+\eta} , 
\enq
where 
\beq
\label{hat_D_and_D}
\hat D / D& = & X = 2^{2-\eta} \pi^{d/2}
\frac{\Gamma(1-\eta/2)}{\Gamma(d/2-(1-\eta/2))} .
\enq
$\hat D$ can be expressed in terms of $B$, $\xi_0^+$, and a universal
number $R$ \cite{Bervillier:76,Privman:91}:
\beq
\label{D_B_xi_0}
\hat D & = & B^2 (\xi_0^+)^{d-2+\eta} R
\enq
with $R=R_c Q_3/(R_\xi^+)^d$. For $(n,d)=(1,3)$ one has $R_c
\simeq 0.066$, $Q_3 \simeq 0.922$, and $R_\xi^+ \simeq 0.27$
\cite{Bervillier:76,Privman:91} leading to $R \simeq 3.09$. 

A Fourier transformation in the $z$-direction of the bulk correlation
function $G_{bulk}(q) = \hat D q^{-2+\eta}$ with $q^2 = p^2 + k^2$ is
leading to its $p$-$z$-representation 
\beq
\label{ft_bulk}
G_{bulk}(p,z_1-z_2) & = & \int_{-\infty}^\infty \frac{d k}{2
\pi} \hat D \frac{e^{i k (z_1-z_2)}}{(p^2+k^2)^{\frac{2-\eta}{2}}}
\esp \\
& = & \frac{\hat D}{2 \pi} p^{-1+\eta} \int_{-\infty}^\infty d \kappa
\frac{e^{i \kappa p(z_1-z_2)}}{(1+\kappa^2)^{\frac{2-\eta}{2}}}
\nonumber .
\enq
For $p(z_1-z_2) \to 0$ this leads to
\beq
\label{G_b_p}
G_{bulk}(p) & = & p^{-1+\eta} \frac{\hat D}{2 \sqrt{\pi}}
\frac{\Gamma(1/2-\eta/2)}{\Gamma(1-\eta/2)} .
\enq
In the limits $L \to \infty$, $z_1+z_2 \to \infty$, $\xi \to \infty$,
and $p(z_1-z_2) \to 0$ the two-point correlation function in the film
reduces to its bulk form. According to Eqs. (\ref{alternatives_4}) and
(\ref{norm_rv}) this implies 
\beq
\label{G_e_hat_d}
{\cal G}_\rv & = & \frac{\hat D}{2 \sqrt{\pi}}
\frac{\Gamma(1/2-\eta/2)}{\Gamma(1-\eta/2)} = B^2 (\xi_0^+)^{d-2+\eta}
\frac{R}{2 \sqrt{\pi}} \frac{\Gamma(1/2-\eta/2)}{\Gamma(1-\eta/2)} \\
& = & B^2 (\xi_0^+)^{d-2+\eta} {\cal U} \nonumber .
\enq
For the three-dimensional Ising model the universal number ${\cal U}$
has the value ${\cal U} \simeq 1.58$.

The knowledge of the perturbative result for $G(p,z_1,z_2,L,t)$ (see
Subappendix \ref{corr_for_z1z2}) enables one to express the nonuniversal
amplitudes ${\cal G}_x$, $x = \mbox{\tt I - IV}$, in terms of 
${\cal G}_\rv$. For example the 
universal ratio ${\cal G}_\rii/{\cal G}_\rv$ is determined by the
normalizations of the scaling functions, i.e., $g_\rii(0,0,0,0)=1$
(Eq. (\ref{norm_rii})) and $g_\rv(\infty,0,\infty,\infty)=1$
(Eq. (\ref{norm_rv})). 
The $\epsilon$-expansion of this ratio is given by 
\beq
\label{G_b_G_e_eps}
{\cal G}_\rii/{\cal G}_\rv & = & 2 \Big( 1 + \epsilon \frac{n+2}{n+8} +
{\cal O}(\epsilon^2) \Big) .
\enq
The amplitudes ${\cal G}_x$, $x = \mbox{\tt I - IV}$, can have bulk, half
space or film character, depending on the normalization limits of the
scaling functions $g_x$. ${\cal G}_\rii$ and ${\cal G}_\rv$ are half
space and bulk amplitudes, respectively. Bulk amplitudes are
independent of the boundary conditions, halfspace amplitudes depend
on the boundary condition of the surface, and film amplitudes
depend on the boundary conditions of both surfaces. Combining
Eqs. (\ref{G_b_G_e_eps}), (\ref{G_e_hat_d}), and (\ref{D_B_xi_0}) we
arrive at  
\beq
\label{G_b_end}
{\cal G}_\rii & = & B^2 (\xi_0^+)^{d-2+\eta} \Big( 1+ \epsilon \frac{n+2}{n+8} + {\cal
O}(\epsilon^2) \Big)  R \frac{1}{\sqrt{\pi}}
\frac{\Gamma(1/2-\eta/2)}{\Gamma(1-\eta/2)} .
\enq
With $R \simeq 3.09$ (Eq. (\ref{D_B_xi_0})) and $\eta \simeq 0.031$
one has for the 3d Ising model 
\beq
\label{G_b_num}
{\cal G}_\rii & = & {\cal R} B^2 (\xi_0^+)^{d-2+\eta} \\
& \simeq & 4.21 B^2 (\xi_0^+)^{d-2+\eta} \nonumber .
\enq
%


\section{One-loop results}
\label{one_loop_results}


\subsection{Correlation functions for $z_1=z_2$}
\label{zz_appendix}

With the abbreviation $\tilde \epsilon = \epsilon \frac{n+2}{n+8}$, so
that $\tilde \epsilon = \frac{1}{3}, \frac{2}{5}, \frac{5}{11}$ for
the Ising, XY, Heisenberg model in $d=3$, the renormalized two-point
correlation function in one-loop order
(Eqs. (\ref{one_loop})-(\ref{renormalization})) is given explicitly as
(see also Ref. \cite{Diehl:83:a}) 
\beq
\label{g_p}
G(p,z,L=\infty,t=0) & = & {\cal G}_\rii z^{1-\eta} g_1(u=pz)
\enq
\beq
&=& \mu^{-\eta} z^{1-\eta} \left( \frac{1-e^{-2u}}{2u} +
\frac{\tilde \epsilon}{4u} \left( -2 Ei(-2u) + e^{2u} Ei(-2u)
+ e^{-2u} Ei(2u) \right) + {\cal O} (\epsilon^2) \right) \nonumber .
\enq
$Ei(x)$ is the exponential integral function. In accordance with the
normalization $g_1(0)=1$ this yields ${\cal G}_\rii = \mu^{-\eta}
(1+\tilde \epsilon + {\cal O} (\epsilon^2))$.

The temperature dependence is described by the scaling function
$g_2(v)$ with $g_2(0)=1$: 
\beq
\label{g_t}
G(p=0,z,L=\infty,t) & = & {\cal G}_\rii z^{1-\eta} g_2(v=z/\xi)
\enq
\beq
& = & \mu^{-\eta} z^{1-\eta} \Big( \frac{1-e^{-2v}}{2v} +
\frac{\tilde \epsilon}{2v} \Big( (e^{-2v}-1)  
K_0(2v) \nonumber \\
&& \qquad \qquad + 2 e^{-2v} \sum_{k=0}^{\infty}
v^{2k+1} \frac{\Psi(k+1) - \ln v +
\frac{1}{2k+1}}{(k!)^2(2k+1)}  
\Big) + {\cal O} (\epsilon^2) \Big) . \nonumber  
\enq
$\Psi(x)$ and $K_0(x)$ denote the psi and Bessel function,
respectively \cite{Gradshteyn:65} (see also
Ref. \cite{Nemirovsky:87}).  

Finally, the dependence of the critical structure factor on the film
thickness is governed by a third scaling function $g_3(w)$,
$g_3(0)=1$, $0 \le w \le 1$:
\beq
\label{g_L}
G(p=0,z,L,t=0) & = & {\cal G}_\rii z^{1-\eta} g_3(w=z/L)
\enq
\beq
& = & \mu^{-\eta} z^{1-\eta} \Big( 1-w + \tilde \epsilon
\Big\{ - \frac{\pi^2}{18} w (1-w)^2 \nonumber \\ 
&& \qquad \qquad
-(1-2w) \Big( 1+C_E+\ln w
+\frac{S^-_{3,2}(w)+I^-_2(w)}{w} \Big) 
\nonumber \\
&& \qquad \qquad 
+ (1-w)\Big( 2+C_E+\ln w -S^+_{2,1}(w)-I^+_1(w)
\Big) \Big\} + {\cal O} (\epsilon^2) \Big) \nonumber
\enq
with the abbreviations
\beq
S^\pm_{k,l} (w) = \sum_{n=k}^{\infty} \frac{B_n(-w) \pm
B_n(w)}{n!(n-l)}, \esp I^\pm_k (w) = \int_1^{\infty}
\frac{dx}{e^x-1} \frac{e^{-xw} \pm e^{xw}}{x^k} .
\enq
$B_n(w)$ are Bernoulli polynomials \cite{Gradshteyn:65}. For the
critical structure factor in the semi-infinite system one has
\beq
\label{normalization}
G(p=0,z,L=\infty,t=0) & = & {\cal G}_\rii z^{1-\eta} = \mu^{-\eta}
z^{1-\eta} \Big( 1 + \te + {\cal O} (\epsilon^2) \Big) .
\enq
From the explicit forms for $g_i$, $i=1,2,3$, in Eqs. (\ref{g_p}) -
(\ref{g_L}) together with $\eta_\| = 2- \tilde \epsilon + {\cal O}
(\epsilon^2)$ one infers the following limiting behaviors:
\beq
\label{g_p_to_0}
g_1(u \to 0) & = & 1 + A_1 u^{-1+\eta_\|} + {\cal O} (u^2) \\
A_1 & = & - \Big[ 1 + \te (1-C_E-\ln 2) + {\cal O}
(\epsilon^2) \Big) \Big] \nonumber ,
\enq
\beq
\label{g_t_to_0}
g_2(v \to 0) & = & 1 + B_1 v^{-1+\eta_\|} + {\cal O} (v^{1/\nu}) \\
B_1 & = & - \Big[ 1 + \te (1-C_E) + {\cal O} (\epsilon^2) \Big]
\nonumber ,
\enq
and
\beq
\label{g_L_to_oo}
g_3(w \to 0) & = & 1 + C_1 w^{-1+\eta_\|} + {\cal O} (w^2) \\
C_1 & = & - \Big[ 1 + \te \Big( \frac{\pi^2}{18}
- C_E +2 (S_2+I_1) - 1 \Big) + {\cal O} (\epsilon^2) \Big] \nonumber 
\enq
where $C_E \approx 0.5772$ is Euler's constant. $S_2$ is given by a
sum over Bernoulli numbers and $I_1$ by an integral:
\beq
\label{S2_I1}
S_2 & = & \sum_{n=2}^{\infty} \frac{B_n}{n!(n-1)} \simeq 8.2877 \cdot
10^{-2} , \quad
I_1 \esp = \esp \int_1^{\infty} dx \frac{1}{e^x-1} \frac{1}{x} \simeq
0.2868 . 
\enq
For the exponentiation of the scaling functions $h_1(u,v)$,
$h_2(u,w)$, and $h_3(v,w)$ we have calculated the amplitude
functions $H_1^{(1)}(\varphi)$, $H_1^{(2)}(\varphi)$, and
$H_1^{(3)}(\varphi)$ (see Eqs. (\ref{h3_transf}) and
(\ref{h3_polar_0})). Their $\epsilon$-expansions are 
\beq
\label{H1_1}
H_1^{(1)} (\varphi) & = & - \Big[ 1 - \tilde \epsilon \Big( \ln \frac{\sin
\varphi}{2} + \frac{\cos \varphi}{2} \ln \frac{1+\cos \varphi}{1-\cos
\varphi} + a_1 \Big) + {\cal O} (\epsilon^2) \Big] \esp , \\
&\varphi& = \arctan ((p \xi)^{-1}) , \nonumber
\enq
\beq
\label{H1_2}
H_1^{(2)} (\varphi) & = & \sin (\varphi) \frac{1+e^{2 \tan
\varphi}}{1-e^{2 \tan \varphi}} + \tilde \epsilon \Big( \sin (\varphi)
\frac{1+e^{2 \tan \varphi}}{1-e^{2 \tan \varphi}} \Big( 1-C_E-\ln (2 \sin
\varphi) -{\cal I}_0(\varphi) \nonumber \\ 
&& + {\cal I}_1(\varphi) \cot \varphi + \frac{1}{12}
\cot^2 \varphi \Big) -\frac{\cos \varphi}{3(1-e^{2 \tan
\varphi})(1-e^{-2 \tan \varphi})} \Big) + {\cal O} (\epsilon^2) \esp ,
\\
&\varphi& = \arctan (p L) , \nonumber
\enq
and
\beq
\label{H1_3}
H_1^{(3)} (\varphi) & = & \sin (\varphi) \frac{1+e^{2 \tan
\varphi}}{1-e^{2 \tan \varphi}} + \tilde \epsilon \Big( \sin (\varphi)
\frac{1+e^{2 \tan \varphi}}{1-e^{2 \tan \varphi}} \Big( 1-C_E-\ln (\sin
\varphi) +{\cal I}^+_1(\varphi) + {\cal I}^-_1(\varphi) \Big) \nonumber \\
&&+\frac{\sin \varphi}{(1-e^{2 \tan
\varphi})(1-e^{-2 \tan \varphi})} (2 \pi + 8 {\cal I}^0_0 (\varphi)
\tan \varphi ) \Big) + {\cal O} (\epsilon^2) \esp , \\
&\varphi& = \arctan (L / \xi) , \nonumber
\enq
with $a_1 \simeq 0.2704$ and the integrals
\beq
\label{I0_of_H1_2}
{\cal I}_0(\varphi) & = & \int_0^\infty \frac{d t}{e^t - 1} \Big(
\frac{1}{t+2\tan \varphi} +
\frac{1}{t-2\tan \varphi} \Big) ,
\enq
\beq
\label{I1_of_H1_2}
{\cal I}_1(\varphi) = \int_0^\infty \frac{d t }{e^t
-1} \Big( \frac{t}{t-2\tan \varphi} - \frac{t}{t+2\tan \varphi}\Big) ,
\enq
\beq
\label{I00_of_H1_3}
{\cal I}^0_0 (\varphi) & = & \int_1^\infty d t \frac{\sqrt{t^2-1}}{e^{2 t
\tan \varphi}-1},
\enq
and
\beq
\label{Ipm_of_H1_3}
{\cal I}^\pm_1 (\varphi) & = & \int_1^\infty d t \frac{\sqrt{t^2-1}}{e^{2 t
\tan \varphi}-1} \frac{t}{t \pm 1}.
\enq


\subsection{Correlation function for $z_1 \not=z_2$}
\label{corr_for_z1z2}

This is the most general case from which all results given above can
be derived. We present $G(p,z_1,z_2,L,t)$ in terms of the scaling
function $g_\ri$ (Eq. (\ref{scaling_form}))
\beq
\label{g_ptL_z1z2}
G(p,z_1,z_2,L,t) & = & {\cal G}_\ri p^{-1+\eta}
g_\ri (x=p\xi,u=z_1/\xi,v=z_2/\xi,y=L/\xi) 
\enq
\beq
& = & \mu^{-\eta} p^{-1+\eta} \Big[ \frac{x}{2a} \Big\{ e^{-a|u-v|} -
e^{-a(u+v)} + \frac{e^{-a(u-v)} + e^{-a(v-u)} -
e^{-a(u+v)} - e^{a(u+v)}}{e^{2ya}-1} \Big\} \nonumber \\
&& +\tilde \epsilon \Big( {\cal J}_0(x,u,v,y) + {\cal J}_\pi(x,u,v,y) +
{\cal J}_1(x,u,v,y) \Big) + {\cal O} (\epsilon^2) \Big] \nonumber
\enq
with
\beq
\label{J0}
&& {\cal J}_0(x,u,v,y) = - \frac{x}{a^3} \int_1^\infty ds
\frac{\sqrt{s^2-1}}{e^{2ys}-1} 
\Big\{ e^{-a|u-v|} (1+a|u-v|) -
e^{-a(u+v)} (1+a(u+v))  \\
&& + \frac{1}{e^{2ya}-1} \Big(
    e^{-a(u-v)} (1+a(u-v)+\frac{2ya}{1-e^{-2ya}}) 
  + e^{-a(v-u)} (1+a(v-u)+\frac{2ya}{1-e^{-2ya}}) \nonumber \\
&& \qquad \qquad
- e^{-a(u+v)} (1+a(u+v)+\frac{2ya}{1-e^{-2ya}})
- e^{ a(u+v)} (1-a(u+v)+\frac{2ya}{1-e^{-2ya}}) \Big) \Big\} 
\nonumber ,
\enq
\beq
\label{Jpi}
{\cal J}_\pi(x,u,v,y) & = & \frac{\pi}{4} \frac{x}{a^2} \frac{1}{(1-e^{-2ya})^2} 
\Big\{ (1+e^{-2ya}) (e^{-a(u+v)} + e^{ a(u+v-2y)}) \\
&&\qquad\qquad\qquad\qquad - 2 e^{-2ya} (e^{-a(v-u)} +e^{-a(u-v)})
\Big\} \nonumber ,
\enq
and
\beq
\label{J1}
{\cal J}_1(x,u,v,y) & = & -\frac{1}{4} \frac{x}{a^2}
\Big\{  e^{-a(u+v)} \frac{J(u,v)}{1-e^{-2ya}} + e^{ a(u+v-2y)}
\frac{J(y-u,y-v)}{1-e^{-2ya}} \\ 
&& - e^{-a(v-u)} \Big( \Theta (v-u) + \frac{1}{e^{2ya}-1} \Big) 
J(y-u,v) \Big) \nonumber \\ 
&& - 
e^{-a(u-v)} \Big( \Theta (u-v) + \frac{1}{e^{2ya}-1} \Big) J(y-v,u)
\Big) 
\Big\} \nonumber
\enq
with $a=\sqrt{1+x^2}$ and 
\beq
J(x_1,x_2) & = &\int_1^{\infty} ds
\frac{\sqrt{s^2-1}}{1-e^{-2ys}} \Big( \Big( \frac{1}{s+a} -
\frac{1}{s} \Big) (e^{-2 x_1 s}+e^{-2 x_2 s}) \\ 
&& - \Big( 
\frac{1}{s-a}-\frac{1}{s} \Big) (e^{-2(y-x_1)s}+e^{-2(y-x_2)s}) \Big)
\esp .
\nonumber 
\enq
%


\subsection{The susceptibility}
\label{excess_app}

The one-loop result of the total susceptibility
(Eq. (\ref{def_exess_sus})) for Dirichlet boundary conditions is
given by $(\gamma_s = \gamma + \nu)$
\beq
\label{fyd}
\chi (L,t) & = & B^2 (\xi_0^+)^{d+1} {\cal R} \Big( \frac{L}{\xi_0^+}
\Big)^{\gamma_s/\nu} f (y=L/\xi) 
\enq
with
\beq
\label{fyd_2}
f(y)&=& y^{-2} \Big[ 1 - \te -
\frac{2}{y} \Big( 1+ \tilde \epsilon \Big( \pi \Big( \frac{1}{2} -
\frac{1}{\sqrt{3}} \Big) -1 \Big) \Big) \\
&& + \frac{4}{y} \frac{1}{e^y+1} - \tilde \epsilon \Big\{ \frac{4}{y}
\frac{1}{e^y+1} + \frac{2}{y}
\frac{e^{-y}}{(1+e^-y)^2} \pi \Big( 1 - \frac{1+e^{-y}}{\sqrt{3}}
\Big) \nonumber \\
&&+\Big( 4+8\frac{e^{-y}}{(1+e^{-y})^2} -\frac{12}{y}
\frac{1-e^{-y}}{1+e^{-y}} \Big) \int\limits_1^\infty d s
\frac{\sqrt{s^2-1}}{e^{2 s y}-1} \nonumber \\
&&+\frac{2}{y}\frac{1-e^{-y}}{1+e^{-y}} \int\limits_1^\infty d s
\frac{\sqrt{s^2-1}}{e^{2 s y}-1} \Big( \frac{1}{s-1} - \frac{1}{s+1} +
\frac{2}{s+1/2} - \frac{2}{s-1/2} \Big) \Big\} \Big] + {\cal O}
(\epsilon^2) \nonumber .
\enq
In the limit $y \to \infty$ the two integrals entering into
Eq. (\ref{fyd_2}) vanish $\sim e^{-2y}$ and therefore they do not
contribute to the terms considered in Eq. (\ref{f_large_y}). However,
in the limit $y \to 0$ these two integrals contribute to the terms
considered in Eq. (\ref{f_small_y}):
\beq
\label{j0}
J_0(y) &=& \int_1^\infty d s \frac{\sqrt{s^2-1}}{e^{2 s y}-1} \\
&=& \frac{\pi^2}{24} y^{-2} - \frac{\pi}{4} y^{-1} + a_2 - \frac{1}{4}
\ln y + {\cal O} (y) \nonumber
\enq
and
\beq
\label{j1} 
J_1 (y) &=&     \int_1^\infty d s \frac{\sqrt{s^2-1}}{e^{2 s y}-1} 
\Big( \frac{1}{s-1} - \frac{1}{s+1} + \frac{2}{s+1/2} -
\frac{2}{s-1/2} \Big) \\
&=& \pi \Big( \sqrt{3} - \frac{3}{2} \Big) y^{-1} -
\frac{\pi}{2\sqrt{3}} + \frac{\sqrt{3}}{12} \pi y + b_2 y^2 -
\frac{\sqrt{3}}{720} \pi y^3 + b_4 y^4 + {\cal O} (y^5) \nonumber 
\enq
with
\beq
\label{a2b2b4}
a_2 & = & \frac{5}{8} - \frac{1}{2} \left( \sum_{n=2}^\infty \frac{B_n}{n!(n-1)}
+ \int_1^\infty \frac{d x}{e^x-1} \frac{1}{x} \right) \esp \simeq
\esp 0.440165 \esp , \\ 
b_2 & = & 6 \left( \sum_{n=0}^\infty \frac{B_n}{n!(n-3)}
+ \int_1^\infty \frac{d x}{e^x-1} \frac{1}{x^3}\right) \esp \simeq
\esp -9.13145 \cdot 10^{-2} \nonumber \esp , \\ 
b_4 & = & 18 \left( \sum_{n=0}^\infty \frac{B_n}{n!(n-5)}
+ \int_1^\infty \frac{d x}{e^x-1} \frac{1}{x^5}\right) \esp \simeq
\esp 5.9879 \cdot 10^{-3} \nonumber \esp .
\enq


\section{Cross section}
\label{details_crosssection}


\subsection{Integration of the asymptotic limits}
\label{s_asypmtotics}

Equation (\ref{film_diffuse}) involves integrals of the following kind:
\beq
\label{fourier_laplace}
\int_0^L dz_1 e^{-\kappa_1 z_1} \int_0^L dz_2 e^{-\kappa_2 z_2} \esp
G(p,z_1,z_2,L,t) ,
\enq
where $\kappa_j \in \{ \pm i( q_2(\alpha_f) \pm q_2(\alpha_i)), \pm i
( q^\ast_2(\alpha_f) \pm q^\ast_2(\alpha_i))\}$, $j=1,2$, (see
Eqs. (\ref{wavefield}) as well as (\ref{film_functions}) and
(\ref{si_functions}) in Appendix \ref{ssss}) and $\kappa_j (\alpha)
\equiv K f_j (\alpha_i, \alpha_f, \alpha_{c12})$ (see above). The
asymptotic behavior of 
Eqs. (\ref{corr_p_z1z2}), (\ref{corr_t_z1z2}), and (\ref{corr_L_z1z2})
can be summarized by the formula 
\beq
\label{g_as}
G_{as}(p,z_1,z_2,L,t) & = & {\cal G}_\rii \Big\{ 
 \Theta(z_2-z_1) z_1^{1-\eta} \Big( \frac{z_2}{z_1}
\Big)^{1-\eta_\perp}  
+\Theta(z_1-z_2) z_2^{1-\eta} \Big( \frac{z_1}{z_2}
\Big)^{1-\eta_\perp}  \\
&+&{\cal C} \Big( \Theta(z_2-z_1) z_1^{1-\eta} \Big( \frac{z_2}{z_1}
\Big)^{1-\eta_\perp} z_2^{-1+\eta_\|}
+\Theta(z_1-z_2) z_2^{1-\eta} \Big( \frac{z_1}{z_2}
\Big)^{1-\eta_\perp} z_1^{-1+\eta_\|} \Big) \Big\} \nonumber \\
&=& {\cal G}_\rii \Big\{ d(z_1,z_2) + a(p,z_1,z_2,L,t) \Big\} \nonumber .
\enq
The expression $d(z_1,z_2)$ corresponds to the leading contribution
${\cal C}=0$. ${\cal C}$ is an abbreviation for the three quantities
$A_1 p^{-1+\eta_\|}$ for $t=0$ and $L=\infty$,
$B_1 (1/\xi)^{-1+\eta_\|}$ for $p=0$ and $L=\infty$, and
$C_1 (1/L)^{-1+\eta_\|}$ for $p=t=0$ in Eqs. (\ref{corr_p_z1z2}),
(\ref{corr_t_z1z2}), and (\ref{corr_L_z1z2}). For the quasi-Laplace
transform of the contribution $d(z_1,z_2)$,
\beq
\label{dd_def}
\bar {\cal D} (\kappa_1,\kappa_2,L) & = & \int_0^L dz_1 e^{-\kappa_1 z_1}
\int_0^L dz_2 e^{-\kappa_2 z_2} \esp d(z_1,z_2) 
\enq
one finds with $f_j = \kappa_j/K$, $j=1,2$,
\beq
\label{dd}
\bar {\cal D}(\kappa_1,\kappa_2,L) & \equiv & 
\bar {\cal D}(f_1,f_2,LK) =  K^{-3+\eta} \Big[ \frac{1}{f_1 f_2 (f_1+f_2)}  
- \frac{e^{-f_1 LK}}{f_2^2 f_1}
- \frac{e^{-f_2 LK}}{f_1^2 f_2} \\
&&+ e^{-(f_1+f_2)LK} \Big( \frac{LK}{f_1 f_2} 
+ \frac{f_1^2+f_1 f_2 + f_2^2}{f_1^2 f_2^2 (f_1 + f_2)} \Big)
\nonumber \\
&& + \frac{\tilde \epsilon}{2} \Big\{ \frac{-2}{f_1 f_2 (f_1 + f_2)} 
- e^{-(f_1 + f_2)LK} \frac{f_1^2 + f_2^2}{f_1^2 f_2^2 (f_1 + f_2)}
\nonumber \\
&& 
+ \frac{1}{f_1^2 f_2} \Big( \ln (f_1/f_2 + 1) 
+ Ei(1,(f_1+f_2)LK) - Ei(1,f_2 LK) \Big) \nonumber \\
&&
+ \frac{1}{f_2^2 f_1} \Big( \ln (f_2/f_1 + 1) 
+ Ei(1,(f_1+f_2)LK) - Ei(1,f_1 LK) \Big) \nonumber \\
&&
+ \frac{e^{-f_1 LK}}{f_2^2 f_1} \Big( 1 - C_E - \ln f_2 - \ln LK -
Ei(1,f_2 LK) \Big) \nonumber \\
&&
+ \frac{e^{-f_2 LK}}{f_1^2 f_2} \Big( 1 - C_E - \ln f_1 - \ln LK -
Ei(1,f_1 LK) \Big) \Big\} + {\cal O} (\epsilon^2) \Big]\nonumber .
\enq
In the limiting case of a semi-infinite film $(L \to \infty)$
Eq. (\ref{dd}) reduces to
\beq
\label{dd_si}
\bar {\cal D}(f_1,f_2,LK=\infty) & = & K^{-3+\eta} \Big[ 
\frac{1}{f_1 f_2 (f_1 + f_2)}  
\Big( 1 + \tilde \epsilon \Big\{ 1 - \\
&&
-\frac{f_1+f_2}{2 f_1} \ln (1+f_1/f_2)  
- \frac{f_1+f_2}{2 f_2} \ln (1+f_2/f_1) \Big\} \Big) + {\cal O}
(\epsilon^2) \Big] \nonumber .
\enq
The corresponding expression for the leading correction term
\beq
\label{aa_def}
\bar {\cal A} (f_1, f_2, LK) & = & \int_0^L d z_1 e^{-\kappa_1 z_1}
\int_0^L d z_2 e^{-\kappa_2 z_2} a(p, z_1, z_2, L, t)
\enq
is
\beq
\label{aa}
\bar {\cal A}(f_1,f_2,LK) & = & K^{-3+\eta} K^{1-\eta_\|} {\cal C}
\Big[ \frac{1}{f_1^2 f_2^2} \Big( 1  
-e^{-f_1 LK}(1+f_1 LK) -e^{-f_2 LK}(1+f_2 LK) \\
&& \qquad \qquad
+e^{-(f_1+f_2)LK}(1+(f_1+f_2)LK+f_1 f_2 (LK)^2) \Big) \nonumber \\
&&
-\frac{\tilde \epsilon}{2} \frac{1}{f_1^2 f_2^2} \Big\{ 2-2C_E - \ln
f_1 f_2 - Ei(1,f_1 LK) - Ei(1,f_2 LK) \nonumber \\
&&
+e^{-f_1 LK} \Big( -1 + (1+f_1 LK)(C_E-1+\ln f_2 -\ln LK + Ei(1,f_2
LK)) \Big) \nonumber \\
&&
+e^{-f_2 LK} \Big( -1 + (1+f_2 LK)(C_E-1+\ln f_1 -\ln LK + Ei(1,f_1
LK)) \Big) \nonumber \\
&&
+e^{-(f_1+f_2)LK} \Big( LK(f_1+f_2)(1+2 \ln LK) \nonumber \\
&& \qquad \qquad \qquad
+ 2(1+ \ln LK) + 2 f_1
f_2 (LK)^2 \ln LK \Big) \Big\} + {\cal O} (\epsilon^2) \Big]
\nonumber 
\enq
with the semi-infinite limit $L \to \infty$
\beq
\label{aa_si}
\bar {\cal A}(f_1,f_2,LK=\infty) & = & K^{-3+\eta} K^{1-\eta_\|} 
{\cal C}' \Big[
\frac{1}{f_1^2 f_2^2} \Big( 1 + \tilde 
\epsilon \Big\{ C_E - 1 + \frac{1}{2} \ln f_1 f_2 \Big\}  \Big) + 
{\cal O} (\epsilon^2) \Big] ,
\enq
where ${\cal C}'$ is an abbreviation for the two quantities
$A_1 p^{-1+\eta_\|}$, for $t=0$ and $L=\infty$, and
$B_1 (1/\xi)^{-1+\eta_\|}$, for $p=0$ and $L=\infty$. Distant wall
corrections to the semi-infinite system vanish exponentially. 
In order to obtain the analytic expressions in Eqs. (\ref{dd}) and
(\ref{aa}) we have expanded $d(z_1,z_2)$ and $a(p,z_1,z_2,L,t)$ in
terms of $\epsilon$ using for the
$\epsilon$-expansion of the exponents $\eta_\| = 2 - \tilde \epsilon +
{\cal O} (\epsilon^2) $, $\eta_\perp = 1 - \tilde \epsilon/2 + 
{\cal O} (\epsilon^2)$, and $\eta = {\cal O} (\epsilon^2)$. 
The function $Ei(1,z)$ is the exponential integral defined by
\beq
\label{ei1z}
Ei(1,z) & = & \int_1^\infty \esp \frac{e^{-z t}}{t} \esp dt = - Ei(-z) .
\enq
This function is numerically more suitable than the exponential
integral $Ei(z)$
\beq
\label{eiz}
Ei(z) & = & - \int_{-z}^\infty \esp \frac{e^{-t}}{t} \esp dt
\enq
appearing in the formulae for the correlation function. 
  

\subsection{Integration of the mean-field correlation function}
\label{mft_crosssection}

Equation (\ref{fourier_laplace}) for the full mean-field correlation
function yields
\beq
\label{mft_cross_pLt}
{\cal S} (b = \sqrt{(p/K)^2+(\xi K)^{-2}},LK,f_1,f_2) & = &\int_0^L dz_1
e^{-\kappa_1 z_1} \int_0^L dz_2 e^{-\kappa_2 z_2} \esp G(p,z_1,z_2,L,t)
\enq
\beq
&=& {\cal G}_\rii K^{-3+\eta} \frac{1}{2 b} \Big\{ 
 \frac{1-e^{-(f_1+b)LK}}{(f_1+b)(f_2-b)} 
-\frac{1-e^{-(f_1+f_2)LK}}{(f_1+f_2)(f_2-b)}
+\frac{1-e^{-(f_1+f_2)LK}}{(f_1+f_2)(f_2+b)} \nonumber \\
&&\qquad \qquad 
-\frac{e^{-(f_2+b)LK}-e^{-(f_1+f_2)LK}}{(f_1-b)(f_2+b)}
-\frac{(1-e^{-(f_1+b)LK})(1-e^{-(f_2+b)LK})}{(f_1+b)(f_2+b)}
\nonumber \\
&&
+\frac{1}{e^{2 b LK}-1} \Big( 
 \frac{(1-e^{-(f_1+b)LK})(1-e^{-(f_2-b)LK})}{(f_1+b)(f_2-b)}
+\frac{(1-e^{-(f_1-b)LK})(1-e^{-(f_2+b)LK})}{(f_1-b)(f_2+b)} 
\nonumber \\
&&\qquad
-\frac{(1-e^{-(f_1+b)LK})(1-e^{-(f_2+b)LK})}{(f_1+b)(f_2+b)}
-\frac{(1-e^{-(f_1-b)LK})(1-e^{-(f_2-b)LK})}{(f_1-b)(f_2-b)} \Big)
\Big\} \nonumber 
\enq
using the notation of Subappendix \ref{s_asypmtotics}. The above
formula exhibits the following limiting expressions:
\beq
\label{mft_cross_p0_Loo_t0}
{\cal S} (b=0,LK=\infty,f_1,f_2) & = & {\cal G}_\rii K^{-3+\eta}
\frac{1}{(f_1+f_2)f_1 f_2} \esp ,
\enq
\beq
\label{mft_cross_p_Loo_t}
{\cal S} (b,LK=\infty,f_1,f_2) & = & {\cal G}_\rii K^{-3+\eta}
\frac{1}{(f_1+f_2)(f_1+b)(f_2+b)} \esp ,
\enq
and
\beq
\label{mft_cross_p0_L_t0}
{\cal S} (b=0,LK,f_1,f_2) & = & {\cal G}_\rii K^{-3+\eta}
\Big\{ \frac{1}{(f_1+f_2)f_1 f_2} - \frac{1}{f_1^2 f_2^2} \frac{1}{LK}
+ \frac{e^{-f_1 LK}+e^{-f_2 LK}}{f_1^2 f_2^2} \frac{1}{LK} \\
&& \qquad \qquad
-\Big( \frac{1}{(f_1+f_2)f_1 f_2} + \frac{1}{f_1^2 f_2^2}
\frac{1}{LK} \Big) e^{-(f_1+f_2)LK} \Big\} \esp . \nonumber
\enq
Equations (\ref{mft_cross_p0_Loo_t0}) - (\ref{mft_cross_p0_L_t0}) lead
to the following three cusp singularities:
\beq
\label{mft_pk_cusp}
\frac{{\cal S} (p \to 0, t=0, LK=\infty, f_1, f_2)}{{\cal S}(p=0, t=0,
  LK=\infty, f_1,f_2)} & = & 1 - \left[ \frac{f_1+f_2}{f_1 f_2}
  \right]^{-1+\eta_\|} \left(
  \frac{p}{K} \right)^{-1+\eta_\|} + {\cal O} (p^2) \esp ,
\enq
\beq
\label{mft_xik_cusp}
\frac{{\cal S} (p = 0, t \to 0, LK=\infty, f_1, f_2)}{{\cal S}(p=0, t=0,
  LK=\infty, f_1,f_2)} & = & 1 - \left[ \frac{f_1+f_2}{f_1 f_2}
  \right]^{-1+\eta_\|} \left(
  \frac{1}{\xi K} \right)^{-1+\eta_\|} + {\cal O} (\xi^{1/\nu}) \esp ,
\enq
and
\beq
\label{mft_Lk_cusp}
\frac{{\cal S} (p = 0, t = 0, LK \to \infty, \tilde f_1, \tilde
  f_2)}{{\cal S}(p=0, t=0, LK=\infty, f_1,f_2)} & = & 
\frac{f_1 f_2 (f_1+f_2)}{\tilde f_1 \tilde f_2 (\tilde f_1 + \tilde
  f_2)} \\
&&
- \left[ \frac{f_1 f_2 (f_1+f_2)}{\tilde f_1^2 \tilde f_2^2}
  \right]^{-1+\eta_\|} \left(
  \frac{1}{L K} \right)^{-1+\eta_\|} + {\cal O} (e^{-L}) \esp .
  \nonumber 
\enq
We note that the last two arguments of the nominator and denominator
on the left hand side of Eq. (\ref{mft_Lk_cusp}) are in general, as
indicated, different from each other. For $L= \infty$ the variables
$f_j$ are given by $-i(q_2(\alpha_f)+q_2(\alpha_i))/K$ or
$i(q^\ast_2(\alpha_i) + q^\ast_2(\alpha_f))/K$ whereas 
for $L < \infty$  the variables $\tilde f_j$ are given by 
$-i(k q_2(\alpha_f) + l q_2(\alpha_i))/K$ or 
$i(m q^\ast_2(\alpha_i) + n q^\ast_2(\alpha_f))/K$ with any
combination of $k,l,m,n = \pm 1$ (see the exponentials in the last
lines of Eqs. (\ref{film_functions}) and (\ref{si_functions}) in
Appendix \ref{ssss}).

For the exponentiation of the $p$-$L$, $\xi$-$L$, and $p$-$\xi$
dependences we introduce polar coordinates (see
 Eq. (\ref{polar_coordinates}))
\beq
\label{pL_polar_coord}
\omega & = & \sqrt{(p/K)^2+(LK)^{-2}} , \esp \varphi = \arctan (p L) 
\esp , \\
\frac{1}{LK} & = & \omega \cos \varphi , \esp \frac{p}{K} = \omega
\sin \varphi \nonumber \esp 
\enq
leading to the scaling function ${\cal S} (\omega, \varphi, \tilde
f_1, \tilde f_2) = {\cal S} (p/K=\omega \sin \varphi, LK = (\omega
\cos \varphi)^{-1}, \tilde f_1, \tilde f_2)$ and to its asymptotic
expansion 
\beq
\label{mft_cross_pL}
\frac{{\cal S} (\omega \to 0,
  \varphi,\tilde f_1,\tilde f_2)}{{\cal S}(\omega=0,f_1,f_2)} & = &  
\frac{f_1 f_2 (f_1+f_2)}{\tilde f_1 \tilde f_2 (\tilde f_1 + \tilde
  f_2)} \\
&&
- \sin (\varphi) \frac{1+e^{-2 \tan \varphi}}{1-e^{-2 \tan \varphi}}
  \left[ \frac{f_1 f_2 (f_1+f_2)}{\tilde f_1^2 \tilde f_2^2} 
  \right]^{-1+\eta_\|}
\esp \omega^{-1+\eta_\|} + \dots \esp . \nonumber
\enq
Because in this section we consider only mean-field scaling functions,
a simple substitution of the scaling variable $p/K$ by $1/\xi K$ in
Eq. (\ref{pL_polar_coord}) leads to the same result for the $\xi$-$L$
dependences. The semi-infinite system is described by the coordinates
\beq
\label{pt_polar_coord}
\omega & = & \sqrt{(p/K)^2+(\xi K)^{-2}} , \esp \varphi = \arctan
\frac{1}{p \xi} \esp , \\
\frac{1}{\xi K} & = & \omega \sin \varphi , \esp \frac{p}{K} = \omega
\cos \varphi \nonumber 
\enq
leading to the asymptotic behavior of the scaling function
\beq
\label{mft_cross_pt}
\frac{{\cal S} (\omega \to 0,f_1,f_2)}{{\cal S}(\omega=0,f_1,f_2)} & = & 1 
- \left[ \frac{f_1+f_2}{f_1 f_2} \right]^{-1+\eta_\|} \esp
\omega^{-1+\eta_\|} + \dots \esp , 
\enq
which is independent of $\varphi$.


\section{Products of wave functions}
\label{ssss}
In order to illustrate the type of transformations appearing in
Eqs. (\ref{film_diffuse}) and (\ref{fourier_laplace}) we present the
explicit expression for the product of wave functions in
Eq. (\ref{wavefield}): 
\beq
\label{film_functions}
&&\psi_f(z_1) \psi_i(z_1) \psi_i^\ast(z_2) \psi_f^\ast(z_2) \\
&=& \Big( s_+(\alpha_f)e^{iq_2(\alpha_f)z_1} +
s_-(\alpha_f)e^{-iq_2(\alpha_f)z_1} \Big) 
    \Big( s_+(\alpha_i)e^{iq_2(\alpha_i)z_1} +
s_-(\alpha_i)e^{-iq_2(\alpha_i)z_1} \Big) \nonumber \\
&& \times \Big( s_+^\ast(\alpha_i)e^{-iq_2^\ast(\alpha_i)z_2} +
s_-^\ast(\alpha_i)e^{iq_2^\ast(\alpha_i)z_2} \Big) 
    \Big( s_+^\ast(\alpha_f)e^{-iq_2^\ast(\alpha_f)z_2} +
s_-^\ast(\alpha_f)e^{iq_2^\ast(\alpha_f)z_2} \Big) \nonumber \\ 
&=& \sum_{k,l,m,n=\pm} s_k(\alpha_f) s_l(\alpha_i) s_m^\ast(\alpha_i)
s_n^\ast(\alpha_f) 
e^{ i(k q_2(\alpha_f)     + l q_2(\alpha_i))z_1}
e^{-i(m q_2^\ast(\alpha_i)+ n q_2^\ast(\alpha_f))z_2} \nonumber 
\enq
where $s$ and $q$ are defined in Eqs. (\ref{ssrt}) and (\ref{q1q2q3}). 
Thus the scattering cross section is proportional to a sum of 16 terms 
involving integrations over $z_1$ and $z_2$. 

For the limiting case of a semi-infinite halfspace one has
\beq
\label{si_functions}
&&\psi_{\infty/2}^f(z_1) \psi_{\infty/2}^i(z_1)
\psi_{\infty/2}^{(i)\ast}(z_2) \psi_{\infty/2}^{(f)\ast}(z_2) \\ 
&=& t_{si}(\alpha_f) e^{iq_2(\alpha_f)z_1}
t_{si}(\alpha_i) e^{iq_2(\alpha_i)z_1}
t^\ast_{si}(\alpha_i)e^{-iq_2^\ast(\alpha_i)z_2}
t^\ast_{si}(\alpha_f) e^{-iq_2^\ast(\alpha_f)z_2} \nonumber \\
&=& |t_{si}(\alpha_f)|^2|t_{si}(\alpha_i)|^2 
e^{i(q_2(\alpha_f)     + q_2(\alpha_i))z_1}
e^{-i(q_2^\ast(\alpha_i)+q_2^\ast(\alpha_f))z_2} 
\nonumber .
\enq
In this limit the above sum reduces to a single term.


%
%
%
%
%
%
%
%
%
%
%
%
%
%
%
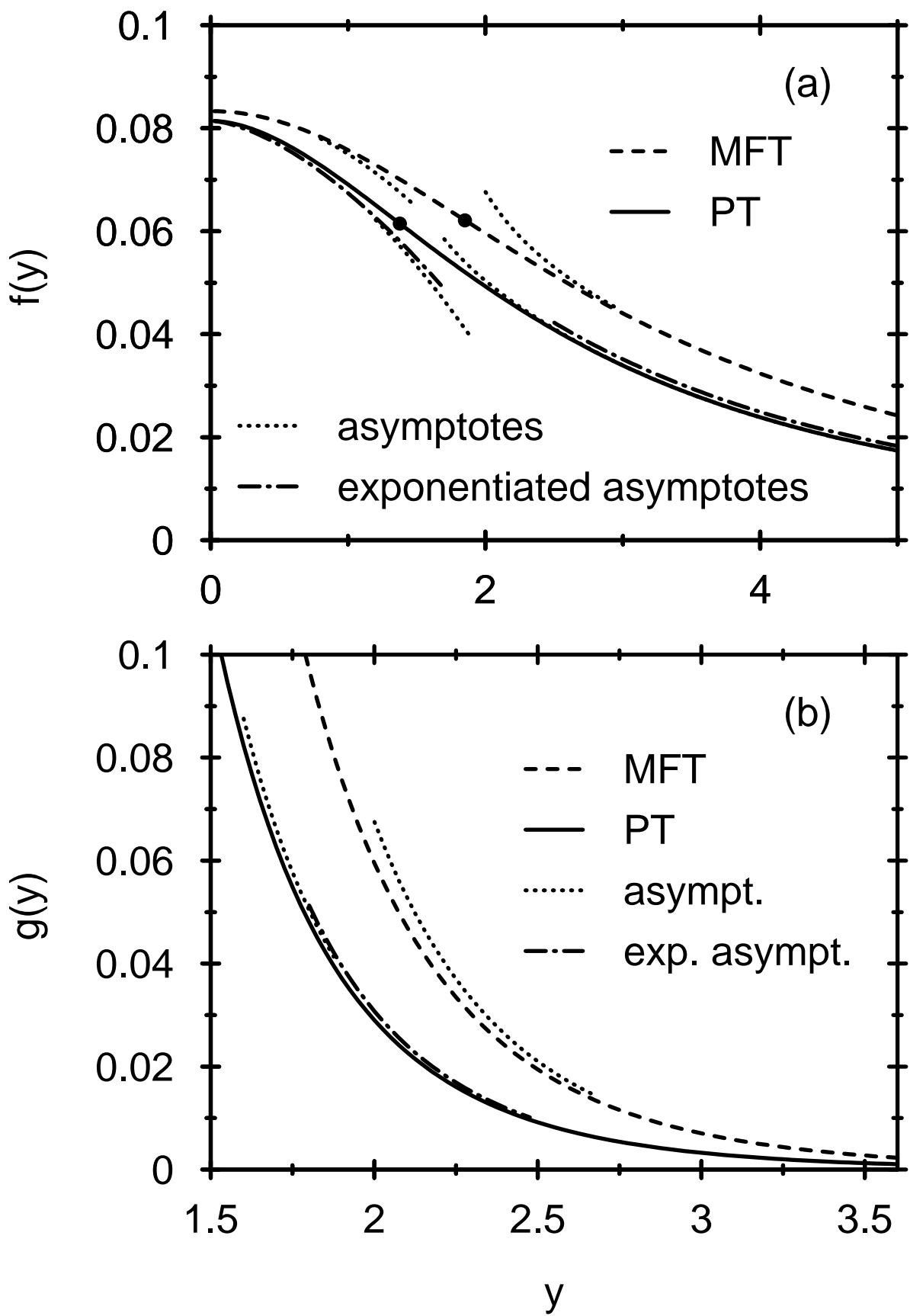
\begin{figure}
\caption{
Universal scaling functions $f(y)$ ((a),
Eq. (\ref{scaling_excess_sus})) and $g(y)$ ((b),
Eqs. (\ref{chi_SD_transparent}) and (\ref{chi_SD_f_g})) of the film
susceptibility for Dirichlet boundary conditions at both surfaces. The
dashed lines are the mean-field (MFT) results whereas the full lines include
non-Gaussian fluctuations obtained by perturbation theory (PT) in first
order $\epsilon$ (Eqs. (\ref{fyd}), 
(\ref{fyd_2}), and (\ref{chi_SD_f_g})). The dotted lines indicate the
asymptotic behaviors of $f(y \to 0)$, $f(y \to \infty)$, $g(y \to 0)$,
and $g(y \to \infty)$ given by Eqs. (\ref{f_small_y}),
(\ref{f_large_y}), (\ref{g_for_small_y}), and (\ref{chi_vanish}),
respectively. The dotted lines correspond to the $\epsilon$-expansion
of these asymptotic behaviors up to ${\cal O}(\epsilon)$ in order to
be compatible with the full scaling functions $f(y)$ and $g(y)$ whose
$\epsilon$-expansions up to ${\cal O} (\epsilon)$ are shown here as
full lines. The dash-dotted curves show the exponentiated forms of the
asymptotic behaviors given by Eqs. (\ref{f_small_y}),
(\ref{f_large_y}), (\ref{g_for_small_y}), and (\ref{chi_vanish}) using
the $\epsilon$-expansion results for the amplitudes but the best
available numbers $\eta = 0.031$ and $\nu = 0.630$ for the critical
exponents. $f(y)$ has a turning point $(\bullet)$ at $y=1.851$ in MFT
and at $y=1.376$ in ${\cal O}(\epsilon)$; $f(0)={\cal D}$
(Eq. (\ref{D_sus})). 
}
\label{fig_f_d}
\end{figure}
%
%
%
%
%
%
%
%
%
%

\end{document}